\documentclass[twocolumn,eqsecnum,aps,floatfix]{revtex4} 

\usepackage{graphicx}
\usepackage{dcolumn}
\usepackage{amsmath}
\usepackage{amssymb}

\makeatletter
\def\btt#1{\texttt{\@backslashchar#1}}%
\DeclareRobustCommand\bblash{\btt{\@backslashchar}}%
\makeatother

\begin{document}

\title{Solar neutrino measurements in Super--Kamiokande--I}

\newcommand{\authoraticrr}{$^{1}$}
\newcommand{\authoratkashiwa}{$^{2}$}
\newcommand{\authoratbu}{$^{3}$}
\newcommand{\authoratbnl}{$^{4}$}
\newcommand{\authoratuci}{$^{5}$}
\newcommand{\authoratcsu}{$^{6}$}
\newcommand{\authoratcnm}{$^{7}$}
\newcommand{\authoratduke}{$^{8}$}
\newcommand{\authoratgmu}{$^{9}$}
\newcommand{\authoratgifu}{$^{10}$}
\newcommand{\authoratuh}{$^{11}$}
\newcommand{\authoratind}{$^{12}$}
\newcommand{\authoratkek}{$^{13}$}
\newcommand{\authoratkobe}{$^{14}$}
\newcommand{\authoratkyoto}{$^{15}$}
\newcommand{\authoratlanl}{$^{16}$}
\newcommand{\authoratlsu}{$^{17}$}
\newcommand{\authoratumd}{$^{18}$}
\newcommand{\authoratduluth}{$^{20}$}
\newcommand{\authoratmiyagi}{$^{21}$}
\newcommand{\authoratnagoya}{$^{22}$}
\newcommand{\authoratsuny}{$^{23}$}
\newcommand{\authoratniigata}{$^{24}$}
\newcommand{\authoratosaka}{$^{25}$}
\newcommand{\authoratokayama}{$^{26}$}
\newcommand{\authoratseoul}{$^{27}$}
\newcommand{\authoratshizuokasc}{$^{28}$}
\newcommand{\authoratshizuoka}{$^{29}$}
\newcommand{\authoratskk}{$^{30}$}
\newcommand{\authorattohoku}{$^{31}$}
\newcommand{\authorattokyo}{$^{32}$}
\newcommand{\authorattokai}{$^{33}$}
\newcommand{\authorattit}{$^{34}$}
\newcommand{\authoratwarsaw}{$^{35}$}
\newcommand{\authoratuw}{$^{36}$}

\newcommand{\addressoficrr}[1]{$^{1}$ #1 }
\newcommand{\addressofkashiwa}[1]{$^{2}$ #1 }
\newcommand{\addressofbu}[1]{$^{3}$ #1 }
\newcommand{\addressofbnl}[1]{$^{4}$ #1 }
\newcommand{\addressofuci}[1]{$^{5}$ #1 }
\newcommand{\addressofcsu}[1]{$^{6}$ #1 }
\newcommand{\addressofcnm}[1]{$^{7}$ #1 }
\newcommand{\addressofduke}[1]{$^{8}$ #1 }
\newcommand{\addressofgmu}[1]{$^{9}$ #1 }
\newcommand{\addressofgifu}[1]{$^{10}$ #1 }
\newcommand{\addressofuh}[1]{$^{11}$ #1 }
\newcommand{\addressofind}[1]{$^{12}$ #1 }
\newcommand{\addressofkek}[1]{$^{13}$ #1 }
\newcommand{\addressofkobe}[1]{$^{14}$ #1 }
\newcommand{\addressofkyoto}[1]{$^{15}$ #1 }
\newcommand{\addressoflanl}[1]{$^{16}$ #1 }
\newcommand{\addressoflsu}[1]{$^{17}$ #1 }
\newcommand{\addressofumd}[1]{$^{18}$ #1 }
\newcommand{\addressofmit}[1]{$^{19}$ #1 }
\newcommand{\addressofduluth}[1]{$^{20}$ #1 }
\newcommand{\addressofmiyagi}[1]{$^{21}$ #1 }
\newcommand{\addressofnagoya}[1]{$^{22}$ #1 }
\newcommand{\addressofsuny}[1]{$^{23}$ #1 }
\newcommand{\addressofniigata}[1]{$^{24}$ #1 }
\newcommand{\addressofosaka}[1]{$^{25}$ #1 }
\newcommand{\addressofokayama}[1]{$^{26}$ #1 }
\newcommand{\addressofseoul}[1]{$^{27}$ #1 }
\newcommand{\addressofshizuokasc}[1]{$^{28}$ #1 }
\newcommand{\addressofshizuoka}[1]{$^{29}$ #1 }
\newcommand{\addressofskk}[1]{$^{30}$ #1 }
\newcommand{\addressoftohoku}[1]{$^{31}$ #1 }
\newcommand{\addressoftokyo}[1]{$^{32}$ #1 }
\newcommand{\addressoftokai}[1]{$^{33}$ #1 }
\newcommand{\addressoftit}[1]{$^{34}$ #1 }
\newcommand{\addressofwarsaw}[1]{$^{35}$ #1 }
\newcommand{\addressofuw}[1]{$^{36}$ #1 }

\author{
J.~Hosaka\authoraticrr,
K.~Ishihara\authoraticrr,
J.~Kameda\authoraticrr,
Y.~Koshio\authoraticrr,
A.~Minamino\authoraticrr,
C.~Mitsuda$^{1,a}$,
M.~Miura\authoraticrr, 
S.~Moriyama\authoraticrr, 
M.~Nakahata\authoraticrr, 
T.~Namba$^{1,b}$
% ...
\\
\smallskip
(The Super-Kamiokande Collaboration) \\ 
\smallskip
\footnotesize
\it
\addressoficrr{Kamioka Observatory, Institute for Cosmic Ray Research, University of Tokyo, Kamioka, Gifu 506-1205, Japan}\\
\addressofkashiwa{Research Center for Cosmic Neutrinos, Institute for Cosmic Ray Research, University of Tokyo, Kashiwa, Chiba 277-8582, Japan}\\
\addressofbu{Department of Physics, Boston University, Boston, MA 02215, USA}\\
\addressofbnl{Physics Department, Brookhaven National Laboratory, Upton, NY 11973, USA}\\
\addressofuci{Department of Physics and Astronomy, University of California, Irvine, Irvine, CA 92697-4575, USA }\\
\addressofcsu{Department of Physics, California State University, Dominguez Hills, Carson, CA 90747, USA}\\
\addressofcnm{Department of Physics, Chonnam National University, Kwangju 500-757, Korea}\\
\addressofduke{Department of Physics, Duke University, Durham NC 27708, USA}\\
\addressofgmu{Department of Physics, George Mason University, Fairfax, VA 22030, USA }\\
\addressofgifu{Department of Physics, Gifu University, Gifu, Gifu 501-1193, Japan}\\
\addressofuh{Department of Physics and Astronomy, University of Hawaii, Honolulu, HI 96822, USA}\\
\addressofind{Department of Physics, Indiana University, Bloomington, IN 47405-7105, USA}\\
\addressofkek{High Energy Accelerator Research Organization (KEK), Tsukuba, Ibaraki 305-0801, Japan }\\
\addressofkobe{Department of Physics, Kobe University, Kobe, Hyogo 657-8501, Japan}\\
\addressofkyoto{Department of Physics, Kyoto University, Kyoto, Kyoto 606-8502, Japan}\\
\addressoflanl{Physics Division, P-23, Los Alamos National Laboratory, Los Alamos, NM 87544, USA }\\
\addressoflsu{Department of Physics and Astronomy, Louisiana State University, Baton Rouge, LA 70803, USA }\\
\addressofumd{Department of Physics, University of Maryland, College Park, MD 20742, USA }\\
\addressofmit{Department of Physics, Massachusetts Institute of Technology, Cambridge, MA 02139, USA}\\
\addressofduluth{Department of Physics, University of Minnesota Duluth, MN 55812-2496, USA}\\
\addressofmiyagi{Department of Physics, Miyagi University of Education, Sendai, Miyagi 980-0845, Japan}\\
\addressofnagoya{Department of Physics, Nagoya University, Nagoya, Aichi 464-8602, Japan}\\
\addressofsuny{Department of Physics and Astronomy, State University of New York, Stony Brook, NY 11794-3800, USA}\\
\addressofniigata{Department of Physics, Niigata University, Niigata, Niigata 950-2181, Japan }\\
\addressofokayama{Department of Physics, Okayama University, Okayama, Okayama 700-8530, Japan }\\
\addressofosaka{Department of Physics, Osaka University, Toyonaka, Osaka 560-0043, Japan}\\
\addressofseoul{Department of Physics, Seoul National University, Seoul 151-742, Korea}\\
\addressofshizuokasc{International and Cultural Studies, Shizuoka Seika College, Yaizu, Shizuoka 425-8611, Japan}\\
\addressofshizuoka{Department of Systems Engineering, Shizuoka University, Hamamatsu, Shizuoka 432-8561, Japan}\\
\addressofskk{Department of Physics, Sungkyunkwan University, Suwon 440-746, Korea}\\
\addressoftohoku{Research Center for Neutrino Science, Tohoku University, Sendai, Miyagi 980-8578, Japan}\\
\addressoftokyo{The University of Tokyo, Bunkyo, Tokyo 113-0033, Japan }\\
\addressoftokai{Department of Physics, Tokai University, Hiratsuka, Kanagawa 259-1292, Japan}\\
\addressoftit{Department of Physics, Tokyo Institute for Technology, Meguro, Tokyo 152-8551, Japan }\\
\addressofwarsaw{Institute of Experimental Physics, Warsaw University, 00-681 Warsaw, Poland }\\
\addressofuw{Department of Physics, University of Washington, Seattle, WA 98195-1560, USA}\\
}


\date{\today}

\begin{abstract}
The details of Super--Kamiokande--I's solar neutrino analysis are given.
Solar neutrino measurement in Super--Kamiokande is a high statistics
collection of $^8$B solar neutrinos via neutrino-electron scattering.
The analysis method and results of the 1496 day data sample are presented.
The final oscillation results for the data are also presented.
\end{abstract}


\maketitle

%
%
\section{Introduction}\label{sec:intro}

 Super--Kamiokande [Super--K, SK] is an imaging water Cherenkov 
detector, which detects $^8$B solar neutrinos by electron scattering.
Due to its unprecedented fiducial size of 22.5 kilotons, Super--K has 
the advantage of making the current highest statistics measurements of
solar neutrinos. It enables us to determine, with high precision,
measurements of the solar neutrino flux,
energy spectrum, and possible time variations of the flux.

 Super--K started taking data in April 1996, and the solar neutrino 
results of the first phase of SK, which ended in July 2001 and is henceforth 
referred to as ``SK--I,'' are described in this paper.
In Section~\ref{sec:detector} and Section~\ref{sec:simulation},
details of the SK detector and its simulation are given.
After the event reconstruction method, detector calibration, and  
sources of background are described in Sections~\ref{sec:reconstruction},
~\ref{sec:calibration}, and~\ref{sec:bg} respectively,
the data analysis method and results are described in Section\ref{sec:red}
and~\ref{sec:result}. Finally, in Section~\ref{sec:osc} the solar neutrino
oscillation analysis is discussed.

%
%
\section{Super--Kamiokande detector}\label{sec:detector}
\subsection{Detector outline}

\begin{figure}[hptb]
 \begin{center}
  \includegraphics[scale=0.3]{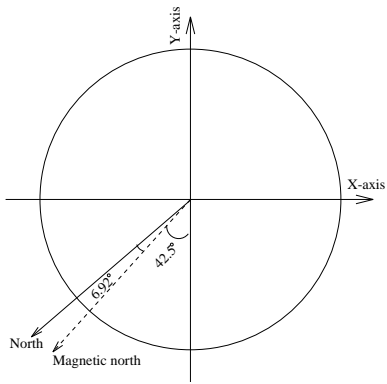}
  \caption{Coordinates of the Super--Kamiokande detector.
  The Z-axis is defined as the upward direction, 
pointing away from the center of the earth.}
  \label{fig:detdir}
 \end{center}
\end{figure}

 As has been discussed in much greater detail elsewhere~\cite{sk_detector},
the Super--Kamiokande  detector consists of 
about 50000 tons of ultra-pure water
in a stainless steel cylindrical water tank
with 11146 20-inch photomultiplier tubes [PMT's] in the inner detector [ID]
and 1885 8-inch PMT's in the outer detector [OD].
The diameter and height of the SK tank are 39.3 meters and 41.4 meters, 
respectively. 
The coordinates of the SK tank are defined in Figure~\ref{fig:detdir}.
In the inner detector, the active photodetector coverage is 40.4\% while 
the remainder is covered with black, polyethylene terephthalate
sheets, generally referred to simply as ``black sheet.'' 
Signals from PMT's are sent through an electronics chain 
which can measure both the arrival time of Cherenkov photons as well as 
the amount of charge they liberate from the phototubes' photocathodes.

 The Super--K detector is located 1000 meters underground 
(2700 meters of water equivalent) in Kamioka Observatory, deep within 
the Kamioka mine in Gifu Prefecture, Japan.  The Observatory is owned and 
operated by the Institute for Cosmic Ray Research [ICRR], a division of the 
University of Tokyo.  The detector's 
latitude and longitude are 36$^{\circ}$ 25'~N and 137$^{\circ}$18'~E,
respectively. Compared to ground level, the intensity of muons is reduced by
about $10^{-5}$ at the depth of the SK detector, and yielding a 
downward-going muon rate through the detector of about 2~Hz.

 As radon concentrations in the Kamioka mine air can exceed 3000~Bq/m$^3$ 
during the summer season, there are air-tight doors between the Super--K 
detector area and the mine tunnel. 
The excavated, domed area above the cylindrical water tank, 
called the ``SK dome,'' is coated with radon-resistant plastic sheets 
to prevent radon in the surrounding rock from entering the air 
above the detector. 
Fresh air from outside the mine is continuously pumped into 
the SK dome area at the rate of $5 \sim 12$ m$^3$/minute.
As a result, the typical radon concentration in the SK 
dome air is 20$\sim$30~mBq/m$^3$.

 As we will see in Section~\ref{sec:bg}, radon can lead to background events in
the solar neutrino data set. In order to keep radon out of the detector itself,
the SK tank is tightly sealed.
Radon-reduced air, produced by a special air purification system in the
mine, is continuously pumped into the space above the water surface
inside the SK tank, maintaining positive pressure. 
The radon concentration of this radon-reduced air is less than 3~mBq/m$^3$.

 Finally, the purified water in the SK tank is continuously circulated 
through the water purification system in the mine at the rate of about
35~tons/hour.  This means that the entire 50 kiloton water volume 
of the detector is passed through the filtration system once every 
two months or so.

\subsection{Photomultiplier tubes}
 The PMT's used in the inner part of SK are the 
20-inch diameter PMT's developed by Hamamatsu Photonics K.K.
in cooperation with members of the original Kamiokande experiment~\cite{pmt1}.
Detailed descriptions of these PMT's, including their 
quantum efficiencies [Q.E.],
single photo-electron distributions, timing resolutions, and so on,
may be found elsewhere~\cite{pmt2}. 

 Good timing resolution for light arrival at each PMT is
essential for event reconstruction. It is around 3~nsec for single
photo-electron light levels.

\begin{figure}[htb]
\begin{center}
\includegraphics[scale=0.35]{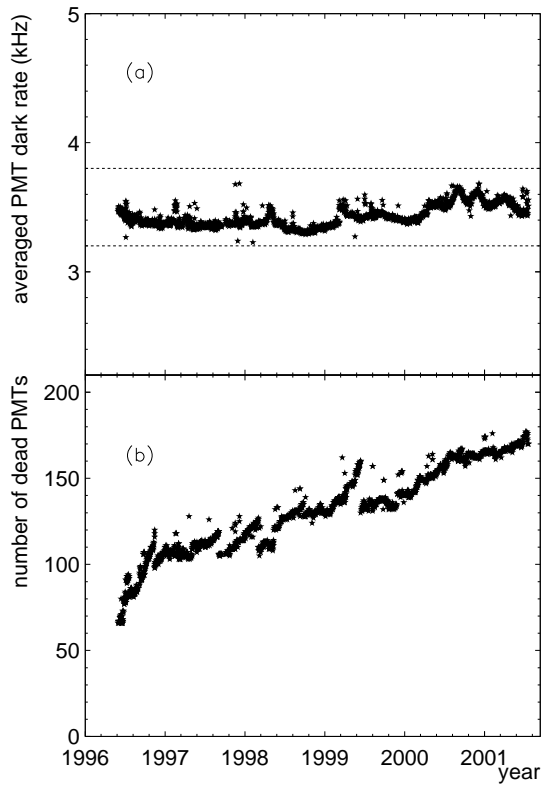}
\caption{(a) The average dark noise rate of the PMT's used in SK--I. 
The dashed lines show the acceptable range for use in the 
solar neutrino analysis. (b) Number of dead PMT's in SK--I.  Note that 
sometimes repairs were possible, usually involving the replacement of 
broken high voltage supplies, leading to sudden drops in the number 
of dead PMT's.}
\label{fig:pmtdarkbad}
\end{center}
\end{figure}

 The dark noise rate of the PMT's is measured to be around 3.5~kHz on 
average and it was stable over the SK--I data taking period as shown in 
Figure~\ref{fig:pmtdarkbad}(a).
The number of accidental hits caused by this dark noise is estimated
to be about 2 hits in any 50~nsec time window.  For the solar neutrino 
analysis, this is corrected during the energy calculation as described later 
in this paper.

 Over time some of the PMT's malfunctioned by producing anomalously 
high dark noise rates or emitting spark-generated light
(a tube which makes its own light is called a ``flasher''). 
The high voltage supplied to these malfunctioning PMT's was 
turned off shortly after the malfunctions arose, 
rendering the tubes in question ``dead.'' The number of dead PMT's 
is shown in Figure~\ref{fig:pmtdarkbad}(b).

 The dark noise rate, plus the numbers of excessively 
noisy and dead PMT's are taken into account in our Monte Carlo [MC] 
detector simulation and also corrected for during event reconstruction.
\subsection{Trigger efficiency}\label{sec:trigger}
 A data acquisition [DAQ] trigger is generated whenever a 
certain number of PMT's are fired within a sliding 200~nsec window. 
Timing and charge information of each fired PMT is
digitized once such a hardware trigger is issued. 
The ID and OD trigger logic are independent of each other.

 When SK--I began data taking in 1996, two threshold levels,
called the low energy [LE] trigger and the high energy [HE] trigger, were 
set in the ID. The threshold number of PMT's for LE and HE triggers were
about 29 and 33 PMTs, respectively. This LE hardware trigger threshold 
corresponded to a 50\% triggering efficiency at 5.7~MeV. In order to avoid 
hardware trigger efficiency issues, the LE {\em analysis} threshold was 
ultimately set to 6.5~MeV, where the LE hardware trigger efficiency 
is essentially 100\%.

 From May of 1997, a third ID hardware trigger threshold 
called the super low energy [SLE] trigger was added.  Originally set 
at 24 PMT's, which provided 50\% triggering efficiency at 4.6~MeV,
the addition of the SLE trigger served to increase the raw trigger rate 
from 10~Hz to 120~Hz.  These SLE events were then passed to an online 
fast vertex fitter, whose operation is described in Section~\ref{sec:vertex}.

 Since almost of these very low energy events are caused by 
$\gamma$s from the rock surrounding the detector and radioactive 
decay in the PMT glass itself, the fast vertex fitter was used to 
reject SLE events with event vertices outside the nominal 22.5~kton 
fiducial volume. This software filtering procedure and its 
associated online computer hardware was called the Intelligent Trigger 
[IT], and it filtered the 110~Hz of SLE triggered events down 
to just 5~Hz of SLE events whose vertices fell within Super--K's 
fiducial volume. Thus a total event rate of 14.6~Hz was transmitted out 
of the Kamioka mine for eventual offline reduction and analysis.

 In 1999, and again in 2000, the IT system was upgraded with additional 
CPU's. By the end of SK--I it provided 100\% triggering efficiency at
4.5~MeV, and 97\% efficiency at 4.0~MeV. Table~\ref{tab:int-trig} shows the
history of the trigger as a function of time.

\begin{figure}[htbp]
\begin{center}
\includegraphics[scale=0.22]{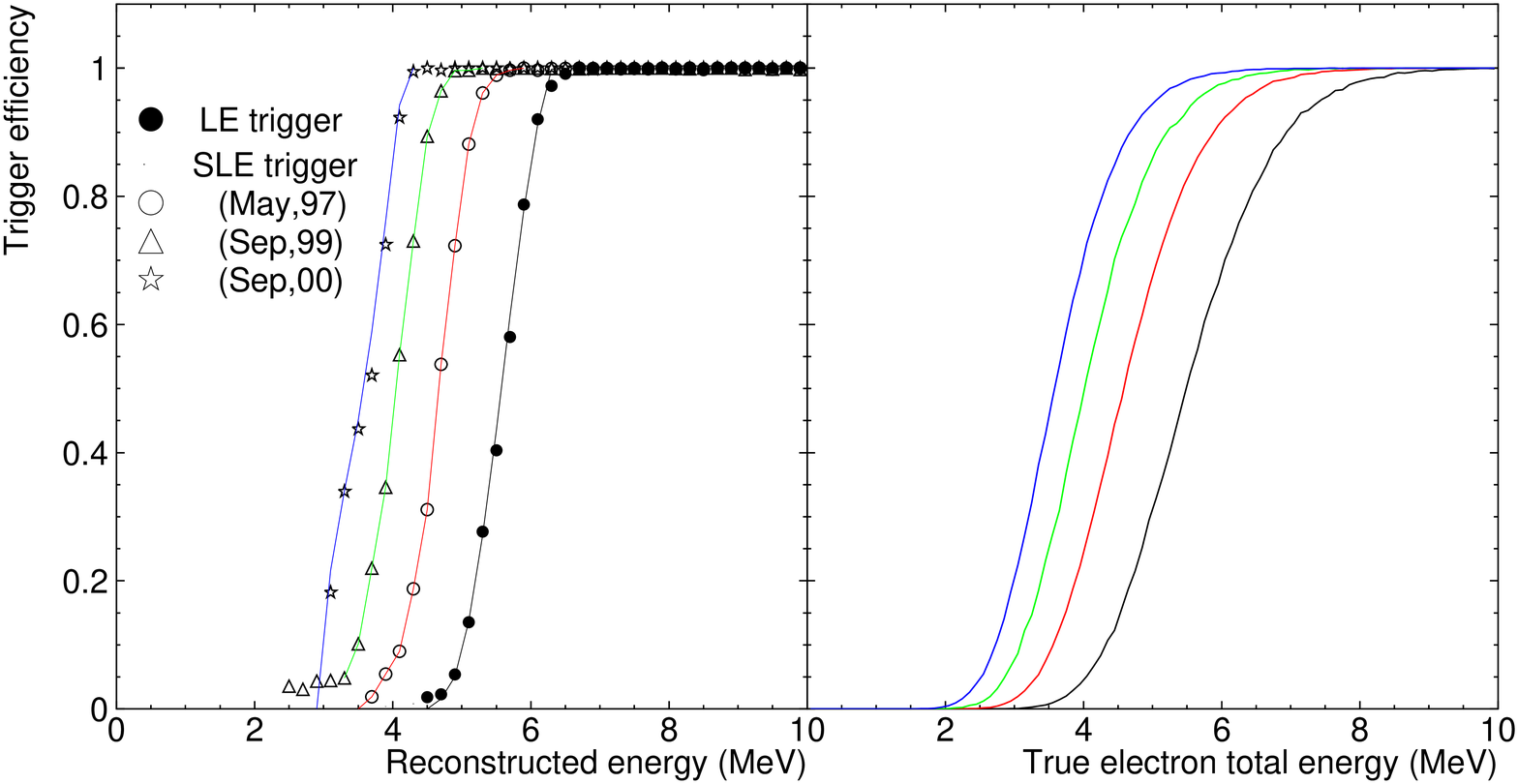}
\caption{Trigger efficiency as a function of energy. The left plot shows
the efficiency as a function of reconstructed energy (see Section~\ref{sec:ene}
for the details of reconstructed energy). The May 1997 LE triggers (black
circles with black line) and SLE triggers (white circles with red line)
were calculated using a Ni(n,$\gamma$)Ni gamma source, while the SLE
triggers on September 1999 (triangles with green line) and September 2000
(stars with blue line) were calculated using $^{16}$N events from a DT
generator. The right plot shows the efficiency as a function of true
electron total energy obtained by a Monte Carlo simulation. 
Identically colored lines represent the same calibration 
data samples in both plots.}
\label{fig:trgeff}
\end{center}
\end{figure}

 The trigger efficiency was checked with both $^{16}$N calibration data
from our DT generator and events from a Ni(n,$\gamma$)Ni gamma source. It
was also continuously monitored using prescaled samples of real,
unfiltered SLE triggered events. The left plot of Figure~\ref{fig:trgeff}
shows the typical trigger efficiency curve as a function of reconstructed
energy, which is obtained from the number of hit PMT's after various
corrections are applied (see Section~\ref{sec:ene} for more details).
A trigger efficiency curve as a function of true electron energy is calculated
by a Monte Carlo simulation -- it is shown in the right plot of
Figure~\ref{fig:trgeff}. This plot is for events whose vertices fall
within the fiducial volume of the detector (i.e., their true vertex
positions are more than 2 meters from the PMT wall).

\begin{table}[htbp]
\begin{center}
\begin{tabular}{| l c c c c c c c |}
\hline
  Start & CPUs & Online  & Filtered & \multicolumn{2}{c}{Hardware}& Analysis & SLE \\
  date  &      & trigger & trigger  & \multicolumn{2}{c}{thres.}  & thres. & live \\
        &  & rate   & rate& (hits) & (MeV) & (MeV) & time \\\hline
  4/96  & 0  & 10   & 10  & 29 & 5.7 & 6.5 & 0 \\
  5/97  & 1  & 120  & 15  & 24 & 4.6 & 5.0 & 96.5 \\
  2/99  & 2  & 120  & 15  & 23 & 4.6 & 5.0 & 99.3 \\
  9/99  & 6  & 580  & 43  & 20 & 4.0 & 5.0 & 99.95 \\
  9/00  & 12 & 1700 & 140 & 17 & 3.5 & 4.5 & 99.99 \\
\hline 
\end{tabular}
\caption{History of the trigger. The units for rate and livetime are
Hz and \%, respectively.}
\label{tab:int-trig}
\end{center}
\end{table}

\label{sec:det}
%
%
\section{Simulation}\label{sec:simulation}
 In the simulation of solar neutrino events in Super--Kamiokande--I 
there are several steps: generate solar neutrinos,
determine recoil electron kinematics, generate and track Cherenkov light in
water, and simulate response of electronics.

\begin{figure}[!htbp]
\begin{center}
\includegraphics[scale=0.35]{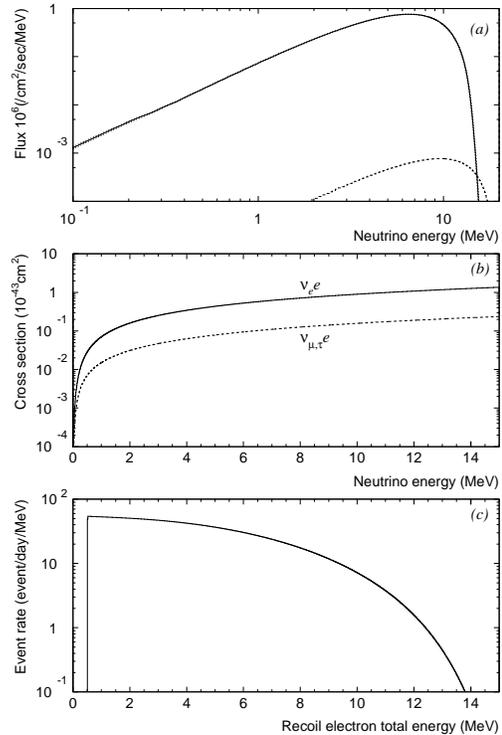}
\caption{(a) Input distributions of $^8$B (solid) and hep (dashed) solar 
neutrino energies.
(b) The cross section of the interaction for $\nu_e$ (solid line) and
$\nu_{\mu,\tau}$ (dashed line) with electrons as a function of neutrino
energy.
 (c) The spectrum of recoil electrons scattered by $^8$B and hep solar
 neutrinos.}
\label{fig:b8nu}
\end{center}
\end{figure}

 In order to generate the $^8$B solar neutrino spectrum in Monte Carlo, 
the calculated spectrum based on the $^8$B decay measurement of  
Ortiz et al.~\cite{ortiz} was used.  Figure~\ref{fig:b8nu}(a) 
shows the input solar neutrino energy distribution
for $^8$B neutrinos. For the uncertainty in the spectrum we 
have adopted the estimation by Bahcall et al~\cite{bpspc}.

 In the next step, the recoil electron energy from the following reaction;
\begin{equation}
   \nu + e \to \nu + e
   \label{eq:nue}
\end{equation}
is calculated. Fig~\ref{fig:b8nu}(b) shows the differential cross section.
The $(\nu_e,e)$ scattering cross section is approximately six times larger
than $(\nu_{\mu,\tau},e)$, because the scattering of $\nu_e$ on an
electron can take place through both charged and neutral current
interactions, while in case of $\nu_{\mu,\tau}$ only neutral current
interactions take place. The radiative corrections in the scattering are
also considered~\cite{rad}. Figure~\ref{fig:b8nu}(c) shows the expected
spectrum of recoil electrons in SK without neutrino oscillations.

 Assuming the $^8$B flux of BP2004 ($5.79\times10^6/cm^2/sec$) and 
before taking into account the detector's trigger requirements, 
the expected number of $\nu_e$ scatter events in SK is
325.6 events per day. MC events are generated assuming a rate of 
10 recoil electron events per minute for the full operation time of 
SK--I, yielding a total of 24,273,070 simulated solar neutrino events.

 We have used GEANT3.21 for simulation of particle tracking through 
the detector. Since the tracking of Cherenkov light is especially important 
in the Super--K detector, parameters related to photon tracking 
are fine tuned through the use of several calibration sources.

\begin{figure}[htbp]
\begin{center}
\includegraphics[scale=0.35]{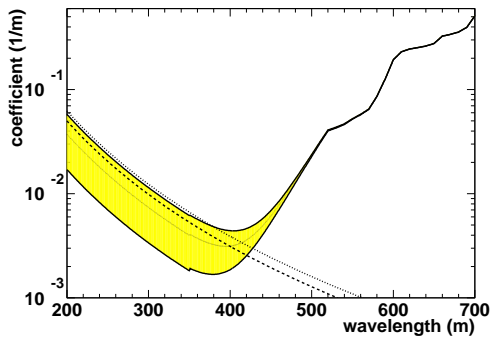}
\caption{Wavelength dependence of the water parameter coefficients: 
         absorption (solid), Rayleigh scattering (dashed) and Mie scattering
         (dotted). The absorption coefficient is also a function of 
water transparency.  The filled region shows the range of this parameter 
as water transparency is changed,  where the two solid lines define the 
SK--I minimum (73~meter) and maximum (98~meter) values.}
\label{fig:wtrans}
\end{center}
\end{figure}

 Figure~\ref{fig:wtrans} shows the wavelength dependence of various 
water coefficients in our MC. Rayleigh scattering is dominant at 
short wavelengths with a 1/$\lambda^4$ dependence. 
The $\lambda$ dependence of absorption and Mie scattering are empirically set
to 1/$\lambda^4$ at shorter wavelengths, while the absorption for longer 
wavelengths are taken from a separate study~\cite{water}. 
The absorption and scattering coefficients in MC
are tuned using LINAC calibration data (see Section~\ref{sec:enelin}) so as to 
match the MC and data energy scale in each position in the detector.

\begin{figure}[htbp]
\includegraphics[scale=0.25]{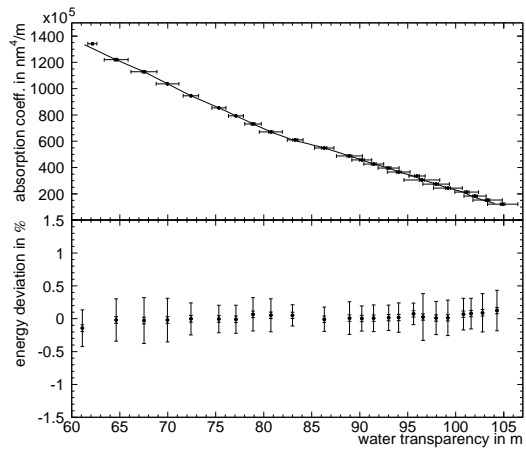}
\caption{Translation of the water transparency parameter into
a short wavelength absorption coefficient.
Absorption is taken to be proportional to $\lambda^{-4}$ for these  
wavelengths. Light scattering is assumed to be constant.
The lower panel shows the energy deviation as a function of input momentum.
The inner error bar of each point is the statistical 
uncertainty, while the outer error bar represents the spread of the 
six samples using 5, 6, 8, 10, 12, and 15~MeV/c as 
the input momenta.}
\label{fig:abstune}
\end{figure}

 The water quality in SK changes as a function of
time as shown in Figure~\ref{wt_final}; this change in water quality was 
taken into account in the MC simulation. By comparing photon arrival timing
distributions using calibration data, it was found that the change in water
attenuation length is mainly due to change in the absorption coefficient.
So, we fixed Rayleigh scattering and Mie scattering coefficients to be 
constant over the entire data taking period and vary only the 
absorption coefficient in the MC simulation. Figure~\ref{fig:abstune} 
shows the result.
The solid line shows the translation of water transparency into the 
absorption coefficient. The lower panel shows the deviation of the peak
of the reconstructed energy from the peak of the input energy as a 
function of water transparency --- the deviation is less than $\pm0.2\%$. 
The coefficients of each
process at shorter wavelength are summarized as follows:
\begin{eqnarray}
8.00 \times 10^7 / \lambda^4 [nm] \quad(1/m) &:& \mbox{Rayleigh scat.,}\nonumber\\
1.00 \times 10^8 / \lambda^4 [nm] \quad(1/m) &:& \mbox{Mie scat.,}\nonumber\\
(2.74 \sim 9.27) \times 10^7 / \lambda^4 [nm] \quad(1/m) &:& \mbox{absorption.}\nonumber
\end{eqnarray}

\label{sec:sim}
%
%
\section{Event reconstruction method}\label{sec:reconstruction}
\subsection{Vertex}\label{sec:vertex}

\begin{figure}[tb]
\includegraphics[scale=0.3]{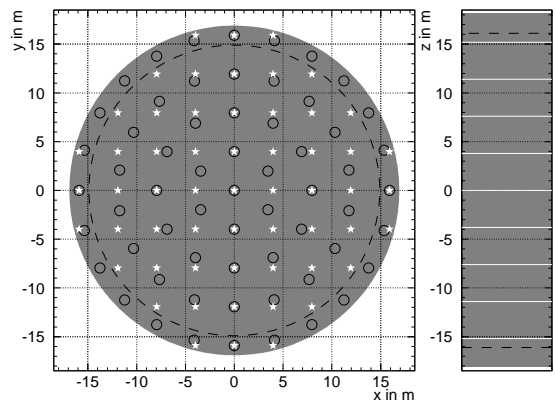}
\caption{Search grid for the vertex reconstruction in the $xy$ plane
and in $z$. The shaded area is the SK inner detector volume. The
dashed line indicates the fiducial volume. The white stars (white
lines) are the grid points used by the standard vertex reconstruction.
The black circles are used by the second vertex reconstruction.}
\label{fig:vertgrid}
\end{figure}

 Electrons in the energy region of interest for solar neutrinos (below 20~MeV)
can travel only a few centimeters in water, so their Cherenkov light
is approximately a point source. The reconstruction of
this vertex relies solely on the relative timing of the
``hits,'' i.e., PMT's struck by one or more Cherenkov photons.
Since the number of observed Cherenkov photons and therefore the
likelihood of a multiply hit PMT is comparatively small, 
about seven recorded photons per MeV of deposited energy, 
the pulse heights of the hits typically follows a one 
photo-electron distribution and yields
no information about the light intensity nor the
distance to the source. For the same reason it is also
impossible to separate reflected or scattered Cherenkov
photons and PMT dark noise from direct light based upon PMT pulse height.
The vertex reconstruction assumes that a photon originating from
vertex $\vec{v}$ and ending in hit $\vec{h}_i$ is traveling
on a straight path and therefore takes the time
$|\vec{v}-\vec{h}_i|/c$ where c is the group velocity
of light in water (about 21.6~cm/ns).
Therefore, assuming only direct light,
the effective hit times $\tau_i=t_i-|\vec{v}-\vec{h}_i|/c$
at vertex $\vec{v}$ of all PMT's should peak around
the time of the event with the width of the PMT timing
resolution for single photoelectrons (about 3~ns).
Light scattering and reflection (as well as dark noise,
pre-, and afterpulsing of the PMT's) introduces tails in the
distribution which will strongly bias the reconstruction 
if a $\chi^2$ is used to evaluate the goodness of fit $g$
for a given vertex. Therefore, we use a ``truncated
$\chi^2$''
\begin{equation}
g(\vec{v})=\sum_{i=1}^N e^{-\frac{1}{2}\left(\frac{\tau_i-t_0}{\sigma_t}\right)^2}
\end{equation}
where $t_0$ is the center of a 10~ns wide search window
in $\tau$ which maximizes the number of hits inside.
The vertex is reconstructed by the maximization of this
goodness varying $\vec{v}$.

 We use three different vertex reconstruction algorithms. Our standard
vertex fit is used to reconstruct event direction
and energy and to compute the likelihood of an event to be due to spallation.
Only events inside the fiducial volume (2~m away from the closest PMT) are
considered. The Super--K background rate below about 7~MeV rises rapidly with
falling energy; most of this background is due to light emanating 
from or near the PMT's themselves and is reconstructed
outside the fiducial volume. However, the rate of misreconstructed
events (background events outside the fiducial volume which get reconstructed
inside) increases rapidly with falling energy due to long resolution tails.
We therefore reconstruct the vertex using a second vertex fitter whose
vertex distribution has smaller tails and then accept only events reconstructed
inside the fiducial volume by {\it both} fits.
To reduce the rate of SK events stored to tape, we remove SLE events 
which reconstruct outside of the fiducial volume online. Unfortunately, 
both offline fitters are too slow to keep up with the trigger rate 
(see Section~\ref{sec:det}) of SLE events and so we are forced to use a fast 
reconstruction to pre-filter these events. If the vertex of the fast 
online fit is inside the fiducial volume the 
event is reconstructed later by the other two fits.

 To ensure convergence
of the maximum search, the standard fit first evaluates the vertex goodness
on a Cartesian grid (see Figure~\ref{fig:vertgrid}).
For a reasonably speedy search
in spite of Super--Kamiokande's large size, a coarse grid of 57 points
with a grid constant of 397.5cm is chosen for the $xy$
plane with nine such layers in $z$ separated by 380~cm. Since the
grid is coarse, the timing resolution $\sigma_t$ is artificially
set to 10~ns to smear out the maximum goodness. After the
single coarse grid point with the largest goodness is identified, the
goodness is calculated at the 27 points of a $3\times3\times3$
cube centered on this grid point with a timing resolution of
$\sigma_t=5$~ns and a spatial grid separation of 156~cm. If the largest
resulting goodness is not in the center, the cube is shifted so that
the largest goodness is in the center of the new cube. Otherwise,
the cube is contracted by a factor of 2.7. If the spatial
grid separation of the cube reaches 5~cm the maximum search is finished
and the reconstructed vertex is the point with the largest
found goodness.

 To limit the bias of the vertex reconstruction due to the tails
in the timing residuals $\tau_i-t_0$, the hits that go in the
vertex fit are selected. Since the event vertex is not known,
the hit selection must be based on the absolute hit times $t_i$
(and the spatial distribution of the hits $\vec{h}_i$). The
hit selection of the standard reconstruction first finds the
start time of a sliding 200~ns window (the time required for a photon to
cross the diagonal of the detector) which contains the largest number
of hits. The rate of late/early hits (background) per ns is determined by
counting the hits outside the window. From this rate, the background
inside the window is estimated. The size of the window is then
reduced in an attempt to optimize the direct light signal divided by
the square root of the late/early hit background. The hits in
the resulting, optimal timing window are those selected to 
compute the vertex goodness.

\begin{figure}[tb]
\includegraphics[scale=0.5]{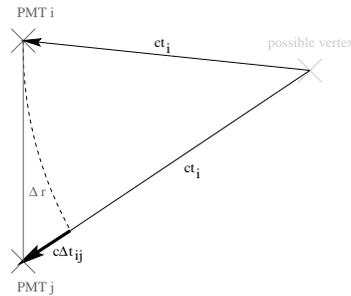}
\caption{Timing constraint for pairs of hits.
From the triangle relation $ct_i+\Delta x>ct_j=ct_i+\Delta ct_{ij}$
follows $\Delta t_{ij}<\Delta x/c=|\vec{h}_i-\vec{h}_j|/c$.}
\label{fig:verttriangle}
\end{figure}

The second vertex reconstruction (to reduce misreconstructed
events originating from the PMT's) modifies
the hit selection and the search grid. For two direct light hits
$i$ and $j$, the timing difference $\Delta t_{ij}$ is limited
to $\Delta t_{ij}<|\vec{h}_i-\vec{h}_j|/c$
(see Figure~\ref{fig:verttriangle}). We select the largest
set of hits whose hit pairs obey $\Delta t_{ij}<|\vec{h}_i-\vec{h}_j|/c$
after eliminating ``isolated hits'' (hits which are further away
than 1250~cm or further away than 35~ns from the nearest neighboring hit).
With these selected hits we perform a grid search with a circular grid of
60 points on the $xy$ plane (grid constant 397.5~cm) and nine such planes
in $z$ separated by 380~cm with a timing resolution of 9.35~ns. The
best fit point is interpolated from the grid points with the
largest goodness. After that, 27 points of a cylindrical section 
around that point with an initial grid constant of 147.1~cm 
and 5~ns resolution are tested. As in the standard reconstruction,
the section is moved (if the center point doesn't have the largest
goodness) or reduced in size by a factor of 0.37 (if it does), and
the search is finished once the grid constant falls below 5~cm.

\begin{figure}[tb]
\includegraphics[scale=0.3]{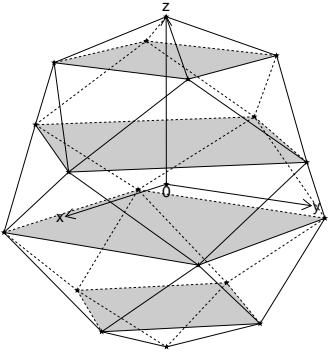}
\caption{Search grid for the fast fit.
18 points around the input vertex are tested,
the orientation is fixed. The grid size
is adjusted between one and eight meters (see Table~\ref{tab:hayai}).
}
\label{fig:hayaigrid}
\end{figure}

\begin{table}[tb]
\centerline{
\begin{tabular}{c cccc}
Step & Size [cm] & $t_0\pm\sigma_t$ [ns] & $a_0\pm\sigma_a$ &
 $t_{\mbox{\tiny win}}$ [ns] \cr
\hline
1   & 800 & $2.1\pm10.2$ & $0.534\pm0.134$ & (-66.7,100.0) \cr
\hline
1   &     & $2.7\pm6.1$  & $0.534\pm0.134$ & (-40.0, 66.7) \cr
2   & 400 & $2.3\pm6.0$  & $0.534\pm0.134$ & (-33.3, 50.0) \cr
3-5 &     & $2.3\pm6.0$  & $0.534\pm0.134$ & (-25.0, 33.3) \cr
\hline
1   &     & $1.5\pm4.7$  & $0.1g$ for $a<0.2, a\ge0.9$ & (-20.0,33.3) \cr
2   & 200 & $2.2\pm4.5$  & $g$ for $0.3\le a<0.85$     & (-20.0,33.3) \cr
3-5 &     & $1.1\pm4.1$  & $0.5g$ otherwise & (-20.0,26.7) \cr
\hline
1-8 & 100 & $0.6\pm3.2$  & none & (-13.3,16.7) \cr
\end{tabular}
}
\caption{Search parameters for the fast fit.}
\label{tab:hayai}
\end{table}

 The fast fit, used to pre-filter SLE events online, also eliminates 
isolated hits (i.e., all other hits separated by either more than 10~m or more 
than 33.3~ns) to reduce the effects of dark noise and reflected or 
scattered light. Then the absolute peak
time is estimated by maximizing the number of PMT hits within a 16.7~ns wide
sliding timing window. PMTs that are in the interval ($-33.3$~ns, 100~ns)
with respect to the peak time are selected. An initial vertex is
calculated by shifting a simple average of the selected PMT positions
2~m toward the detector's center. The time at the event vertex is also 
determined;  this time must be smaller than the absolute peak time yet 
not differ from it by
more than 117~ns. The initial vertex time is chosen to be 58.3~ns before
the absolute peak time.  It is then corrected to the average of all
time-of-flight subtracted PMT times that are in the interval
($-133.3$~ns, 200~ns) around the initial time. The
magnitude of the anisotropy
\begin{equation}
\vec{a}=
\frac{\sum_{\mbox{\tiny selected hits}} q_i (\vec{h}_i-\vec{v})/|\vec{h}_i-\vec{v}|}
{\sum_{\mbox{\tiny selected hits}} q_i}
\end{equation}
is also calculated at this stage.  Next, 
18 points around the initial vertex are tested with an initial search radius
of 8~m (see Table~\ref{fig:hayaigrid}), while the goodness $g$ is modified 
to take into account the anisotropy.
At first, the goodness is multiplied by the factor 
$e^{-0.5((a-a_0)/\sigma_a)^2}$. Later in the search, this factor 
is replaced by a table.  The times are shifted by the expected 
mismatch between the average vertex time
and the vertex peak time $t_0$. Only hits inside a time interval around the
shifted vertex time are considered. The search radii, time shifts, 
time resolutions, anisotropy factors, and time intervals are listed 
in Table~\ref{tab:hayai}.

\begin{figure}[tb]
\includegraphics[scale=0.3]{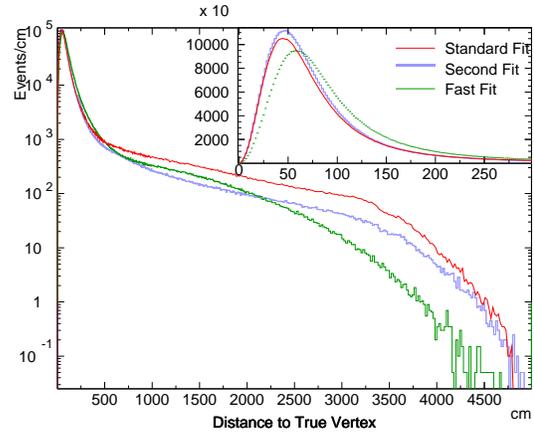}
\caption{Distance from reconstructed to correct vertex for $^8$B Monte
Carlo events with total recoil electron energy above 4.5~MeV.
The inserted panel magnifies the distributions near zero on a linear scale.}
\label{fig:vdist}
\end{figure}

\begin{figure}[tb]
\includegraphics[scale=0.3]{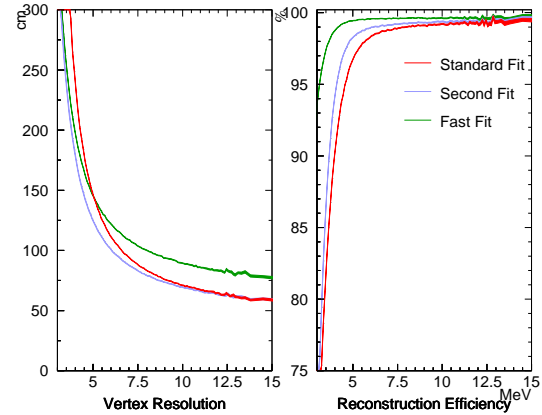}
\caption{Vertex resolution (68.2\% of reconstructed
events reconstruct inside a
sphere of radius $\sigma$ from correct vertex) of $^8$B Monte
Carlo events as a function of total recoil electron energy.
The curves on the right-hand side show the fraction of reconstructed
events as a function of total recoil electron energy.}
\label{fig:vres}
\end{figure}

As shown in Figures~\ref{fig:vdist} and~\ref{fig:vres}, the 
performance of all three fitters is evaluated using $^8$B Monte Carlo.
The distribution of the distance between the reconstructed and correct
vertex is analyzed for each fit (see Figure~\ref{fig:vdist}); 
the vertex resolution is extracted as the distance which contains
68.2\% of all reconstructed vertices.
For recoil electrons above 4.5~MeV, the fast fit shows the worst resolution
(115~cm). However, it has the lowest rate of very distant misfits ($>30$~m).
The resolution of the standard fit is 102~cm, while the second fit's 
resolution is 94~cm.
Figure~\ref{fig:vres} shows the vertex resolution
as a function of generated total recoil electron energy. Since the vertex 
reconstruction occasionally fails (due to a number of selected
hits that is too small), the efficiency to reconstruct 
the event anywhere at all is also plotted.


\subsection{Direction}\label{sec:dir}
\begin{figure}[!hbp]
  \includegraphics[scale=0.4]{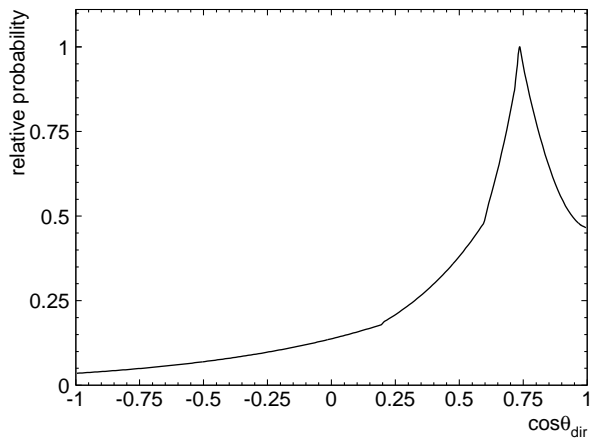}
  \caption[Likelihood function for the direction and acceptance]
          {The distribution of the opening angle between the
           direction of the generated particle and the vector from
           the reconstructed vertex to the hit PMT position.
           This plot is made for 10~MeV electron MC events.}
  \label{fig:dirdeg}
\end{figure}

 Since the recoil electron preserves the direction of the solar neutrino,
directional reconstruction is important for solar neutrino analysis.
This characteristic directionality is used to extract the 
solar neutrino signal in Super--K.
To calculate the direction, a maximum likelihood method using the Cherenkov
ring pattern is adopted. The likelihood function is
\begin{equation}
  L(\vec d ) \equiv \sum_{i}^{N_{30}} \log (f{(\cos \theta_{dir})})_i \times
               \frac{\cos\theta_i}{a(\theta_i)}.
  \label{eq:dirlike}
\end{equation}
where $N_{30}$ is the number of hit PMT's with residual times within a 30~nsec
window and \(f{(\cos \theta_{dir})}\) is the function that represents
the distribution of the opening angle between the particle direction
and the vector from the reconstructed vertex 
to the hit PMT position made by MC.
A plot of $f{(\cos \theta_{dir})}$ for 10~MeV electrons is shown in
Figure~\ref{fig:dirdeg}. The distribution is broad with the peak at $42^\circ$
because of the effects of electron multiple scattering and Cherenkov light
scattering in water. $\theta_i$ is the opening angle between the direction
of the vector from the reconstructed vertex to the i-th hit PMT position
and the direction that PMT is facing. $a(\theta_i)$ is the acceptance
of PMT's as a function of $\theta_i$ which is made by MC.
The direction is reconstructed as Eq.~(\ref{eq:dirlike}) becomes maximum
using a grid search method whose step sizes are
\( 20^\circ, 9^\circ, 4^\circ, 1.6^\circ \).

\begin{figure}[!hpbt]
  \includegraphics[scale=0.4]{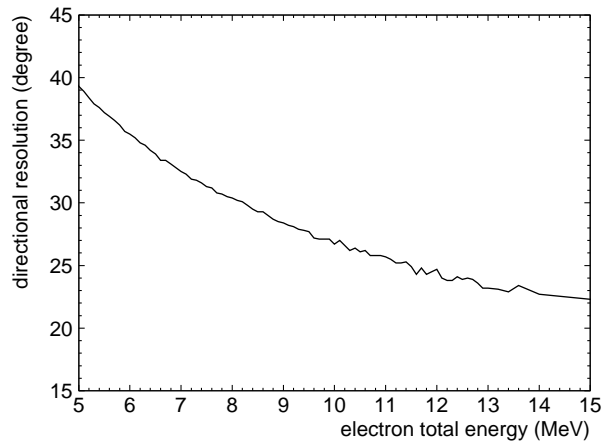}
  \caption{The directional resolution's dependence on energy, 
determined using Monte Carlo events.}
  \label{fig:dir}
\end{figure}

 The quality of the directional reconstruction is estimated from the
differences between the generated and reconstructed directions using MC.
Figure~\ref{fig:dir} shows the directional resolution dependence on energy
within the fiducial volume; it uses directions which are generated uniformly.
Here, the resolution is defined as the 68\% point for the angle difference
distribution. The angular resolution for 10~MeV electrons is about 25~degrees.

\subsection{Energy}\label{sec:ene}
 The energy of a fully-contained charged particle in Super--Kamiokande 
is approximately proportional to the number
of generated Cherenkov photons, and thus is also proportional to 
the total number of photo-electrons in the resulting hit PMT's. 
When the particle's energy is low it is also proportional to 
number of hit PMT's because in such a case the number of Cherenkov 
photons collected by any given PMT is almost always zero or one. 
In order to avoid the effect of noise hits with higher charge,
the number of hit PMT's $(N_{hit})$ with some corrections is used for energy
determination for the solar neutrino analysis.
In order to reject accidental hits due to dark noise in the PMT's,
only the hit PMT's with residual times within a 50~nsec 
window are used for calculating $N_{hit}$.
Moreover, we applied several corrections to $N_{hit}$, yielding an effective
number of hits ($N_{eff}$) which has the same value at every position 
in the detector for a given particle energy. These corrections account 
for the variation of the water transparency, the geometric acceptance of  
each hit PMT, the number of bad PMT's, the PMT dark noise rate, and so on. 
The equation for $N_{eff}$ is:
\begin{eqnarray}
  N_{eff} & = & \sum_{i=1}^{N_{hit}}\{
    \left( X_i + \epsilon_{tail} - \epsilon_{dark} \right)
    \times \frac{N_{all}}{N_{normal}} \times \nonumber \\
    && \frac{R_{\mbox{cover}}}{S(\theta_i,\phi_i)}
    \times \exp\left(\frac{r_i}{\lambda(\mbox{run})}\right)
    \times G_i(t)
    \}
  \label{eq:neff}
\end{eqnarray}
where $X_i$ is an occupancy used to estimate the effect of multiple  
photo-electrons, $\epsilon_{tail}$ is the correction for late hits 
outside the 50~ns window, and $\epsilon_{dark}$ is for dark noise correction.
The definition of $X_i$ is as follows:
\begin{equation}
  X_i = \left\{
   \begin{array}{ll}
     \displaystyle{\frac{\log\frac{1}{1-x_i}}{x_i}}&
         \qquad x_i < 1 \\[0.5cm]
     3 & \qquad x_i = 1
   \end{array} \right.
  \label{eq:occ}
\end{equation}
where $x_i$ is the ratio of the number of hit PMT's to the total 
number of PMT's in a $3 \times 3$ patch around the i-th hit PMT. 
This correction estimates the number of photons which arrived at 
the i-th hit PMT by using the number of hit PMT's 
surrounding it.  Here, the ratio of the unhit PMT's in this patch of nine 
tubes is $1-x_i$.

 The second factor is the bad PMT correction, where $N_{all}$ is the total
number of PMT's, 11146, and $N_{normal}$ is the number of properly operating
PMT's for the relevant subrun.

\begin{figure}[htbp]
\begin{center}
\includegraphics[scale=0.4]{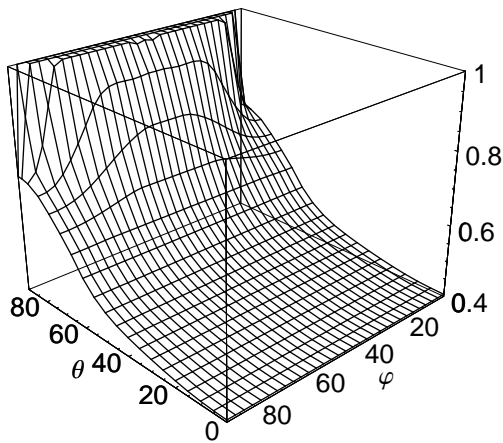}
\caption{Function of the effective photo-coverage's dependence on the 
incident angle to a PMT.}
\label{fig:cov}
\end{center}
\end{figure}

 The third factor is the effective photo coverage. The average of the photo
coverage, which is the ratio of the area covered by PMT's to all 
inner detector area, is $R_{\mbox{cover}} = 0.4041$ . 
However, the effective photo coverage changes with the
incident angle of the photon to the PMT. Therefore, we applied a coverage
correction, where $S(\theta_i,\phi_i)$ is the photo coverage from the 
directional vantage point of
$\theta_i,\phi_i$. Figure~\ref{fig:cov} shows this function.

 The fourth factor is the water transparency correction, where $r_i$ is
the distance from the reconstructed vertex to the i-th hit PMT position,
and $\lambda$ is water transparency.

 The fifth factor, $G(t)_i$, is the gain correction at the single
photo-electron level as a function of the time of the manufacture of each PMT.

\begin{figure}[htbp]
\begin{center}
\includegraphics[scale=0.4]{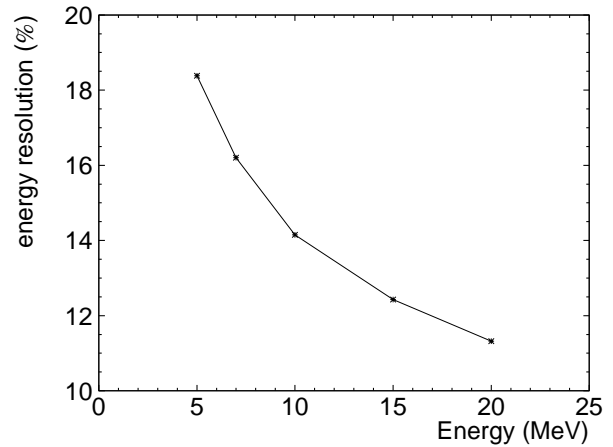}
\caption{The energy resolution's dependence on energy as determined using 
Monte Carlo events.}
\label{fig:eres}
\end{center}
\end{figure}

 Finally, the conversion function from $N_{eff}$ to energy is 
determined by uniformly generated Monte Carlo vertices.
This is total energy; all mention of ``energy'' hereafter 
refers to total energy, i.e. including the scattered electron's 
rest mass and momentum. The relation to energy is calculated 
by the following equation;
\begin{equation}
  \mbox{E} = \alpha + \beta N_{eff}(1- \gamma N_{eff} (1- \delta N_{eff} (1- \epsilon N_{eff}))),
  \label{eq:enelf}
\end{equation}
here $\alpha = 0.735, \beta = 0.134, \gamma = 6.049\times10^{-4},
\delta = 6.441\times10^{-3},$ and $\epsilon = 1.541\times10^{-3}.$
The typical conversion factor from $N_{eff}$ to energy is 6.97.
The energy resolution is also estimated by the same MC described above.
Figure~\ref{fig:eres} shows the energy resolution as a function of energy;
it is 14.2\% for 10~MeV electrons.

\subsection{Muon}\label{sec:muon}
 Precise reconstruction of cosmic ray muon tracks which penetrate 
the Super--K detector is needed for the solar neutrino analysis.  This is
because  nuclear spallation events induced by these cosmic ray muons 
are a dominant background to the solar neutrino signal, and the correlation in
time and space between spallation events and their parent muons is 
extremely useful in rejecting the spallation background. 
Hence, the track position precision requirement for our muon fitter 
is about 70~cm since the vertex resolution for low energy events
with energies around 10~MeV is also about 70~cm.

 The muon fitter has three components; an ``initial fitter,'' a 
``TDC fitter,'' and a ``geometric check.''

 The initial fitter assumes the position of the first fired PMT is the
entrance point of a cosmic ray muon, and the center of gravity of the
saturated tubes' positions as the exit point.  The PMT at the exit point 
has the most photo-electrons [p.e.'s] of all PMT's; the expected 
value is up to 500~$p.e.$.
However, our front end electronics saturate at $\sim 230$ p.e. in a single 
tube, and so typically several dozen PMT's near the exit point are saturated.
Therefore, the reconstructed exit point is not the position of the most-hit 
PMT, but rather the center of gravity of the saturated tubes' positions.

 For the entrance position search, we consider the dark noise of the PMT's.
For SK--I the dark noise rate of a typical PMT was $\sim$ 3.5 kHz, 
though it could rise to $>10$ kHz in the case of a ``noisy PMT.'' 
In the initial fitter, the following methods
are used to reject dark noise hits:
\begin{enumerate}
	\item 	Charge information of PMT\\
                The fitter requires that the PMT at the entrance point have
                more than 2 $p.e.$, because typical dark hit PMT's have 
                less than 1 $p.e.$ while the amount of charge which they 
		receive from a muon is several $p.e.$'s.
	\item	Timing information of eight neighbor PMT's\\
		Cosmic ray muons emit lots of Cherenkov light ($\sim
		340$ photons/cm), so several PMT's near the entrance 
		point should record photons at the same time. 
		The fitter therefore 
		requires that the PMT at the entrance point have more than
		five fired neighbor PMT's within 5~ns.
\end{enumerate}
The results of the ``initial fitter'' are then used as 
initial values by the next stage.

 The TDC fitter is a fitter based on a grid search method using timing
information. In this fitter, a goodness of fit is also defined to obtain the
best track of the muon:
\begin{equation}
   g(\vec{v})=\sum_{i=1}^N e^{-\frac{1}{2}\left(\frac{\tau_i-t_0}{f\times\sigma_t}\right)^2},
\end{equation}
where the definitions are the same as for the ``truncated
$\chi^2$'' in Section~\ref{sec:vertex} except for the factor ($f$),
which is found to be 1.5 by Monte Carlo.  The TDC fitter surveys a circle of
radius 5.5~m from the exit point obtained by initial fitter and then obtains
the direction which maximizes the goodness.

 The muon fitter determines the muon track by using two 
independent parameters, timing and charge, and the two results 
are sometimes different.
For example, in the case of muon bundle events, PMT's near the exit point fire
at early times, because the velocity of a cosmic ray muon is faster than 
that of light in water. As a result of these early hits, the exit 
point can sometimes be mistaken for an entrance
point. In order to reject this kind of misfit, the geometric check 
fitter estimates the consistency of both previous 
fitters by requiring the following after the TDC fit:
\begin{itemize}
 \item There is no saturated PMT within 3~m from the entrance point.
 \item There should be some saturated PMT within 3~m from the exit point.
\end{itemize}
The muon events which do not satisfy these requirements are regarded as
failed reconstructions.

\begin{figure}[hbtp]
\includegraphics[scale=0.4]{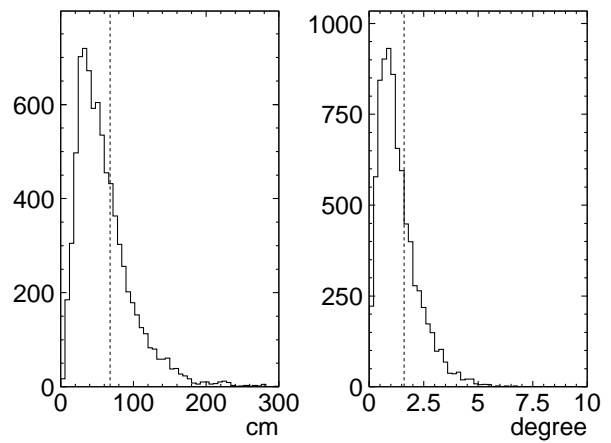}
\caption{Performance of the muon fitter. Left (right) figure shows the
vertex (angular) correlation between generated and reconstructed Monte Carlo 
muon tracks.  The area to the left of the dashed line in both plots 
shows the 1-$\sigma$ (= 68\%) included region.}
\label{fig:muonper}
\end{figure}

 We estimated the performance of the muon fitter by using Monte Carlo events.
Figure~\ref{fig:muonper} shows the vertex and angular resolutions.
For the entrance point, the 1-$\sigma$ vertex resolution 
is estimated to be 68~cm.  This value corresponds closely to the spacing 
interval between each PMT, 70.7~cm. 
The vertex resolution of exit point is 40~cm,
Therefore, the distance between the actual muon track and the fitted one is 
estimated to be less than 70~cm, which means that this muon fitter 
satisfies our precision requirement. The angular resolution is 
estimated to be 1.6$^\circ$.

%

%
%
\section{Calibration}\label{sec:calibration}
\subsection{Water transparancy measurement}\label{sec:wtr}
 Cherenkov photons can travel up to 60~m before reaching PMT's in Super--K,
and so light attenuation and scattering in water directly affects the number of
photons that are detected by the PMT's. Since the energy of an event is mainly
determined from the number of hit PMT's, the water transparency [WT] must be
precisely determined for an accurate energy measurement.

 The water transparency in SK is monitored continuously by using
the decay electrons (and positrons) from cosmic ray $\mu$
events that stop in the detector:
$\mu^{-} \to e^{-} + \bar{\nu}_e + \nu_{\mu},$
and
$\mu^{+} \to e^{+} + \nu_e + \bar{\nu}_{\mu}.$
At SK's underground depth (2700 m.w.e.), cosmic ray muons reach the
detector at a rate of $\sim$ 2 Hz.  Approximately 6000 $\mu$'s per day
stop in the inner detector and produce a decay electron (or positron).
In order to monitor the WT effectively, it is important to 
have a pure sample of $\mu$-e decay events. Several criteria are applied
to select these events:
\begin{itemize}
 \item The time difference [$\Delta T$] between the stopping $\mu$ event
and the $\mu$-e decay candidate must satisfy: 2.0~$\mu$sec $< \Delta T <$ 
8.0~$\mu$sec.
 \item The reconstructed vertex of the $\mu$-e decay candidate must be
contained within the 22.5 kton fiducial volume.
 \item $N_{hit}$ must be at least 50.
\end{itemize}
These criteria select $\sim$1500 $\mu$-e decay events daily, which 
is sufficient statistics to search for variation in the WT in one week.
The average energy of the $\mu$-e decay events is $\sim$ 37~MeV, which is 
much higher than that of solar neutrinos ($<$ 20 MeV).
Therefore, the $\mu$-e decays cannot be used for the absolute energy
calibration.  However, they can be used to monitor the stability of
water transparency and energy scale over time.

 In order to remove the effects of scattered and reflected light,
hit PMT's are selected by the following criteria:
(1) PMT's must have timing that falls within the 50~nsec timing window, 
after the time-of-flight [TOF] subtraction, (2) PMT's must be 
within a cone of opening
angle $\rm 32^{\circ} \sim 52^{\circ}$ with respect to the direction of the
$\mu$-e decay event. A plot of the number of hit PMT's vs. distance from the
vertex to the PMT is made using the selected PMT's, fit with a linear 
function, and the inverse of the slope gives the water transparency.

\begin{figure}[htbp]
\begin{center}
\includegraphics[width=8.5cm]{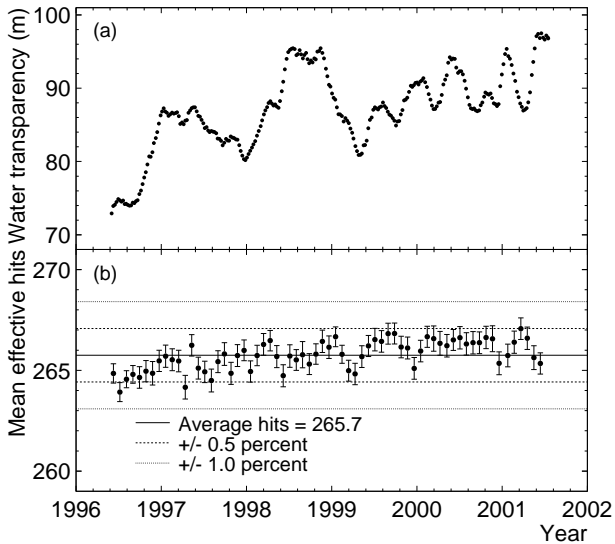}
\caption{(a) Time variation of the measured water transparency during SK--I.
Stability of the SK--I energy scale as a function of time.
(b) Each point represents the mean $N_eff$ value for 28 days of $\mu$-e data.
The solid line represents the average $N_eff$ for all $\mu$-e events in SK--I;
the dashed and dotted lined represent, respectively, $\pm$ 0.5\% and
$\pm$ 1.0\% deviations from the average.
 }
\label{wt_final}
\end{center}
\end{figure}

 Figure \ref{wt_final}(a) shows variations in the WT as a
function of time. Each point on the plot represents WT for one week.
In order to reduce the effects of statistical fluctuations
on the weekly measurement, the WT for a given week is 
defined as a running average over five weeks of data.
Figure \ref{wt_final}(b) plots the mean
$N_{eff}$ value for $\mu$-e decay events in SK--I.
From this figure, it can be seen that the energy scale has 
remained stable to within $\pm$ 0.5\% during the SK--I runtime.

\subsection{Energy calibration}
 The energy scale calibration is necessary to correctly convert
the effective number of hits ($N_{eff}$) to the total energy of the 
recoil electrons induced by solar neutrinos. LINAC and DT calibration are
discussed in this section.
\subsubsection{LINAC}
\begin{figure}[htb]
\begin{center}
\includegraphics[scale=0.4]{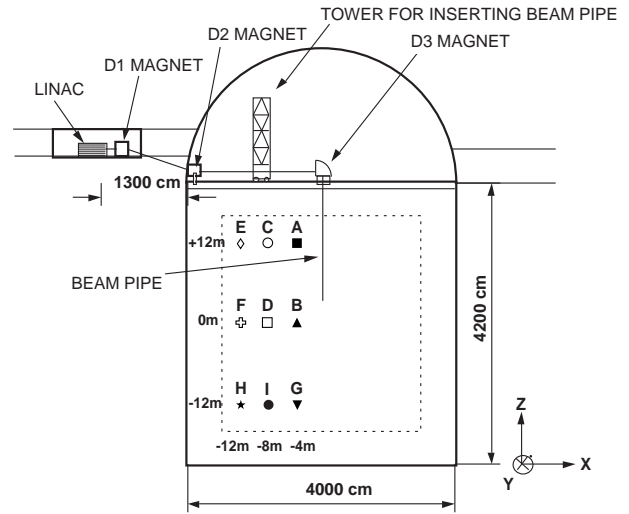}
\caption{The LINAC system at SK. The dotted line represents the fiducial
         volume of the detector and the black dots indicate the positions
         where LINAC data were taken.}
\label{fig:hw-all}
\end{center}
\end{figure}

 The primary instrument for energy calibration is an electron linear
accelerator [LINAC]. The LINAC calibration of Super--K has been discussed 
in detail elsewhere~\cite{linac}.
The LINAC is used to inject downward-going electrons
of known energy and position into the SK tank.
The momentum of these electrons is tunable, with a range of 5.08~MeV to 
16.31~MeV; this corresponds well with the energies of interest to 
solar neutrino studies.
LINAC data is collected at nine different positions in the SK tank,
as shown in Figure~\ref{fig:hw-all}. The data are compared to MC,
and MC is adjusted until its absolute energy scale agrees well with data 
in all the positions and momenta. Once adjusted, this MC is extrapolated
to cover events in all directions throughout the entire detector.

 The electrons are introduced into SK via a beam pipe whose 
endcap's exit window is a 100~$\mu$m-thick sheet of titanium.  This 
allows electrons to pass through without significant momentum loss
but prevents water from entering the beam pipe.

 Seven different momentum data sets are taken at each LINAC position.
The absolute energy in the LINAC system is measured with a germanium [Ge] 
detector. Table \ref{tab:mom} shows the different momenta used in the LINAC
calibration. The average occupancy at the beam pipe endcap is set to about
0.1~electrons/spill. The reason for this low occupancy is to reduce the 
number of spills which include multiple electrons.

\begin{table}
\begin{center}
\begin{tabular}{|c|c|c|}  \hline
beam momentum & Ge energy & in-tank energy \\
(MeV/c)       & (MeV)     & (MeV) \\ \hline
5.08          & 4.25      & 4.89 \\
6.03          & 5.21      & 5.84 \\
7.00          & 6.17      & 6.79 \\
8.86          & 8.03      & 8.67 \\
10.99         & 10.14     & 10.78 \\
13.65         & 12.80     & 13.44 \\
16.31         & 15.44     & 16.09 \\ \hline
\end{tabular}
\end{center}
\caption{LINAC beam momentum. The second column gives 
the energy measured in the Ge calibration system. The last column lists 
the total energy of the electrons after leaving the beam pipe. }
\label{tab:mom}
\end{table}

\begin{figure}
\includegraphics[width=8.0cm]{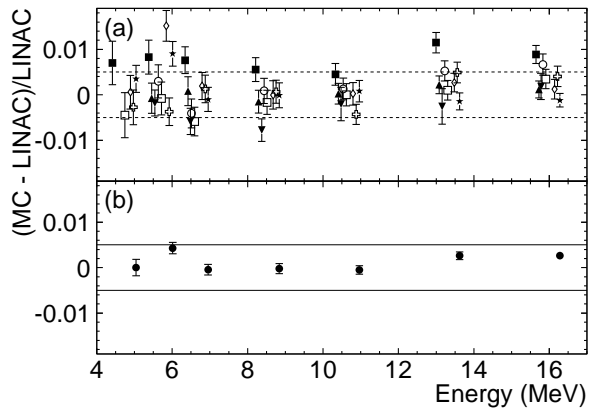}
\caption{The relative difference between the reconstructed energy of 
         LINAC data and the corresponding MC. (a) The results are shown 
         for all positions and beam momenta. (b) Averaged over all
         positions.
 The dashed lines show $\pm$ 0.5\%.
 See Fig.~\ref{fig:hw-all} for the positions.}
\label{fig:lin-edep}
\end{figure}

 The energy scale and resolution obtained by LINAC is compared to MC
at each position and momentum. Figure~\ref{fig:lin-edep}(a) shows the deviation
in the energy scale between data and MC. 
Figure~\ref{fig:lin-edep}(b) shows the average over all the positions
from (a). The deviation at each position is less than 1\% and the position
averaged momentum dependence of the deviation is less than 0.5\%.

\begin{figure}
\includegraphics[width=8.0cm]{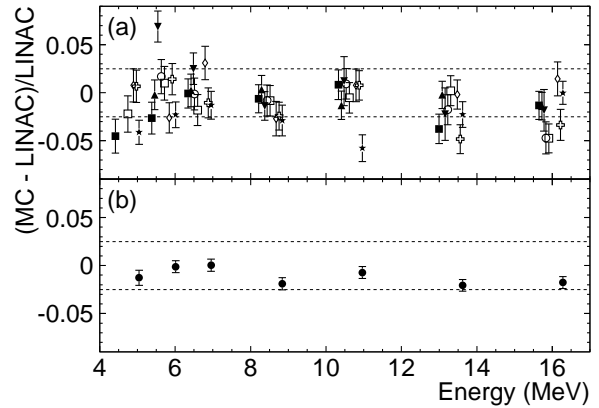}
\caption{The reconstructed energy resolution from LINAC data and MC.
         (a) shows the deviation of the resolution 
         in each energy and position in LINAC calibration
         between data and MC. (b) Averaged over all positions.
 The dashed lines show $\pm$ 2.5\%. 
 See Fig.~\ref{fig:hw-all} for the positions.}
\label{fig:lin-eres}
\end{figure}

Figure~\ref{fig:lin-eres} (a) shows the deviation of the energy resolution
at each energy and position in LINAC calibration between data and MC,
and (b) shows these deviations averaged over all positions.
The difference in reconstructed energy resolution between data and MC
is less than 2.5\%.

 The absolute energy scale of the detector was tuned by using the LINAC 
calibration data. The uncertainty of the absolute energy scale comes from
the uncertainties of water transparency measurement when the LINAC calibration
data was taken (0.22\%), position dependence of the energy scale (0.21\%),
time variation of the energy scale obtained by LINAC calibration (0.11\%),
tuning accuracy of MC simulation (0.1\%), electron beam energy determination 
in LINAC calibration (0.21\%), and directional dependence of energy scale
since the LINAC beam is only downward-going (0.5\%).
Adding those contributions in quadrature, the total uncertainty of 
Super--K--I's absolute energy scale is estimated to be 0.64\%.

\label{sec:enelin}
\newcommand{\nsixteen}{\mbox{$^{16}N$}}
\newcommand{\osixteen}{\mbox{$^{16}O$}}
\newcommand{\epeak}{\mbox{$E_{peak}$}}
\newcommand{\rup}{\mbox{$R_{up}$}}
\newcommand{\rdown}{\mbox{$R_{down}$}}

\subsubsection{\nsixteen\ from the DT Generator}
 Even though the energy scale is determined from LINAC calibration and is
fixed, \nsixteen\ was also used as a calibration source. 
Since events from \nsixteen\ decay are isotropic, they are useful in probing
the directional dependence of the energy scale. Also, because of the 
portability of the DT generator we were able to probe the energy scale 
at many positions in the detector on a monthly basis.

 With a half-life of 7.13 seconds, the Q-value of the decay of 
\nsixteen\ is 10.4~MeV, and the most probable
decay mode produces a 6.1~MeV $\gamma$ ray together with a $\beta$
decay electron of maximum energy 4.3~MeV.
Man-made \nsixteen\ was obtained using a deuterium-tritium neutron
generator [``DT generator''].  The DT generator uses the fusion
reaction $^2H + ^3H \rightarrow ^4He + n$ to produced 14.2~MeV neutrons.
A fraction of the neutrons collide with \osixteen\ to produce \nsixteen.
The DT generator thus provided us with a large sample of essentially
background free \nsixteen\ data for use in energy scale calibration.
More details about the DT generator are given elsewhere~\cite{nim_dt}.

\begin{figure}[htb]
\includegraphics[width=8cm,clip]{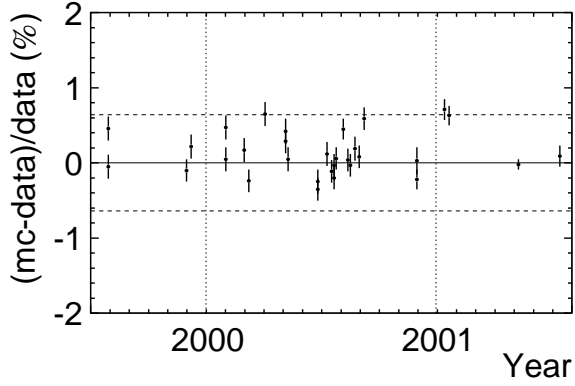}
\caption{The deviation between data and MC for the volume-averaged 
DT energy scale as a function of time. The DT generator was 
available for the last two years of SK--I, starting in July, 1999.
The dashed line shows the $\pm$0.64\% of energy scale uncertainty, which
is estimated by LINAC calibration as discussed in Section\ref{sec:enelin}.}
\label{fig:escale_vavg}
\end{figure}

 The \nsixteen\ decay energy spectrum is a superposition of several
$\gamma$ ray lines and $\beta$ continua of various end points.
The reconstructed energy spectrum has a peak around $\epeak = 6.9 \sim
7.0$~MeV.  The shape of the peak region ($5.5 \sim 9.0$~MeV) is
approximated by a Gaussian with a width of $1.6 \sim 1.7$~MeV.
The deviation of this energy peak between data and MC is measured.
Figure~\ref{fig:escale_vavg} shows 31 data sets of DT results
during two years of running, and it shows the deviation is 
within $\pm 1\%$ when averaged over all positions.

\begin{figure}
\includegraphics[width=8.0cm,clip]{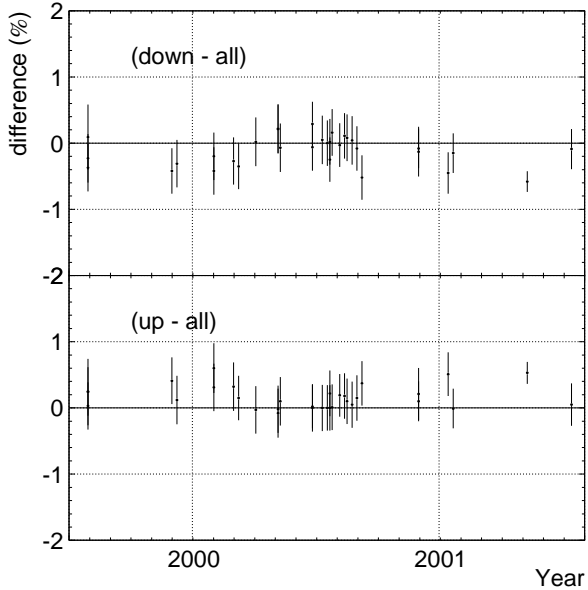}
\caption{Directional dependence of energy scale by DT calibration.
         Upper figure shows $\rdown - R$ as a function of time,
         lower figure shows the same for $\rup - R$. Note that $\rup
         \approx -\rdown$.}
\label{fig:direction_dep}
\end{figure}

 The direction dependence of the energy scale is measured as follows.
First, the deviation between data and MC is calculated for all directions.
Then it is calculated for two divided data samples: events in the  
upward or downward directions. It should be noted that in all cases, 
the energy scale is averaged over the detector volume. 
Figure~\ref{fig:direction_dep} shows plots 
comparing the energy scale obtained from using just downward- or 
upward-going events to that determined by using the entire  
sample. The difference is within $\pm 0.6\%$.

%
%

\subsection{Vertex calibration}

\begin{figure}[hptb]
  \begin{center}
    \includegraphics[width=8.0cm]{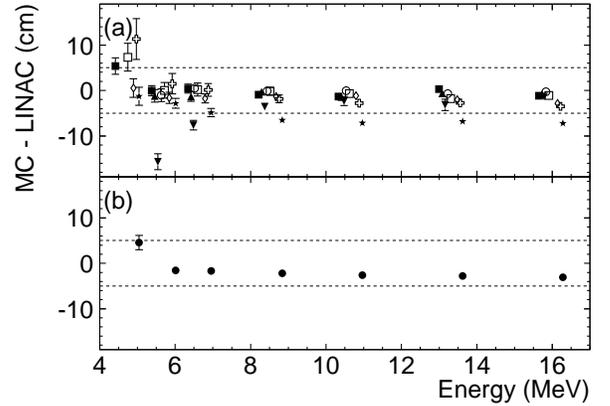}
    \caption{The reconstructed vertex resolution from LINAC data and MC.
         (a) shows the deviation of the resolution in each energy 
   and position in LINAC calibration between data and MC. 
         (b) averaged over all positions.
   The dashed lines show $\pm$ 5 cm.
   See Fig.~\ref{fig:hw-all} for the positions.}
    \label{fig:lin-vres}
  \end{center}
\end{figure}

 Figure~\ref{fig:lin-vres} shows the vertex resolution in each energy and
position in LINAC and Monte Carlo. 
The differences of vertex resolution between data and MC
is less than $\pm5$ cm.

 The vertex shift in the reconstruction is estimated using a Ni(n,$\gamma$)Ni
gamma source,~\cite{sk_detector} because the gamma ray is emitted in almost
uniform directions. The vertex shift is defined as a vector from an averaged
position of the reconstructed vertex of the data to that of a corresponding MC.
Table~\ref{tab:vershift} shows the vertex shift at several source positions.
The systematic error for the solar neutrino flux as a result of vertex 
shift (which could move events in or out of the fiducial volume) 
is evaluated from these values, and it is $\pm1.3\%$.
\begin{table}[hptb]
   \begin{center}
     \begin{tabular}{|c||c|c|c|} \hline
      Position & $\Delta x$ & $\Delta y$ & $\Delta z$ \\ \hline
      (35.3,-70.7,-1200)   & -1.8 & -1.9 & -2.8 \\
      (35.3,-70.7,0)       &  0.6 & -0.5 & -2.8 \\
      (35.3,-70.7,+1200)   & -1.1 & -1.7 &  0.6 \\
      (35.3,-70.7,+1600)   &  0.0 & -3.2 & -3.5 \\
      (35.3,-1555.4,-1200) & -5.5 &  8.6 & -9.1 \\
      (35.3,-1555.4,0)     & -5.5 & 23.4 & -3.0 \\
      (35.3,-1555.4,+1200) & -0.5 &  7.0 &  6.0 \\
      (35.3,-1555.4,+1600) & -3.3 &  7.6 &  2.7 \\
      (35.3,-1201.9,-1200) & -4.2 &  5.4 & -9.1 \\
      (35.3,-1201.9,0)     & -1.4 & 15.5 & -2.2 \\
      (35.3,-1201.9,+1200) &  2.5 &  8.9 &  4.2 \\
      (1520.0,-70.9,-1200) & -5.7 & -1.4 &-10.3 \\
      (1520.0,-70.9,0)     &-18.6 & -3.0 & -4.0 \\
      (1520.0,-70.9,+1200) &-12.6 & -1.0 &  6.2 \\
      (-35.3,-1555.4,-1200) & 3.7 &-11.4 & -8.6 \\
      (-35.3,-1555.4,0)     & 6.6 &-19.5 & -2.0 \\
      (-35.3,-1555.4,+1200) & 1.6 &-20.3 &  7.2 \\ \hline
     \end{tabular}
   \end{center}
   \caption{The vertex shift measured by Ni(n,$\gamma$)Ni gamma source
            calibration. The unit is cm.}
   \label{tab:vershift}
\end{table}

\subsection{Direction calibration}
\begin{figure}[htbp]
  \begin{center}
    \includegraphics[width=8.0cm]{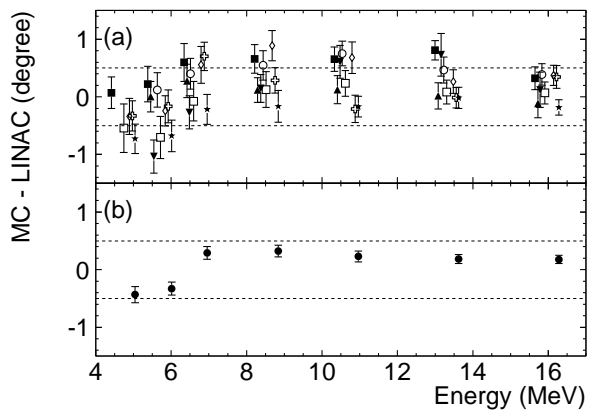}
    \caption{The reconstructed angular resolution from LINAC data and MC.
          (a) shows the deviation of the resolution in each energy 
   and position in LINAC calibration between data and MC.
          (b) is averaged over all positions.
         The dashed lines show $\pm$ 0.5 degree.
   See Fig.~\ref{fig:hw-all} for the positions.}
    \label{fig:lin-ares}
  \end{center}
\end{figure}

 The angular resolution is calibrated by LINAC data and MC.
Figure~\ref{fig:lin-ares} shows the angular resolution variation 
in each energy and position between LINAC and Monte Carlo events.
The difference in angular resolution between LINAC data and MC
is less than $\pm0.5$ degree.

 We have applied this difference as a correction factor to the expected signal 
shape, then taken the same amount of the correction as our systematic error
due to angular resolution. This systematic error is 1.2\%.
%

%
%
\section{Background}\label{sec:bg}
\subsection{Low energy backgrounds}
 The main background sources below about 6.5~MeV for the solar neutrino
events are (1) events coming from outside fiducial volume, and (2) $^{222}$Rn.

\begin{figure}
\center
\includegraphics[width=7.5cm,clip]{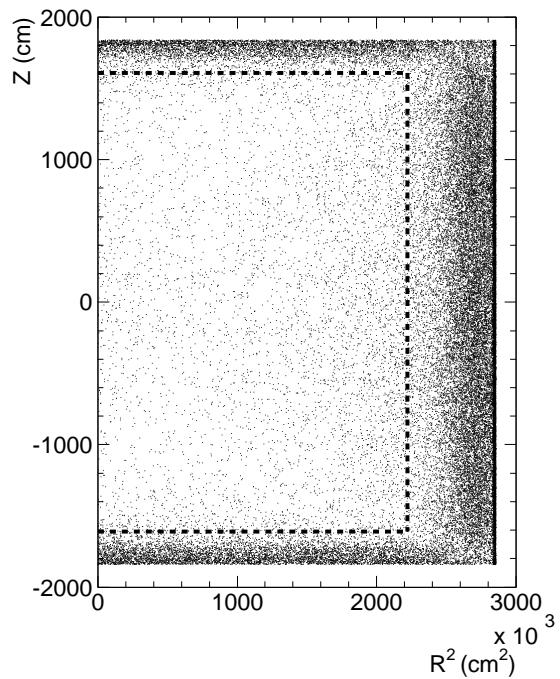}
\caption{Typical vertex distribution of the low energy events
 before the Intelligent Trigger selection.
 The analysis energy threshold for this plot is 5.0~MeV.
 The dashed line shows the fiducial volume edge.
}
\label{external-events}
\end{figure}

 Figure~\ref{external-events} shows a typical vertex distribution of
the low energy events before the Intelligent Trigger selection.
Most of the low energy events occur near the inner detector's wall region. 
They originate from radioactivity of the PMT's, black sheet, 
PMT support structures, and mine rocks surrounding the SK detector. 
Though the fiducial volume is selected to be 2~m from this inner detector wall,
there are a lot of remaining events around the edge of the fiducial volume.
The reconstructed directions of these remaining, externally-produced events 
in the fiducial volume are pointed, on average, strongly inward.
We have eliminated most of these external events by using an event
selection based upon vertex and direction information.
The details of this event selection will be explained 
in Section~\ref{sec:gamcut}.
Although most of these external events are eliminated by this cut, 
some quantity of external events still remain. 
These remaining external events are one of the major backgrounds in
this energy region.

 Another major source of background in the low energy region are 
radioactive daughter particles from the decay of $^{222}$Rn. 
$^{214}$Bi, which is one of these daughter particles, undergoes  
beta decay with resulting electron energies up to 3.26~MeV.
Due to the limited energy resolution of the detector, these electrons 
can be observed in this energy region.

 We have reduced $^{222}$Rn in the water by our water purification
system to contribute less than 1~mBq/m$^3$~\cite{sk_detector}, and monitor 
the radon concentration in real-time by several radon 
detectors~\cite{radon1,radon2,radon3}. 
However, the water flow from the water inlets, located 
at the bottom of the SK detector, stirs radon emanated from the inner 
detector wall into the fiducial volume~\cite{radon4}.  
Therefore, there is an event excess after the external event cut 
in the detector bottom region as compared to the top region. 
We also supply radon-reduced air~\cite{radon4} into the space 
above the water surface in the SK tank to prevent radon 
in the mine air from dissolving into the purified water.
The radon concentration in this radon-reduced air is less than 3~mBq/m$^3$
and the measured radon level in the air in the tank is stable at
20$\sim$30~mBq/m$^3$. Periods of anomalously high radon concentration 
due to water system troubles were removed from this analysis.

\subsection{High energy backgrounds}
 Above about 6.5~MeV, the dominant background source is radioactive
isotopes produced by cosmic ray muons' spallation process with oxygen nuclei. 
Some fraction of downward-going cosmic ray muons interact with 
oxygen nuclei in the water and produce various radioactive isotopes. 
These radioisotopes are called ``spallation products.'' 
Table~\ref{tab:spaeve} shows a summary of possible spallation products in SK.
The $\beta$ and/or $\gamma$ particles from the radioactive 
spallation products are observed in Super--K, causing 
what are called ``spallation events.''
The spallation events retain some correlation in time and space 
with their parent muons.
Using this correlation, we have developed a cut to remove
these spallation background events efficiently.
The detail of this cut will be explained in Section~\ref{sec:spacut}.
\begin{table}
 \begin{center}
  \begin{tabular}{cccc}
   \hline
   \hline
   Isotope & $\tau_{\frac{1}{2}}$(sec) & Decay mode & Kinetic Energy(MeV) \\
   \hline
   ${}^8_2$He & 0.119 & $\beta^-$ & 9.67 + 0.98 ( $\gamma$ ) \\
   & & $\beta^-$ n & ( 16 \% ) \\
   ${}^8_3$Li & 0.838 & $\beta^-$ & $\sim$13 \\
   ${}^8_5$B & 0.77 & $\beta^+$ & $\sim$13.9 \\
   ${}^9_3$Li & 0.178 & $\beta^-$ & 13.6 ( 50.5 \% ) \\
   & & $\beta^-$ n & ( $\sim$50 \% ) \\
   ${}^9_6$C & 0.127 & $\beta^+$ p & 3 $\sim$ 15 \\
   ${}^{11}_3$Li & 0.0085 & $\beta^-$ & 16 $\sim$ 20 ($\sim$50\%) \\
   & & $\beta^-$ n & $\sim$16 ($\sim$50\%) \\
   ${}^{11}_4$Be & 13.8 & $\beta^-$ & 11.51 ( 54.7 \% ) \\
   & & & 9.41 + 2.1 ( $\gamma$ ) ( 31.4 \% ) \\
   ${}^{12}_4$Be & 0.0236 & $\beta^-$ & 11.71 \\
   ${}^{12}_5$B & 0.0202 & $\beta^-$ & 13.37 \\
   ${}^{12}_7$N & 0.0110 & $\beta^+$ & 16.32 \\
   ${}^{13}_5$B & 0.0174 & $\beta^-$ & 13.44 \\
   ${}^{13}_8$O & 0.0086 & $\beta^+$ & 13.2, 16.7 \\
   ${}^{14}_5$B & 0.0138 & $\beta^-$ & 14.55 + 6.09 ( $\gamma$ ) \\
   ${}^{15}_6$C & 2.449 & $\beta^-$ & 9.77 ( 36.8 \% ) \\
   & & & 4.47 + 5.30 ( $\gamma$ ) \\
   ${}^{16}_6$C & 0.747 & $\beta^-$ n & $\sim$4 \\
   ${}^{16}_7$N & 7.13 & $\beta^-$ & 10.42 ( 28.0 \% ) \\
   & & & 4.29 + 6.13 ( $\gamma$ ) ( 66.2 \% ) \\
   \hline
   \hline
  \end{tabular}
  \caption{Possible radioactive spallation products in Super--Kamiokande.}
  \label{tab:spaeve}
 \end{center}
\end{table}
%

%
%
\section{Data analysis}\label{sec:red}
\subsection{Noise reduction}
\begin{figure}[hbpt]
\begin{center}
\includegraphics[scale=0.35]{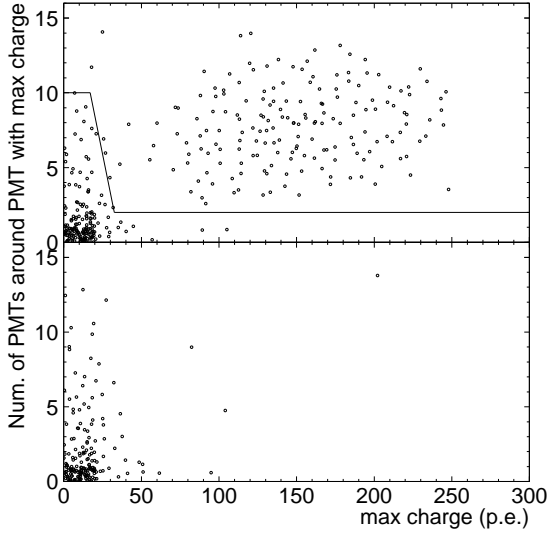}
\caption{The relation between max charge and the number of hit PMT's 
         in a $5 \times 5$ patch surrounding the PMT with the maximum charge.
	 At most 24 tubes in this patch (plus the one at the center) can be
	 activated. The upper plot shows a
         typical distribution including active 
         flasher events, and the lower plot 
         shows a typical good data set. The cut region is the area above 
	 the line in the upper figure.}
\label{fig:fla}
\end{center}
\end{figure}

 The first step of the data reduction is an elimination of the noise and
obvious background events. First of all, the events with total photoelectrons
less than 1000, which corresponds to $\sim 100$~MeV, are selected. 
Next, the following data reductions are applied:
(a) Events whose time difference since a previous event was less 
than 50~$\mu$sec were removed in order to eliminate decay electrons 
from stopping muons.
(b) Events with an outer detector trigger and over 20 outer PMT hits
were removed in order to eliminate entering events like cosmic ray muons.
(c) A function to categorize noise events is defined by the ratio
of the number of hit PMT's with $|Q| \leq 1.0$~p.e. and the 
total number of hit
PMT's. Since typical noise events should have many hits with low charge,
events with this ratio larger than 0.4 were removed.
(d) A function which can recognize an event where most of its hits are 
clustered in one electronics module [ATM] is defined.
It is the ratio of the maximum number of hits in any one module 
to the total number of channels (usually 12) in one ATM module.
Events with a larger ratio than 0.95 were removed as they generally arose 
due to local RF noise in one ATM module.
(e) Events produced by flasher PMT's must be removed. 
These flasher events often have a relatively larger charge than normal events,
therefore they are recognized using the maximum charge value and the number of
hits around the maximum charge PMT. The slightly involved criteria is shown in
Figure~\ref{fig:fla}. (f) An additional cut to remove noise and flasher events
is applied using a combination of a tighter goodness cut($goodness \leq 0.6$)
and a requirement of azimuthal uniformity in the Cherenkov ring pattern. Good
events have a uniform azimuth distribution of hit-PMT's along the
reconstructed direction, while flasher events often have clusters of hit
PMT which lead to a non-uniform azimuth distribution.

 In this noise reduction, the number of candidate solar neutrino 
events went from $3.43\times10^7$ to $1.81\times10^7$ after applying the  
2~m fiducial volume cut and constraining the energy region from 5.0 to
20.0~MeV. Table~\ref{tab:reduc} shows the reduction step summary.

 The loss of solar neutrino signal by the noise reduction is
evaluated to be $0.8\%$ using a Ni(n,$\gamma$)Ni gamma source at several edge
positions of the inner detector. The difference between data and MC for this
source yields the 
systematic error for this series of reductions; it's $\pm 1.0\%$ for our 
solar neutrino flux
measurement. The loss is mainly due to the flasher cut.

\subsection{Spallation cut}\label{sec:spacut}
 The method to reject spallation products shown in Section~\ref{sec:bg}
is described in this section. In order to identify spallation events,
likelihood functions are defined based on the following parameters: 
\begin{itemize}
 \item $\Delta L$ : Distance from a low energy event to the track of
the preceding muon.
 \item $\Delta T$ : Time difference from muon event to the low energy event.
 \item $Q_{res}$ : Residual charge of the muon event,
$Q_{total} - Q_{unit} \times L_{\mu}$, where $Q_{total}$ is the total charge,
$Q_{unit}$ is the total charge per track length and $L_{\mu}$ is reconstructed
track length of the muon track.
\end{itemize}
Some fraction of muons deposit very large amounts of energy in the detector
and in such cases vertex position reconstruction is not reliable.
Therefore, the spallation likelihood functions are defined for the following
two cases,\\
in the case of a successful muon track fit;
\begin{eqnarray}
\lefteqn{L_{spa}(\Delta L,\Delta T, Q_{res}) = }\nonumber\\
&&L_{spa}^{\Delta L}(\Delta L, Q_{res}) 
\times L_{spa}^{\Delta T}(\Delta T) 
\times  L_{spa}^{Q_{res}}(Q_{res}),
\end{eqnarray}
in the case of a failure fit the muon track;
\begin{equation}
L_{spa}(\Delta T, Q_{total}) =
L_{spa}^{\Delta T}(\Delta T) 
\times  L_{spa}^{Q_{total}}(Q_{total}).
\end{equation}
where $L_{spa}^{\Delta L}(\Delta L, Q_{res})$, 
$L_{spa}^{\Delta T}(\Delta T)$, $L_{spa}^{Q_{res}}(Q_{res})$ and 
$L_{spa}^{Q_{total}}(Q_{total})$ are likelihood functions 
for $\Delta L$, $\Delta T$ and $Q_{res}$.

\begin{figure}[htbp]
  \begin{center}
  \includegraphics[scale=0.46]{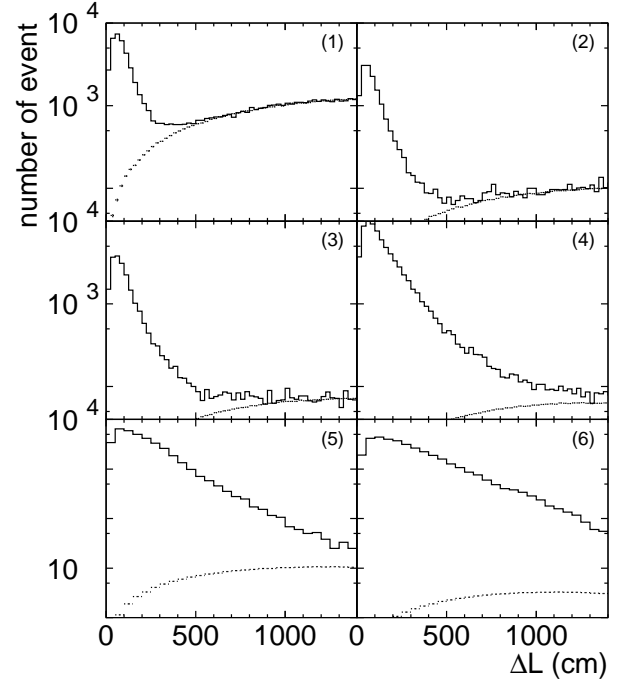}
  \caption{$\Delta L$ distribution for each $Q_{res}$ ranges,
           (1) $Q_{res}<2.4\times 10^4$~p.e.
           (2) $2.4\times 10^4 < Q_{res} < 4.8\times 10^4$~p.e.
           (3) $4.8\times 10^4 < Q_{res} < 9.7\times 10^4$~p.e.
           (4) $9.7\times 10^4 < Q_{res} < 4.8\times 10^5$~p.e.
           (5) $4.8\times 10^5 < Q_{res} < 9.7\times 10^5$~p.e.
           (6) $9.7\times 10^5$~p.e.$ < Q_{res}$.
           The solid line shows the data, and the dashed line shows the
           random sample.}
  \label{dl}
  \end{center}
\end{figure}

 Figure~\ref{dl} shows the $\Delta L$ distributions 
from spallation candidates for six $Q_{res}$ ranges,
and the spallation-like function is made from these plots.
Here, the selection criteria is $\Delta T < 0.1~sec$ 
and $N_{eff} \ge 50$ (equivalent to 7.2~MeV). The peak around 0 $\sim$ 100~cm
is caused by spallation events and that around 1500~cm is due to chance
coincidence. The distribution of non-spallation events is calculated using
a sample which shuffles the event times time randomly; 
it is the dashed line in the figure, and the non-spallation-like 
function is also made by these plots.
After subtracting the non-spallation from the spallation function,
and taking a ratio with a random coincidence distribution, 
$ L_{spa}^{\Delta L}(\Delta L, Q_{res}) $
was obtained.

\begin{figure}[htbp]
  \begin{center}
  \resizebox{8.0cm}{!}{
  \includegraphics{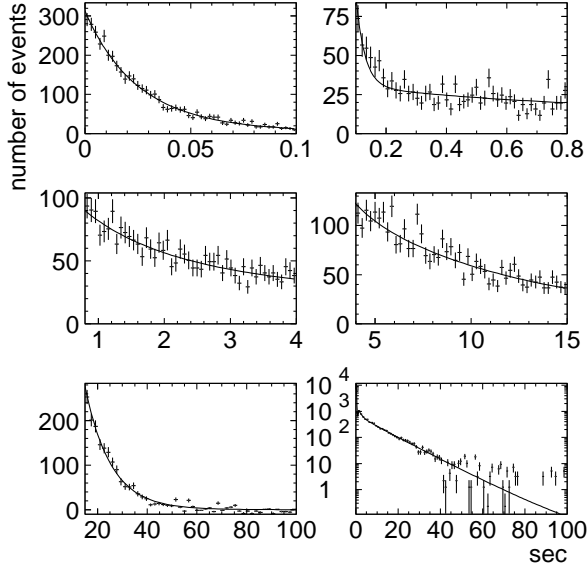}
  \vspace{-1cm}
  }
 \caption{$\Delta T$ distribution for each time range,
          (1) $0.0 < \Delta T < 0.1$~sec
          (2) $0.1 < \Delta T < 0.8$~sec
          (3) $0.8 < \Delta T < 4.0$~sec
          (4) $4.0 < \Delta T < 15$~sec
          (5) $15 < \Delta T < 100$~sec
          (6) $0 < \Delta T < 100$~sec.
   Cross marks are data
  and lines are fitted likelihood function $L_{spa}^{\Delta T}(\Delta T)$}.
  \label{dt}
  \end{center}
\end{figure}

\begin{figure}[htbb]
 \includegraphics[scale=0.44]{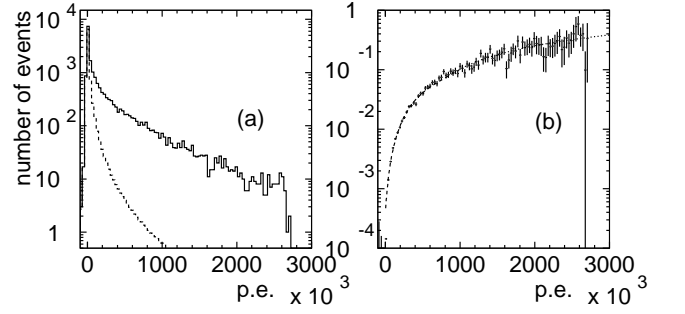}
 \caption{(a) $Q_{res}$ distribution for the spallation candidate events 
	  (solid line) and the non-candidate events (dashed line).
          (b) Cross marks show the result of subtracting 
          dashed from solid line in (a),
          and dotted line shows the likelihood function
          $L_{spa}^{Q_{res}}(Q_{res})$.}

  \label{qres}
\end{figure}

 Figure~\ref{dt} shows the $\Delta T$ distributions
from spallation candidates for each time range.
Here, the selection criteria is $\Delta L <$ 300~cm
and $N_{eff} \ge 50$ and $Q_{res} < 10^6$~p.e..
These distributions are fitted with the following function:
\begin{equation}
L_{spa}^{\Delta T}(\Delta T) = \sum_{i=1}^{7} A_i \left( \frac{1}{2} \right)^{-\frac{\Delta T}{\tau^i_{1/2}}}
\end{equation}
where $\tau^i_{1/2}$ are half-life times of typical radioactive elements 
produced by spallation. The evaluated half-life times and 
radioisotopes are summarized in Table~\ref{tab:dttable}.
\begin{table}
 \begin{center}
  \begin{tabular}{|c|c|c|c|}
   \hline
   $i$ & radioactivity & $\tau^i_{1/2}$ & $A_i$ \\
   \hline
   1 & $^{12}_{5}$B & $2.02\times 10^{-2}$ & $1.20\times 10^5$ \\
   \hline
   2 & $^{12}_{7}$N & $1.10\times 10^{-2}$ & $3.39\times 10^4$ \\
   \hline
   3 & $^{9}_{3}$Li & $1.78\times 10^{-1}$ & $3.39\times 10^2$ \\
   \hline
   4 & $^{8}_{3}$Li & $8.40\times 10^{-1}$ & $1.25\times 10^3$ \\
   \hline
   5 & $^{15}_{6}$C & 2.45 & $1.35\times 10^2$ \\
   \hline
   6 & $^{16}_{7}$N & 7.13 & $6.76\times 10^2$ \\
   \hline
   7 & $^{11}_{4}$Be & 13.83 & 7.79 \\
   \hline
  \end{tabular}
  \caption[]{The parameters of the likelihood $L^{\Delta T}_{spa}(\Delta T)$}
  \label{tab:dttable}
 \end{center}
\end{table}
 In order to obtain the likelihood function for residual charge 
$L_{spa}^{Q_{res}}(Q_{res})$, time correlated events with 
low energy events ($\Delta T <$ 0.1sec., $N_{eff} \ge$ 50) and 
non-correlated events are selected.
Figure~\ref{qres}(a) shows the $Q_{res}$ distribution for 
spallation and non-spallation candidates.
After subtracting the non-spallation from the spallation function,
the resulting distribution's fit by a polynomial function 
is shown in Figure~\ref{qres}(b).

\begin{figure}[htbp]
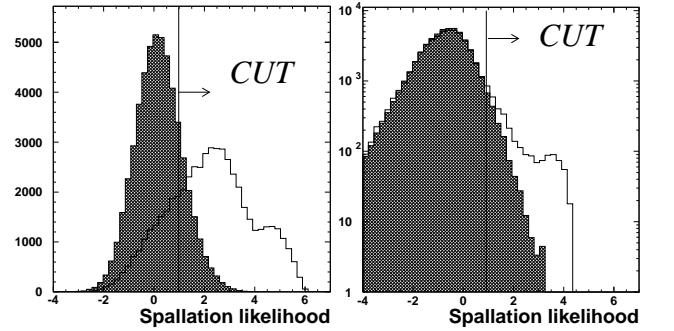

  \begin{center}
  \includegraphics[width=4.2cm,clip]{./data-analysis-spacut/likemax3.epsi}
  \includegraphics[width=4.0cm,clip]{./data-analysis-spacut/likemax2.epsi}
  \caption{The maximum likelihood value distribution for
      success (left) and failure (right) of muon track reconstruction.
      The blank histogram shows all data while the 
      hatched histogram shows the plot for the random sample; the latter
      is used to evaluate dead time.}
  \label{likemax}
  \end{center}
\end{figure}

 To employ the spallation cut, the likelihood values are calculated for 
all muons in the previous 100 seconds before a low energy event, and 
a muon is selected which gives the maximum likelihood value($L_{max}$). 
Figure~\ref{likemax} shows $L_{max}$ distributions for both the 
successful muon track reconstruction case and the failed reconstruction case.
To be considered a spallation event the selection criteria 
are $L_{max} > 0.98$ (when fit succeeded) and $L_{max} > 0.92$ (when 
fit failed).

\begin{figure}[htbp]
  \includegraphics[scale=0.44]{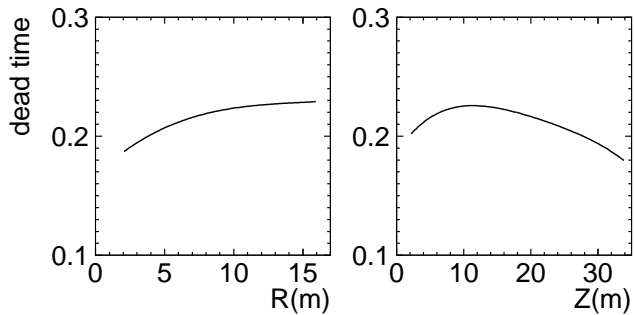}
  \caption{Position dependence of the spallation cut's dead time. 
           The horizontal axis shows the distance from the barrel (left), 
	   and the minimum distance from the top or bottom (right).}
  \label{dead}
\end{figure}

 The dead time for low energy events caused by the spallation cut is
estimated to be 21.1\%. This estimation is calculated using a low energy sample
which is shuffled in time and position. It has some position dependence because
of the SK tank geometry. Figure~\ref{dead} shows the position dependence
of the dead time as a function of the distance from the inner barrel 
wall and the top or bottom walls. The systematic error due to position 
dependence is estimated by comparing between MC and data, and 
it is estimated to be $\pm0.2\%$ for flux, and $\pm0.1\%$ for day-night flux
difference and other time variations.

\subsection{Ambient background reduction}\label{sec:2nd}
 Even after the fiducial volume cut, some fraction of the 
ambient background still
remains. It is mainly due to mis-reconstruction of the vertex position.
In order to remove the remaining background, several cuts to evaluate the
quality of the reconstructed vertex were applied.

\begin{figure}[htbp]
  \begin{center}
  \includegraphics[scale=0.4]{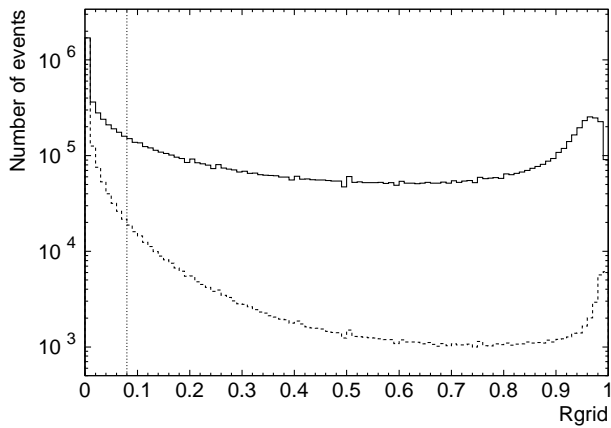}
  \caption{$R_{grid}$ distribution for data (solid line) and MC (dashed line).
           The normalization is done by scaling to the first bin. 
           The dotted line
           shows the reduction criteria.}
  \label{fig:gringo}
  \end{center}
\end{figure}

\paragraph{Fit stability cut}
The goodness value of the vertex was calculated for points in the 
region of the reconstructed vertex position and the sharpness of 
the goodness distribution as a function of detector coordinates was evaluated. 
First of all, the goodness at 
$\sim300$ grid points around the original vertex are calculated, and 
their difference
from the goodness at the reconstructed vertex ($\Delta_g$) is determined.
The number of test points which give a $\Delta_g$ more than some threshold
was counted; the threshold is as a function of energy and vertex position.
The ratio of the number of points over this threshold to the total number of 
grid points 
is defined as $R_{grid}$. The $R_{grid}$ distribution of the data
is shown in Figure~\ref{fig:gringo} along with simulated 
$^8$B Monte Carlo events.
Events with $R_{grid} > 0.08$ are rejected as background.
The systematic error of this reduction was evaluated using LINAC data and MC
and the systematic error for flux measurement is estimated to be $\pm1.0\%$.

\begin{figure}[htbp]
  \begin{center}
  \includegraphics[scale=0.4]{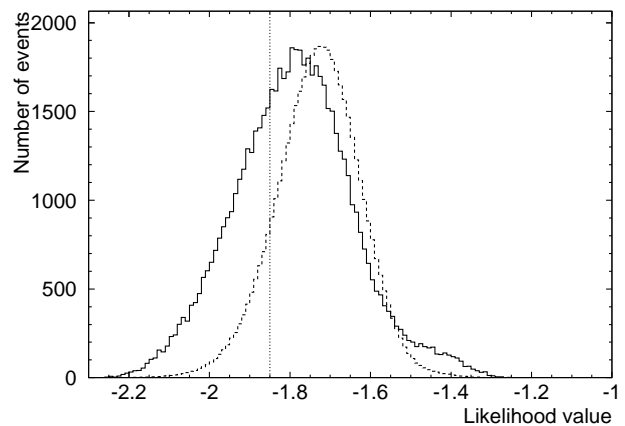}
  \caption{Likelihood distribution for data (solid line) and MC (dashed line).
           The MC is normalized to the peak bin. The dotted line
           shows the reduction criteria.}
  \label{fig:patlik}
  \end{center}
\end{figure}

\paragraph{Hit pattern cut}
Mis-reconstructed events often do not have the expected Cherenkov ring pattern 
when the hit PMT's are viewed from the reconstructed vertex. Also, some 
fraction
of spallation products like $^{16}$N emit multiple gammas in addition to 
an electron and so do not fit a single Cherenkov ring pattern very well. 
In order to remove those kind of events, the following likelihood 
function is defined:
\begin{equation}
  L(E, \vec x ) \equiv \frac{\sum_{i=1}^{N_{hit}} \log (f{(\cos \theta_{dir},E, \vec x )})_i}
               {N_{hit}},
  \label{eq:patlik}
\end{equation}
here, \(f{(\cos \theta_{dir}E,\vec x)}\) is the same function as
that derived from MC in Eq.~(\ref{eq:dirlike}), but here it depends on  
energy and vertex position. Figure~\ref{fig:patlik} shows 
the likelihood distribution
Eq.~(\ref{eq:patlik}) for data and $^8$B MC. The events with a likelihood less
than $-1.85$ are rejected. The systematic error of this cut is estimated using
solar neutrino MC and a sample of very short lived spallation products; 
it is $+1.0\%$ and $-0.5\%$ for the flux measurement.

\begin{figure}[htbp]
  \begin{center}
  \includegraphics[scale=0.4]{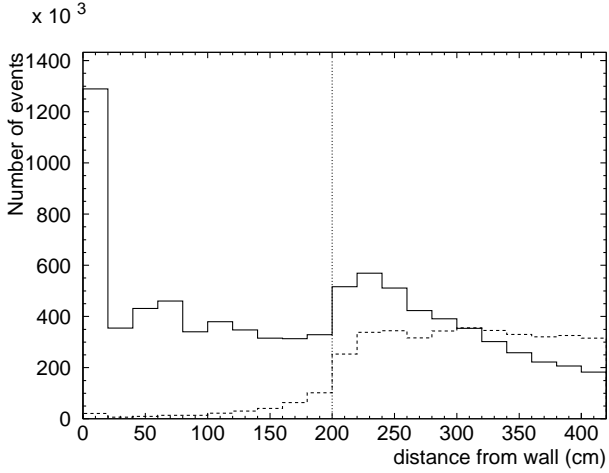}
  \caption{The distance from the wall distribution using the second vertex
           reconstruction for data (solid) and MC (dashed).
           The normalization of MC is done by the total number of events.
           The dotted line shows the reduction criteria.}
  \label{fig:cluscut}
  \end{center}
\end{figure}

\paragraph{Fiducial volume cut using the second vertex fitter}
The 2~m fiducial volume cut using the second vertex reconstruction described
in Section~\ref{sec:vertex} is applied. Figure~\ref{fig:cluscut} shows the
distance from the wall distribution of data and MC using the second vertex
reconstruction. Table~\ref{tab:reduc} also shows the reduction step summary
described in this section.

\subsection{Gamma ray cut}\label{sec:gamcut}
 The gamma-ray cut is used for reduction of external gamma rays which 
mainly come from the surrounding rock, PMT glass, and stainless steel 
support structure of the detector.
These gamma rays are one of the major background especially for solar 
neutrino data.

\begin{figure}[htbp]
  \begin{center}
  \includegraphics[scale=0.45]{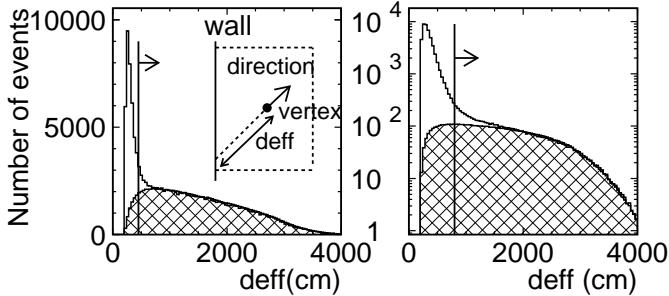}
  \caption{Effective distance (d$_{eff}$) of $E$ $\ge$ 6.5~MeV (left)
           and 5.0MeV $\le$ $E <$ 6.5~MeV (right).
           Blank histogram shows data and hatched area shows solar neutrino MC.}
  \label{effwall}
  \end{center}
\end{figure}

 The distinctive feature of external gamma rays is that they travel from 
the outside edge of the SK volume and continue on to the inside.
In order to remove this kind of event, reconstructed direction and vertex 
information is used to define the effective distance from the wall,
d$_{eff}$, as shown in the inset of the left plot in Figure~\ref{effwall}.
As they emanate from the wall itself, the value d$_{eff}$ for external 
gamma ray events should be small.
Figure~\ref{effwall} also shows d$_{eff}$ distributions for data and MC.
The criterion of the gamma ray cut is determined by maximizing its 
significance; here the number of remaining events after reduction
in data and MC are used to define the significance, $ N_{MC}/\sqrt{N_{data}}$.
The gamma ray cut criteria are determined to be:
\begin{enumerate}
 \item d$_{eff}$ $\ge$ 450~cm ( for $E \ge$ 6.5~MeV )
 \item d$_{eff}$ $\ge$ 800~cm ( for 5.0MeV $\le$ $E <$ 6.5~MeV)
\end{enumerate}

\begin{figure}[htbp]
  \begin{center}
  \includegraphics[scale=0.45]{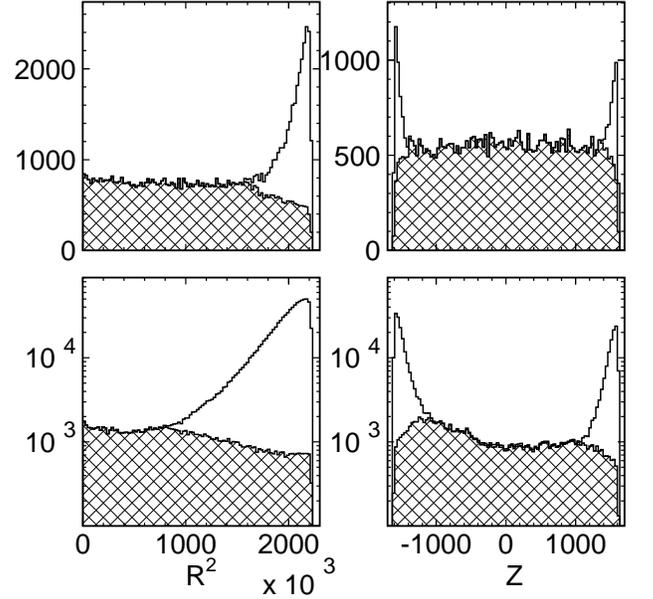}
  \caption{Vertex distribution before (blank) and after
	   (hatched) applying the gamma ray cut. The left plots 
            are R$^2$ for
            $|$Z$|<$1000~cm, and the right ones are Z for $|$R$|<$1000~cm.
            The upper plots show $E$ $\ge$ 6.5~MeV, and lower ones show 
            5.0~MeV $\le$ $E <$ 6.5~MeV.}
  \label{fig:gamver}
  \end{center}
\end{figure}

\begin{figure}[htbp]
  \includegraphics[scale=0.38]{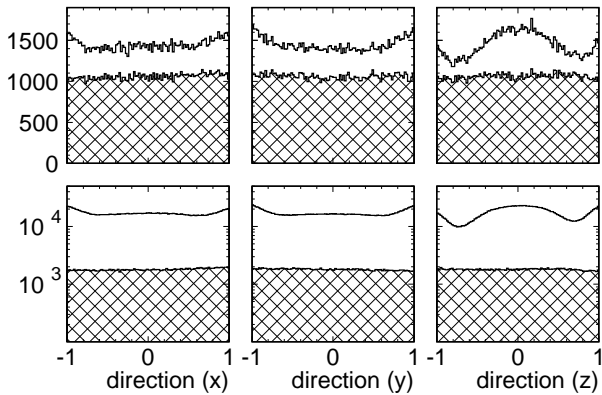}
  \caption{Directional distribution (X, Y, Z) before (blank) and after
	   (hatched) applying the gamma ray cut. The upper plots  
	   show $E$ $\ge$ 6.5~MeV,
           and the lower plots show 5.0~MeV $\le$ $E <$ 6.5~MeV.}
  \label{fig:gamdir}
\end{figure}

 Figure~\ref{fig:gamver} and \ref{fig:gamdir} show the vertex and direction
distributions before and after the gamma ray cut for real data in the  
energy regions $E$ $\ge$ 6.5~MeV and 5.0~MeV $\le$ $E <$ 6.5~MeV.
The dead times introduced by this cut in the energy regions 
$E$ $\ge$ 6.5~MeV and 5.0~MeV $\le$ $E <$ 6.5~MeV are estimated 
to be 6.9\% and 22.0\%.

 Since the gamma ray cut uses reconstructed vertex and direction, the 
differences of vertex and angular resolution between data and MC 
can introduce systematic errors. In order to estimate this effect, we
shift the reconstructed vertex and direction of events 
within the difference of data and MC, which are measured by LINAC,
and apply the gamma ray cut to these events.
The difference of reduction efficiency before and after this shift
is taken to be the systematic error for the gamma ray cut.
The systematic error of this cut for the flux measurement is estimated 
to be $\pm 0.5\%$ , while for the energy spectrum it is estimated 
to be $\pm 0.1\%$.

\subsection{$^{16}$N cut}
 $^{16}$N background is generated by the capture of $\mu^-$
on $^{16}$O nucleus in water:
$$^{16}O + \mu^- \to ^{16}N + \nu_{\mu}.$$
The most probable decay mode of $^{16}$N produces a 6.1~MeV $\gamma$ ray
together with a $\beta$ decay electron of maximum energy 4.3~MeV;
its half life is 7.13~seconds.

 To tag this kind of event, spatial and time correlations with
captured stopping muons are used. The selection method is as follows: 
1) in order to select only a captured muon sample, 
collect a sample of stopping muons which are not followed by a 
decay event in 100~$\mu$sec, then 
2) select a low energy event sample within 335~cm from the stopping point 
of the muon as well as in a time window of 100~msec to 30~seconds following the
stopping muon. The number of such candidate low energy events is 
9843 in 1496~days of low energy data. The dead time for this 
reduction is estimated by using a so-called random sample, where the time and 
vertex positions of the events have been mixed randomly.  The same $^{16}$N 
event selection as described above is then 
applied to the random sample, yielding a dead time for the 
$^{16}$N cut of 0.49\%.

\subsection{Event reduction summary}
\begin{figure}[htbp]
 \begin{center}
  \resizebox{8.5cm}{!}{
  \includegraphics{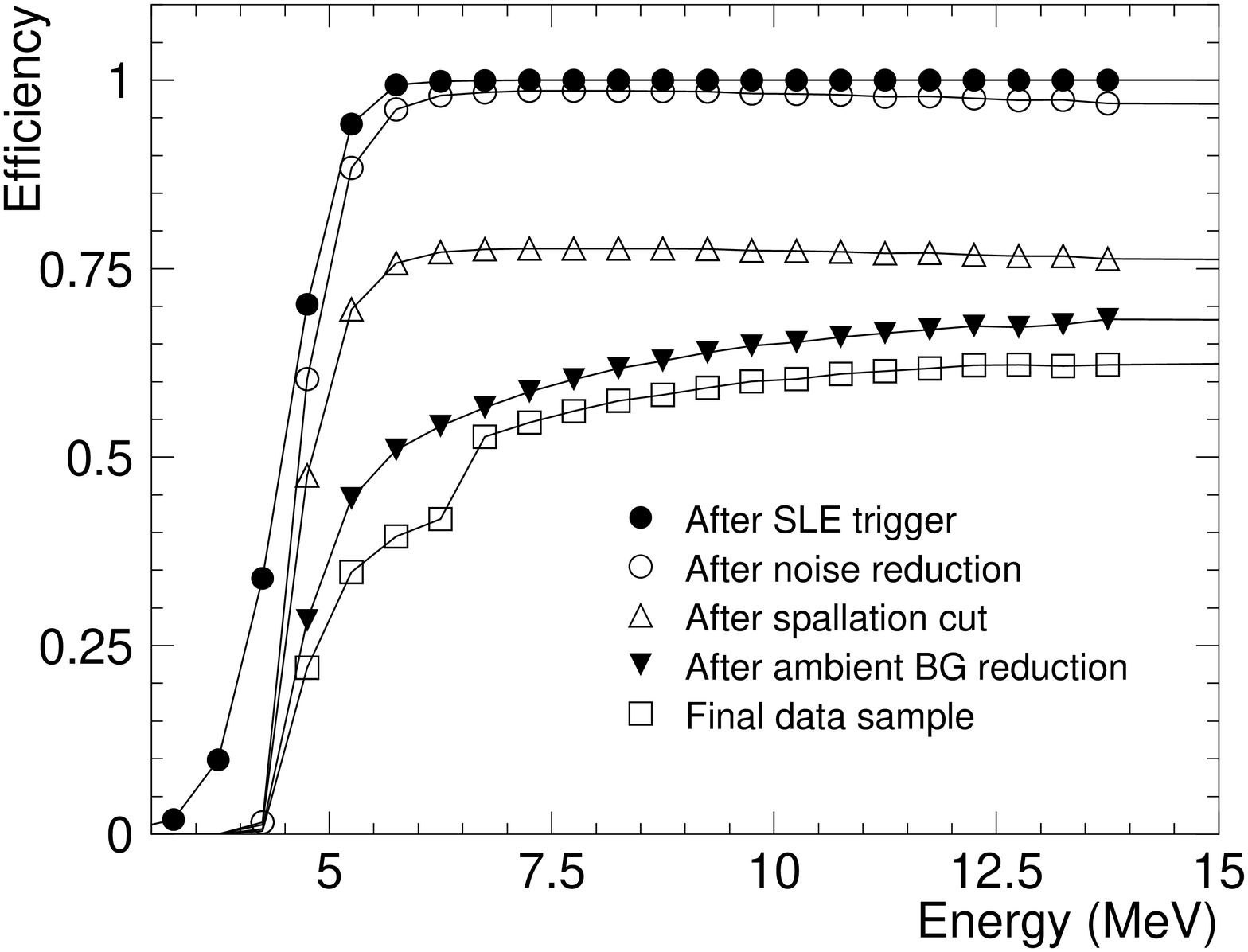}
  \vspace{-1cm}
  }
  \caption{Summary of the reduction efficiencies on MC.}
  \label{mceff}
 \end{center}
\end{figure}

\begin{figure}[htbp]
 \begin{center}
  \resizebox{8.5cm}{!}{
  \includegraphics{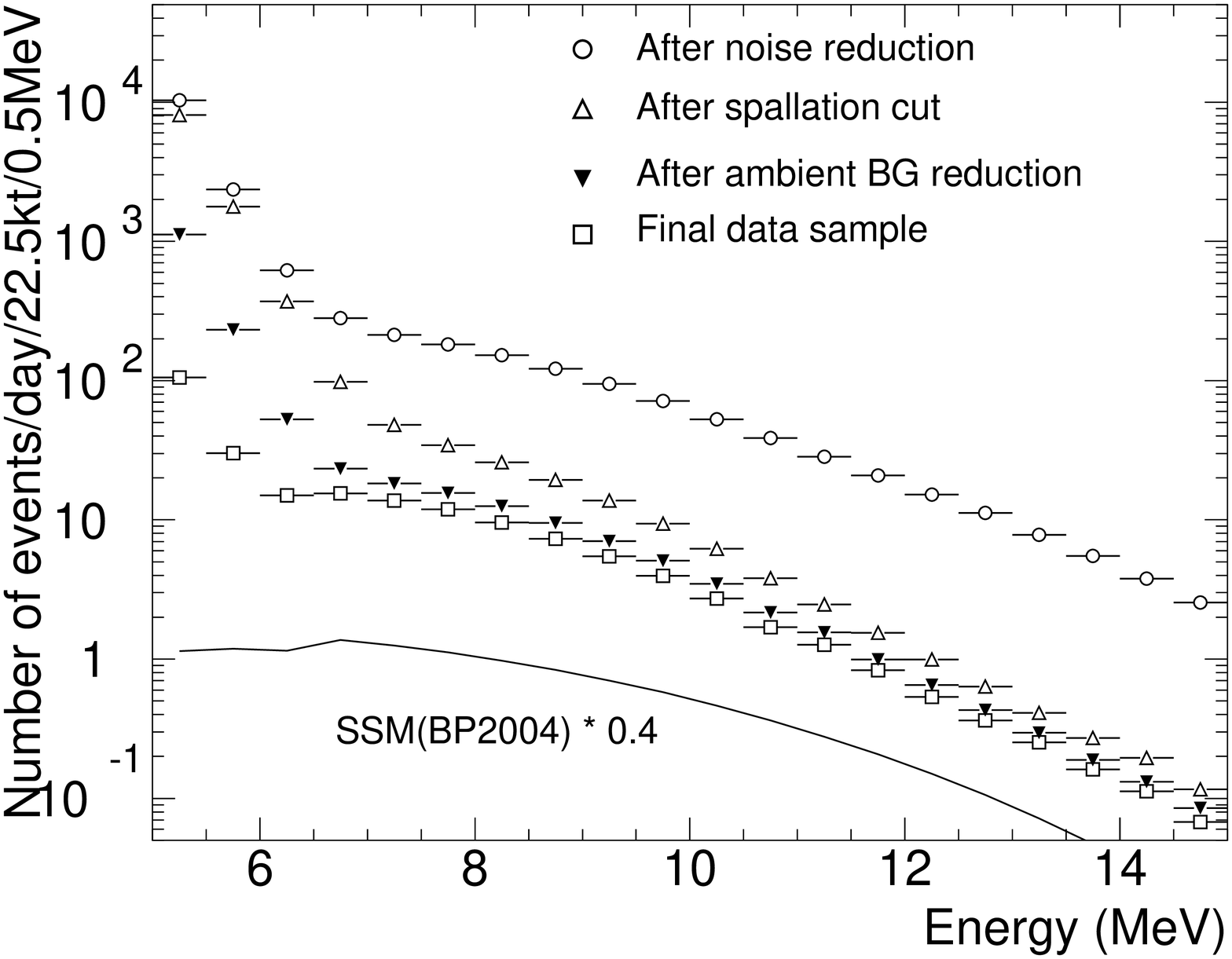}
  \vspace{-1cm}
  }
  \caption{Summary of the data reduction steps.}
  \label{redstep}
 \end{center}
\end{figure}

 Figure~\ref{mceff} shows the reduction efficiencies after each step
as a function of energy using MC simulation events. The efficiency is used
as a correction when the energy spectrum is calculated as described in
Section~\ref{sec:spec}. Figure~\ref{redstep} shows the event rate
after each reduction step for data and also shows the predicted 
solar neutrino spectrum.  Table~\ref{tab:reduc} summarizes the 
numerical results of the reduction steps. The number of events after all the
reduction steps is 286557.
\begin{table}[hbpt]
 \begin{center}
  \begin{tabular}{cc}
   \hline
   Total & $3.43\times10^7$ \\
   \hline
    A.   &Noise reduction\\
   (a)   & $2.66\times10^7$ \\
   (b)   & $2.51\times10^7$ \\
   (c)   & $2.50\times10^7$ \\
   (d)   & $2.50\times10^7$ \\
   (e)   & $2.48\times10^7$ \\
   (f)   & $1.81\times10^7$ \\
   \hline
    B.   &Spallation cut\\
         & $1.29\times10^7$ \\
   \hline
    C.   &Ambient B.G. cut\\
   (a)   & $3.61\times10^6$ \\
   (b)   & $2.72\times10^6$ \\
   (c)   & $1.86\times10^6$ \\
   \hline
    D.   &Gamma cut \\
         & $2.96\times10^5$ \\
   \hline
    E.   &$^{16}$N cut \\
         & $2.87\times10^5$ \\
   \hline
  \end{tabular}
  \caption{The summary of number of events remaining after each 
reduction step}
  \label{tab:reduc}
 \end{center}
\end{table}
%

%
%
\section{Results}\label{sec:result}
\subsection{Solar neutrino signal extraction}
\begin{figure}[htbp]
\centerline{\includegraphics[width=8.6cm]{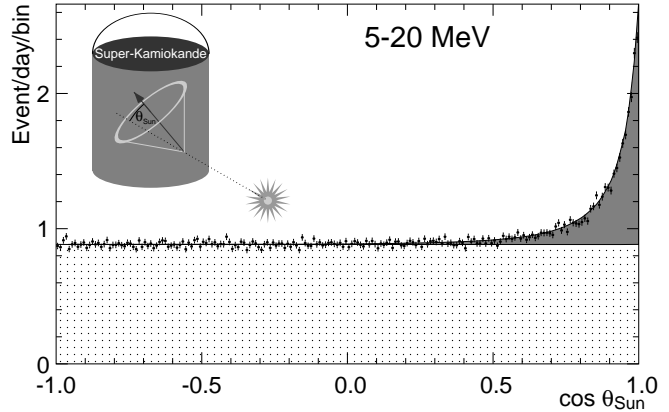}}
\caption{Angular distribution of solar neutrino event candidates.
The shaded area indicates the elastic scattering peak. The dotted
area is the contribution from background events.}
\label{fig:cossun}
\end{figure}

The solar neutrino signal is extracted from the directional correlation
of the recoiling electrons with the incident neutrinos in $\nu$-e scattering.
Figure~\ref{fig:cossun} shows $\cos\theta_{\mbox{\small sun}}$ where
$\theta_{\mbox{\small sun}}$ is the angle between the reconstructed
recoil electron direction and the expected neutrino direction (calculated
from the position of the sun at the event time).

The solar neutrino flux is extracted by a likelihood fit of
the solar neutrino signal and the background to this distribution.
This likelihood function is defined as follows:
\begin{equation}
{\cal L}=e^{-\left(\sum_i B_i+S\right)}\prod_{i=1}^{N_{\mbox{\tiny bin}}}\prod_{j=1}^{n_i}
\left(B_i\cdot b_{ij}+S\cdot Y_i \cdot s_{ij}\right)
\label{eqn:likelihood}
\end{equation}
We define $N_{\mbox{\tiny bin}}=21$ energy bins: 18 energy bins of 0.5~MeV
between 5.0 and 14.0~MeV, two energy bins of 1~MeV between 
14.0 and 16.0~MeV, and one
bin between 16.0 and 20.0~MeV. $S$ is the total number of solar neutrino signal
events, and $n_i$, $B_i$, and $Y_i$ represent the number of observed events,
the number of background events, and the expected fraction of signal events in
the $i$-th energy bin, respectively. We use two types of probability density
functions: $p(\cos\theta{\mbox{\tiny sun}},E)$ describes the angular shape
expected for solar $\nu_e$'s of recoil electron energy $E$ (signal events) and
$u_i(\cos\theta{\mbox{\tiny sun}})$ is the background shape in energy bin $i$.
Each of the $n_i$ events in energy bin $i$ is assigned the background factor
$b_{ij}=u_i(\cos\theta_{ij})$ and the signal factor
$s_{ij}=p(\cos\theta_{ij},E_j)$.

The signal shape $p(\cos\theta{\mbox{\tiny sun}},E)$
is obtained from the known, strongly forward-peaked angular
distribution of neutrino-electron elastic scattering with smearing due to 
multiple scattering and the detector's angular resolution.
The background shape $u_i(\cos\theta{\mbox{\tiny sun}})$
has no directional correlation with the neutrino direction,
but deviates from a flat shape due to the cylindrical shape
of the SK detector: the number of PMT's per solid angle depends
on the SK zenith angle.
In order to calculate the expected background shape, we use
the angular distribution of data itself. The presence of
solar neutrinos in the sample biases mostly the azimuthal distribution,
so at first we fit only the zenith angle distribution
and assume the azimuthal distribution to be flat.
We generate toy Monte Carlo directions according to this fit
and calculate $\cos\theta{\mbox{\tiny sun}}$. We also
fit both zenith and azimuthal distributions, approximately
subtracting the solar neutrino events from the sample and
repeat the toy Monte Carlo calculation.
We compare the obtained number of solar neutrino events from
both background shapes and assign the difference as a 
systematic uncertainty.
Since the azimuthal distributions don't deviate very significantly
from flat distributions, we quote the solar neutrino events obtained
from the first shape (assuming a flat azimuthal distribution).
The dotted area in Figure~\ref{fig:cossun} shows this background shape.
The systematic uncertainty due to the background shape is 0.1\%
for the entire data sample (5.0-20.0 MeV).
If the data sample is divided into a day and a night sample,
the systematic uncertainty is 0.4\%.
The amount of background contamination is much less above 10~MeV
than it is near the SK--I energy threshold (5.0~MeV), so small differences
in background shape between the two methods become important only
in the lowest energy bins: between 5.0 and 5.5~MeV, the systematic
uncertainty is estimated to be 1.2\%, between 5.5 and 6.0~MeV
0.4\%, and above 6.0~MeV 0.15\%.

\subsection{Observed solar neutrino flux}

Figure~\ref{fig:cossun} shows the $\cos \theta_{sun}$ distribution for
1496 days of SK--I data. The best fit value for the  number of signal events 
due to solar neutrinos between 5.0~MeV and 20.0~MeV is calculated by the
maximum likelihood method in Eq.~(\ref{eqn:likelihood}), and the result for 
SK--I is $22,404 \pm 226~({\rm stat.})^{+784}_{-717}~({\rm sys.})$.
The corresponding $^8B$ flux is:
\begin{displaymath}
 2.35 \pm 0.02~({\rm stat.}) \pm 0.08~({\rm sys.})
 \times 10^6~{\rm cm}^{-2} {\rm s}^{-1}.
\end{displaymath}
%
%

 The systematic errors for the solar neutrino flux, seasonal variation
and day-night differences for the energy range 5.0~MeV to 20.0~MeV are shown
in Table~\ref{tab:sysflux}. The detailed explanations are written in 
each topic's section, but the total systematic error for 
the solar neutrino flux measurement is estimated to be 
$^{+3.5\%}_{-3.2\%}$.
\begin{table}[htbp]
  \begin{tabular}{l|ccc}
   \hline
   \hline
   & Flux & Seasonal & day-night \\
   \hline
   Energy scale, resolution & $\pm 1.6$ & $^{+1.2}_{-1.1}$ & $^{+1.2}_{-1.1}$ \\
   Theoretical uncertainty & $^{+1.1}_{-1.0}$ & & \\
   \hspace*{5mm}for $^8$B spectrum & & & \\
   Trigger efficiency & $^{+0.4}_{-0.3}$ & $\pm 0.1$ & \\
   Reduction & $^{+2.1}_{-1.6}$ & $\pm 0.5$ & \\
   Spallation dead time & $\pm 0.2$ & $\pm 0.1$ & $\pm 0.1$ \\
   Gamma ray cut & $\pm 0.5$ & $\pm 0.25$ &  \\
   Vertex shift & $\pm 1.3$ & &  \\
   Background shape & $\pm 0.1$ & & $\pm 0.4$ \\
   \hspace*{5mm}for signal extraction &  & & \\
   Angular resolution & $\pm 1.2$ & & \\
   Cross section of $\nu$-e scattering& $\pm 0.5$ & & \\
   Livetime calculation & $\pm 0.1$ & $\pm 0.1$ & $\pm 0.1$ \\
   \hline
   Total & $^{+3.5}_{-3.2}$ & $\pm 0.3$ & $^{+1.3}_{-1.2}$ \\
   \hline
   \hline
  \end{tabular}
  \caption{Systematic error of each item (in \%).}
  \label{tab:sysflux}
\end{table}

\subsection{Time variations of solar neutrino flux}
\subsubsection{Day-Night difference}
\begin{figure}[htbp]
\includegraphics[width=8cm,clip]{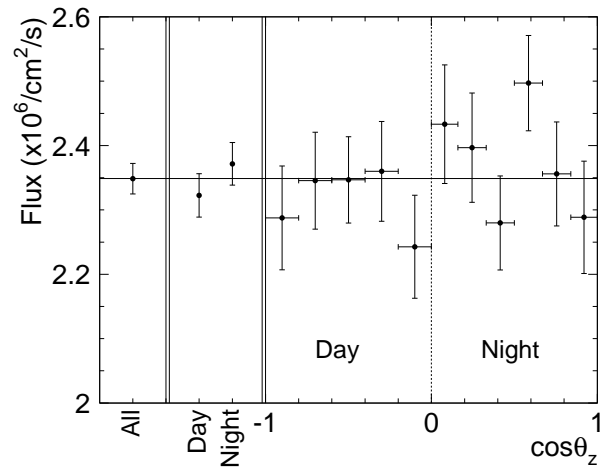}
\caption{The solar zenith angle dependence of the solar neutrino flux
(error bars show statistical error).  The width of the night-time bins was
chosen to separate solar neutrinos that pass through the Earth's dense core
(the rightmost Night bin) from those that pass through the mantle. 
The horizontal line shows the flux for all data.}
\label{fig:dn}
\end{figure}

 The day time flux and night time flux of solar neutrinos in SK--I 
are calculated using events which occurred when the solar zenith 
angle cosine was less than and greater than
zero, respectively. The observed flux are:
\begin{eqnarray*}
\Phi_{day} & = & 2.32 \pm 0.03~({\rm stat.}) ^{+0.08}_{-0.07}~({\rm sys.})
\times 10^6~{\rm cm}^{-2} {\rm s}^{-1} \\
\Phi_{night} & = & 2.37 \pm 0.03~({\rm stat.}) ^{+0.08}_{-0.08}~({\rm sys.})
\times 10^6~{\rm cm}^{-2} {\rm s}^{-1}
\end{eqnarray*}
Their difference leads to a day-night asymmetry, defined as ${\cal A} =
(\Phi_{day} - \Phi_{night})/( \frac{1}{2} (\Phi_{day} + \Phi_{night}))$.
We find:
\begin{displaymath}
{\cal A} = -0.021 \pm 0.020~(\rm{stat.})^{+0.013}_{-0.012}~(\rm{sys.})
\end{displaymath}
 Including systematic errors, this is less than $1-\sigma$ from zero asymmetry.
The largest sources of systematic error in the asymmetry
are energy scale and resolution ($^{+0.012}_{-0.011}$) and the non-flat
background shape of the $\cos \theta_{sun}$ distribution ($\pm 0.004$).
As described in the neutrino oscillation analysis section, we
can reduce the statistical uncertainty if we assume two-neutrino
oscillations within the Large Mixing Angle region. The day-night
asymmetry in that case is
\begin{displaymath}
{\cal A} = -0.017 \pm 0.016~(\rm{stat.})^{+0.013}_{-0.012}~(\rm{sys.})
\pm 0.0004~(\rm{osc.})
\end{displaymath}
with the final, tiny additional uncertainty due to the 
uncertainty of the oscillation parameters themselves.
Figure~\ref{fig:dn} shows the solar neutrino flux as a function of the solar
zenith angle cosine.

\subsubsection{Seasonal variation}
\begin{figure}[htbp]
\includegraphics[width=8cm,clip]{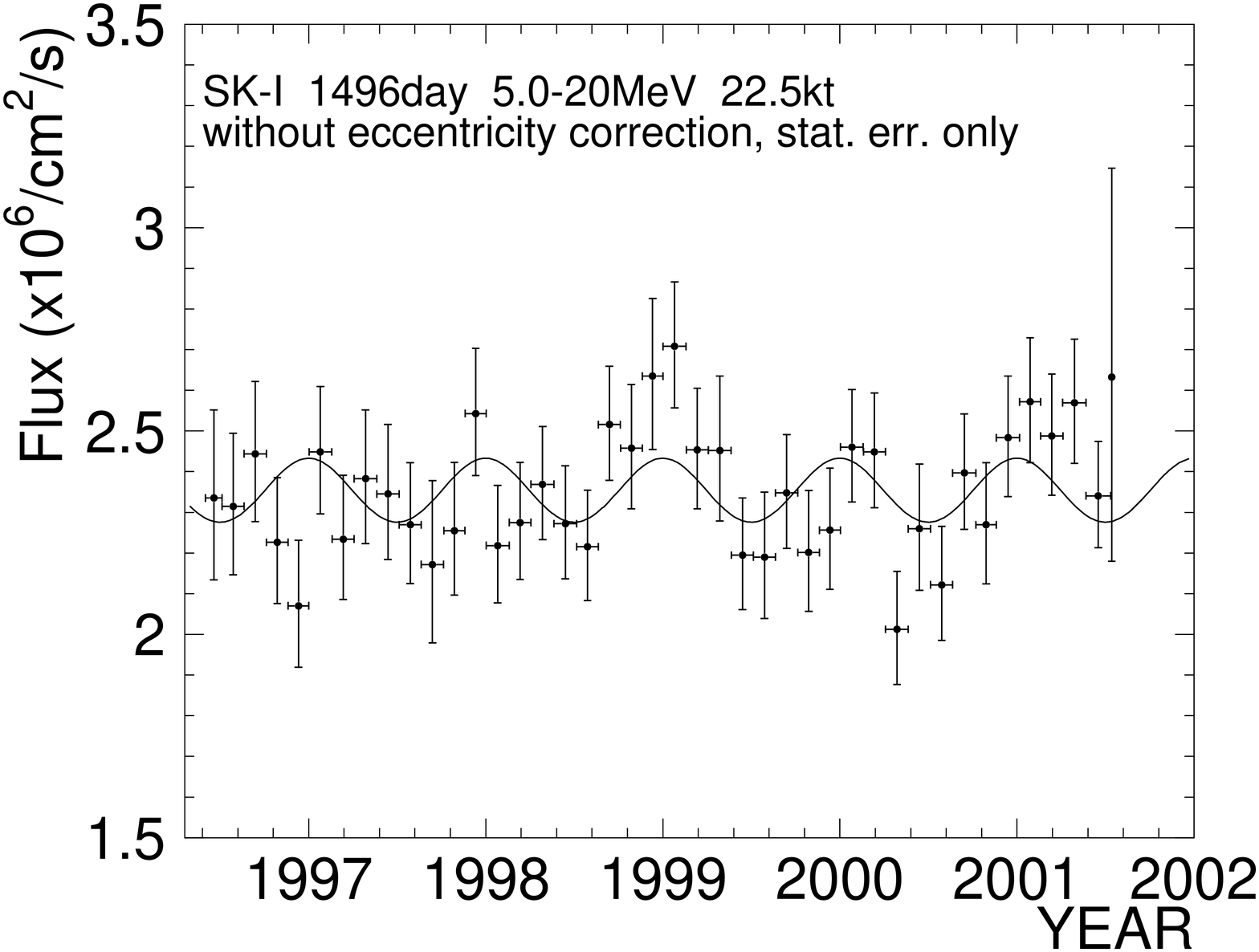}
\caption{Solar neutrino flux as a function of time. The binning of the 
horizontal axis is 1.5 months.}
\label{fig:mod}
\end{figure}

 Figure~\ref{fig:mod} shows the monthly variation of the flux,
which each horizontal bin covers 1.5 months. The figure shows that the 
experimental operation is very stable.

\begin{figure}[htbp]
\includegraphics[width=8cm,clip]{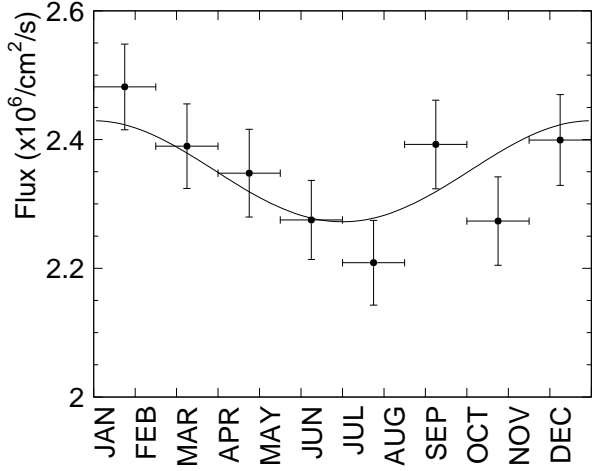}
\caption{The seasonal variation of the solar neutrino flux.  The solid line
is the prediction based on the eccentricity of the Earth's orbit.}
\label{fig:seas}
\end{figure}

 Figure~\ref{fig:seas} shows the seasonal variation of solar neutrino flux.
As in Figure~\ref{fig:mod}, 
each horizontal time bin is 1.5 months wide, but in this figure data taken at 
similar times during the year over the entire course of SK--I's data taking 
has been combined into single bins.  The 1.7\% orbital eccentricity
of the Earth, which causes about a 7\% flux variation simply due to
the inverse square law, is included in the flux prediction (solid line). 
The observed flux variation is consistent with
the predicted annual modulation. Its $\chi^2$/d.o.f. is 4.7/7, which is
equivalent to 69\% C.L.. If we fit the eccentricity to the Earth's orbit to the
observed SK rate variation, the perihelion shift is $13\pm18$ days (with
respect to the true perihelion) and the eccentricity is
$2.1\pm0.3\%$~\cite{dnpaper}.  This is the world's first observation
of the eccentricity of the Earth's orbit made with neutrinos. 
The total systematic error on the relative flux values in each seasonal bin
is estimated to $\pm 1.3\%$. The largest sources come from energy scale
and resolution ($^{+1.2\%}_{-1.1\%}$) and reduction cut efficiency
($\pm 0.5\%$), as shown in Table~\ref{tab:sysflux}.

\subsection{Energy spectrum}\label{sec:spec}
\begin{figure}[htbp]
\includegraphics[width=8.5cm,clip]{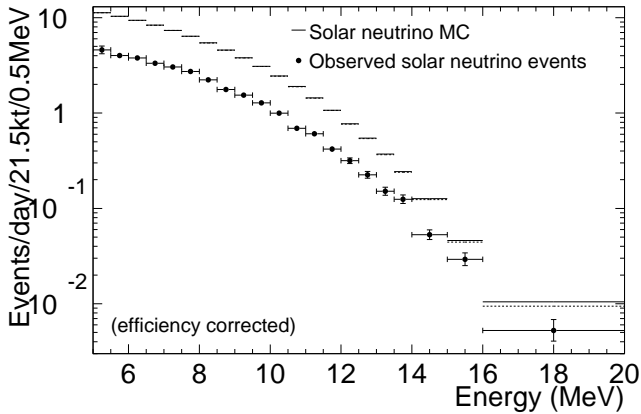}
\caption{Energy spectrum of the solar neutrino signal. 
 The horizontal axis is the total energy of the recoil electrons.
 The vertical axis is the event rate of the observed
 solar neutrino signal events.
The error bars are a quadrature of the statistical and uncorrelated
errors.
The reduction efficiencies in Fig.~\protect{\ref{mceff}} are corrected.
 BP2004 SSM flux values are used for the $^8$B and hep MC fluxes  
 in this plot. The dashed line shows the contribution of only $^8$B.
}
\label{fig:spec}
\end{figure}

Figure~\ref{fig:spec} shows the expected and measured recoil electron energy
spectrum. The expected spectrum is calculated by the detector 
simulation described
in Section~\ref{sec:sim}, and BP2004 is used as a solar model. The solid line
shows the expected spectrum of $^8$B and hep neutrinos, and the dashed 
line shows the contribution of only $^8$B neutrinos. 

The observed and expected event rates are summarized in 
Table~\ref{tab:rate21bin} and Table~\ref{tab:rate8bin}.
In these tables the reduction efficiencies listed in
Fig.~\ref{mceff} are corrected.

The uncorrelated and correlated systematic errors for each energy bin are shown
in Table~\ref{tab:unsysspe} and Table~\ref{tab:cosysspe}, respectively.
\begin{table*}[htbp]
 \begin{center}
  \begin{tabular}{c|c|c|c|c|c}
   \hline
   \hline
   Energy    & \multicolumn{3}{c|}{Observed rate} & \multicolumn{2}{c}{Expected rate} \\
   (MeV)     &  ALL & DAY & NIGHT                 &  $^8$B   &  hep \\
             & $ -1 \leq \cos\theta_{\rm z} \leq 1 $ 
             & $ -1 \leq \cos\theta_{\rm z} \leq 0 $ 
             & $  0 <    \cos\theta_{\rm z} \leq 1 $ & & \\ 
   \hline
   \hline
 $ 5.0- 5.5$ & $ 74.7^{+  6.6}_{-  6.5}$ & $ 72.1^{+  9.5}_{-  9.4}$ & $ 77.1^{+  9.2}_{-  9.0}$ & 182.9 & 0.312 \\ 
 $ 5.5- 6.0$ & $ 65.0^{+  3.3}_{-  3.2}$ & $ 64.8^{+  4.7}_{-  4.6}$ & $ 65.1^{+  4.6}_{-  4.5}$ & 167.7 & 0.309 \\ 
 $ 6.0- 6.5$ & $ 61.5^{+  2.4}_{-  2.3}$ & $ 60.2^{+  3.4}_{-  3.3}$ & $ 62.6^{+  3.3}_{-  3.2}$ & 151.9 & 0.294 \\ 
 $ 6.5- 7.0$ & $ 54.1^{+  1.7}_{-  1.7}$ & $ 54.2^{+  2.4}_{-  2.4}$ & $ 53.9^{+  2.4}_{-  2.3}$ & 135.3 & 0.284 \\ 
 $ 7.0- 7.5$ & $ 49.4^{+  1.5}_{-  1.5}$ & $ 49.2^{+  2.2}_{-  2.1}$ & $ 49.6^{+  2.1}_{-  2.1}$ & 119.2 & 0.266 \\ 
 $ 7.5- 8.0$ & $ 44.3^{+  1.4}_{-  1.4}$ & $ 44.8^{+  2.0}_{-  1.9}$ & $ 43.8^{+  1.9}_{-  1.9}$ & 103.5 & 0.249 \\ 
 $ 8.0- 8.5$ & $ 36.3^{+  1.2}_{-  1.2}$ & $ 35.7^{+  1.7}_{-  1.7}$ & $ 36.8^{+  1.7}_{-  1.7}$ &  88.3 & 0.236 \\ 
 $ 8.5- 9.0$ & $ 28.7^{+  1.0}_{-  1.0}$ & $ 26.6^{+  1.5}_{-  1.4}$ & $ 30.6^{+  1.5}_{-  1.4}$ &  74.1 & 0.221 \\ 
 $ 9.0- 9.5$ & $ 25.0^{+  0.9}_{-  0.9}$ & $ 25.4^{+  1.4}_{-  1.3}$ & $ 24.6^{+  1.3}_{-  1.2}$ &  61.4 & 0.196 \\ 
 $ 9.5-10.0$ & $ 20.8^{+  0.8}_{-  0.8}$ & $ 20.7^{+  1.2}_{-  1.1}$ & $ 20.8^{+  1.2}_{-  1.1}$ &  49.9 & 0.185 \\ 
 $10.0-10.5$ & $ 16.2^{+  0.7}_{-  0.7}$ & $ 15.7^{+  1.0}_{-  0.9}$ & $ 16.7^{+  1.0}_{-  0.9}$ &  39.6 & 0.167 \\ 
 $10.5-11.0$ & $ 11.2^{+  0.6}_{-  0.5}$ & $ 10.9^{+  0.8}_{-  0.7}$ & $ 11.5^{+  0.8}_{-  0.8}$ &  30.7 & 0.149 \\ 
 $11.0-11.5$ & $9.85^{+ 0.51}_{- 0.49}$ & $ 9.65^{+ 0.74}_{- 0.68}$ & $10.03^{+ 0.73}_{- 0.68}$ & 23.28 & 0.130 \\ 
 $11.5-12.0$ & $6.79^{+ 0.42}_{- 0.40}$ & $ 7.14^{+ 0.63}_{- 0.58}$ & $ 6.47^{+ 0.58}_{- 0.53}$ & 17.27 & 0.118 \\ 
 $12.0-12.5$ & $5.13^{+ 0.36}_{- 0.33}$ & $ 5.05^{+ 0.52}_{- 0.47}$ & $ 5.21^{+ 0.51}_{- 0.46}$ & 12.45 & 0.098 \\ 
 $12.5-13.0$ & $3.65^{+ 0.30}_{- 0.28}$ & $ 3.96^{+ 0.46}_{- 0.41}$ & $ 3.38^{+ 0.41}_{- 0.36}$ &  8.76 & 0.090 \\ 
 $13.0-13.5$ & $2.46^{+ 0.25}_{- 0.23}$ & $ 2.56^{+ 0.38}_{- 0.33}$ & $ 2.37^{+ 0.35}_{- 0.30}$ &  5.94 & 0.073 \\ 
 $13.5-14.0$ & $2.02^{+ 0.22}_{- 0.20}$ & $ 1.95^{+ 0.32}_{- 0.27}$ & $ 2.09^{+ 0.32}_{- 0.27}$ &  3.88 & 0.060 \\ 
 $14.0-15.0$ & $1.72^{+ 0.21}_{- 0.19}$ & $ 1.60^{+ 0.31}_{- 0.25}$ & $ 1.85^{+ 0.31}_{- 0.26}$ &  4.01 & 0.094 \\ 
 $15.0-16.0$ & $0.949^{+0.157}_{-0.133}$ & $0.750^{+0.218}_{-0.165}$ & $1.136^{+0.238}_{-0.192}$ & 1.439 & 0.057 \\ 
 $16.0-20.0$ & $0.341^{+0.103}_{-0.077}$ & $0.240^{+0.148}_{-0.093}$ & $0.423^{+0.156}_{-0.109}$ & 0.611 & 0.068 \\ 
   \hline
   \hline
  \end{tabular}
  \caption{Observed and expected event rates in each energy bin at 1 AU.
The unit of the rates is events/kton/year.
The errors in the observed rates are statistical only.
The reduction efficiencies in Fig.~\protect{\ref{mceff}} are corrected, and 
the expected event rates are for the BP2004 SSM flux values 
($^8$B flux is 5.79 $\times 10^6$, hep flux is 7.88 $\times 10^3$/cm$^2$/sec).
$\theta_{\rm z}$ is the angle between the z-axis of the detector 
and the vector from the Sun to the detector.}
  \label{tab:rate21bin}
 \end{center}
\end{table*}
\begin{table*}[htbp]
 \begin{center}
  \begin{tabular}{c|c|c|c|c|c|c|c|c|c}
   \hline
   \hline
   Energy    & \multicolumn{7}{c|}{Observed rate} & \multicolumn{2}{c}{Expected rate} \\
   (MeV)     &  DAY & MANTLE1 & MANTLE2 & MANTLE3 & MANTLE4 &MANTLE5 &CORE   &  $^8$B   &  hep \\
             & $\cos\theta_{\rm z}: -1 \sim 0 $ 
             & $  0    \sim 0.16 $ 
             & $  0.16 \sim 0.33 $ 
             & $  0.33 \sim 0.50 $ 
             & $  0.50 \sim 0.67 $ 
             & $  0.67 \sim 0.84 $ 
             & $  0.84 \sim 1 $ & & \\ 
   \hline
   \hline
 $ 5.5- 6.5$ & $ 127.^{+   6.}_{-   6.}$ & $ 124.^{+  15.}_{-  15.
 }$ & $ 106.^{+  14.}_{-  14.}$ & $ 132.^{+  13.}_{-  12.}$ & $ 146.^{+  13.}_{-  12.
 }$ & $ 140.^{+  14.}_{-  13.}$ & $ 119.^{+  15.}_{-  14.}$ &  320. & 0.603\\ 
 $ 6.5- 8.0$ & $ 149.^{+   4.}_{-   4.}$ & $ 166.^{+  11.}_{-  10.
 }$ & $ 158.^{+  10.}_{-   9.}$ & $ 137.^{+   8.}_{-   8.}$ & $ 150.^{+   8.}_{-   8.
 }$ & $ 141.^{+   9.}_{-   9.}$ & $ 137.^{+  10.}_{-   9.}$ &  358. & 0.799\\ 
 $ 8.0- 9.5$ & $ 87.8^{+  2.6}_{-  2.6}$ & $ 90.7^{+  7.2}_{-  6.8
 }$ & $ 92.1^{+  6.7}_{-  6.4}$ & $ 90.5^{+  5.8}_{-  5.5}$ & $ 99.8^{+  5.9}_{-  5.6
 }$ & $ 90.3^{+  6.4}_{-  6.0}$ & $ 88.5^{+  7.0}_{-  6.6}$ & 223.8 & 0.653\\ 
 $ 9.5-11.5$ & $ 57.1^{+  1.9}_{-  1.8}$ & $ 56.5^{+  5.2}_{-  4.8
 }$ & $ 63.3^{+  5.0}_{-  4.6}$ & $ 56.8^{+  4.1}_{-  3.9}$ & $ 59.6^{+  4.2}_{-  3.9
 }$ & $ 60.1^{+  4.6}_{-  4.3}$ & $ 60.9^{+  5.2}_{-  4.8}$ & 143.4 & 0.631\\ 
 $11.5-13.5$ & $ 18.7^{+  1.0}_{-  0.9}$ & $ 20.0^{+  2.8}_{-  2.4
 }$ & $ 13.8^{+  2.3}_{-  2.0}$ & $ 15.3^{+  2.0}_{-  1.8}$ & $ 19.5^{+  2.2}_{-  2.0
 }$ & $ 17.0^{+  2.3}_{-  2.0}$ & $ 20.4^{+  2.7}_{-  2.3}$ &  44.4 & 0.379\\ 
 $13.5-16.0$ & $4.28^{+ 0.48}_{- 0.43}$ & $ 4.78^{+ 1.45}_{- 1.08
 }$ & $ 6.97^{+ 1.56}_{- 1.24}$ & $ 5.82^{+ 1.22}_{- 0.98}$ & $ 5.58^{+ 1.19}_{- 0.95
 }$ & $ 3.70^{+ 1.14}_{- 0.85}$ & $ 3.93^{+ 1.27}_{- 0.93}$ &  9.33 & 0.211\\ 
   \hline
   \hline
  \end{tabular}
  \caption{Observed and expected event rates in each zenith-spectra 
data set at a one astronomical unit [AU] distance from the Sun. 
The unit of the rates is events/kton/year.
The errors in the observed rates are statistical only.
The reduction efficiencies in Fig.~\protect{\ref{mceff}} are corrected, and 
the expected event rates are for the BP2004 SSM flux values.
$\theta_{\rm z}$ is the angle between z-axis of the detector 
and the vector from the Sun to the detector.}
  \label{tab:rate8bin}
 \end{center}
\end{table*}
\begin{table}[htbp]
 \begin{center}
  \begin{tabular}{l|ccccc}
   \hline
   \hline
   Energy (MeV) & $5-5.5$ & $5.5-6$ & $6-6.5$ & $6.5-7$ & $7-$ \\
   \hline
   Trigger efficiency & $^{+2.5}_{-1.5}$ & $^{+0.8}_{-0.6}$ & $\pm 0.1$ & $\pm 0.2$ & \\
   Reduction & $\pm 0.9$ & $\pm 0.9$ & $\pm 0.9$ & $\pm 0.9$ & $\pm 0.9$ \\
   Gamma ray cut & $\pm 0.1$ & $\pm 0.1$ & $\pm 0.1$ & $\pm 0.1$ & $\pm 0.1$ \\
   Vertex shift & $\pm 0.2$ & $\pm 0.2$ & $\pm 0.2$ & $\pm 0.2$ & $\pm 0.2$ \\
   Background shape& $\pm 0.6$ & $\pm 0.5$ & $\pm 0.1$ & $\pm 0.1$ & $\pm 0.1$ \\
   \hspace*{2mm}for signal extraction & & & & & \\
   Angular resolution & $\pm 2.3$ & $\pm 1.0$ & $\pm 1.0$ & $\pm 1.0$ & $\pm 1.0$ \\
   Cross section of & $\pm 0.2$ & $\pm 0.2$ & $\pm 0.2$ & $\pm 0.2$ & $\pm 0.2$ \\
   \hspace*{5mm}$\nu$-e scattering& & & & & \\
   \hline
   Total & $^{+3.5}_{-29}$ & $^{+1.7}_{-1.6}$ & $\pm 1.4$ & $ \pm 1.5$ & $\pm 1.4$ \\
   \hline
   \hline
  \end{tabular}
  \caption{Energy uncorrelated systematic errors for each energy bin.}
  \label{tab:unsysspe}
 \end{center}
\end{table}
\begin{table}[htbp]
 \begin{center}
  \begin{tabular}{c||cc|cc|cc}
   \hline
   \hline
   Energy     & \multicolumn{2}{c|}{Scale(\%)} & \multicolumn{2}{c|}{Resolution(\%)} & \multicolumn{2}{c}{Theory(\%)} \\
   \hline
   \hline
   $5.0-5.5$   & $+0.1$ & $ 0.0$  & $+0.2$ & $-0.2$ & $+0.1$ & $ 0.0$ \\
   $5.5-6.0$   & $-0.1$ & $+0.1$  & $+0.2$ & $-0.2$ & $ 0.0$ & $+0.1$ \\
   $6.0-6.5$   & $-0.3$ & $+0.2$  & $+0.2$ & $-0.2$ & $-0.1$ & $+0.1$  \\
   $6.5-7.0$   & $-0.5$ & $+0.4$  & $+0.2$ & $-0.2$ & $-0.3$ & $+0.2$  \\
   $7.0-7.5$   & $-0.7$ & $+0.6$  & $+0.2$ & $-0.2$ & $-0.4$ & $+0.4$  \\
   $7.5-8.0$   & $-0.9$ & $+0.9$  & $+0.2$ & $-0.2$ & $-0.5$ & $+0.5$  \\
   $8.0-8.5$   & $-1.1$ & $+1.1$  & $+0.2$ & $-0.2$ & $-0.7$ & $+0.7$  \\
   $8.5-9.0$   & $-1.4$ & $+1.4$  & $+0.1$ & $-0.1$ & $-0.9$ & $+0.9$  \\
   $9.0-9.5$   & $-1.7$ & $+1.8$  & $+0.1$ & $-0.1$ & $-1.1$ & $+1.1$   \\
   $9.5-10.0$  & $-1.9$ & $+2.1$  & $ 0.0$ & $ 0.0$ & $-1.3$ & $+1.4$   \\
   $10.0-10.5$ & $-2.3$ & $+2.5$  & $-0.1$ & $+0.1$ & $-1.5$ & $+1.7$   \\
   $10.5-11.0$ & $-2.6$ & $+2.8$  & $-0.3$ & $+0.2$ & $-1.8$ & $+2.0$   \\
   $11.0-11.5$ & $-3.0$ & $+3.2$  & $-0.5$ & $+0.4$ & $-2.1$ & $+2.3$   \\
   $11.5-12.0$ & $-3.4$ & $+3.6$  & $-0.8$ & $+0.7$ & $-2.4$ & $+2.6$   \\
   $12.0-12.5$ & $-3.8$ & $+4.1$  & $-1.1$ & $+0.9$ & $-2.7$ & $+3.0$   \\
   $12.5-13.0$ & $-4.3$ & $+4.5$  & $-1.4$ & $+1.3$ & $-3.1$ & $+3.3$   \\
   $13.0-13.5$ & $-4.8$ & $+5.0$  & $-1.9$ & $+1.7$ & $-3.4$ & $+3.8$   \\
   $13.5-14.0$ & $-5.4$ & $+5.5$  & $-2.4$ & $+2.2$ & $-3.8$ & $+4.2$   \\
   $14.0-15.0$ & $-6.3$ & $+6.3$  & $-3.3$ & $+3.1$ & $-4.4$ & $+5.0$   \\
   $15.0-16.0$ & $-7.7$ & $+7.6$  & $-4.9$ & $+4.7$ & $-5.1$ & $+6.2$   \\
   $16.0-20.0$ & $-9.9$ & $+10.2$ & $-8.1$ & $+8.2$ & $-5.6$ & $+8.7$   \\
   \hline
   \hline
  \end{tabular}
  \caption{Energy correlated systematic errors for each of the 21 energy bins.}
  \label{tab:cosysspe}
 \end{center}
\end{table}
%

\subsection{Hep solar neutrino}

The expected hep neutrino flux is three orders of magnitude 
smaller than the $^8$B solar neutrino flux.
However, since the end-point of the hep neutrino spectrum is about 18.8~MeV 
compared to about 16~MeV for the $^8$B neutrinos,
the high energy end of the $^8$B spectrum should be relatively 
enriched with hep neutrinos. 
In order to discuss the flux of hep neutrinos,
the most sensitive energy region for hep neutrino was estimated, then, 
assuming all the signal events in this energy region were due to 
hep neutrinos, an upper limit of the hep solar neutrino flux was obtained.
Any possible effects from neutrino oscillations were not considered 
in this analysis.
 
\begin{figure}
 \includegraphics[width=8.5cm,clip]{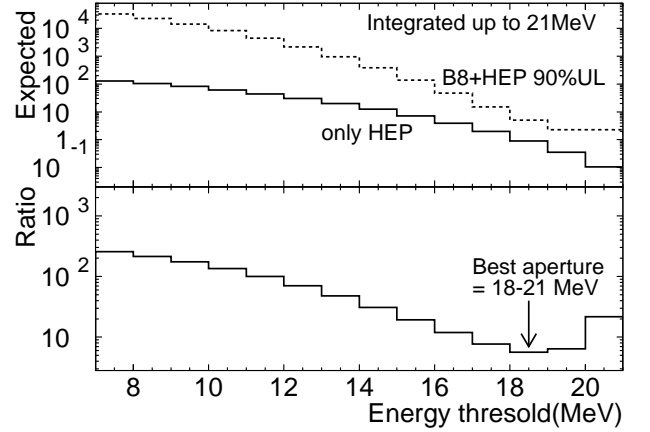}
 \caption{Integral energy distributions for the hep solar neutrino analysis. 
 The horizontal axis is the energy threshold of the recoil electrons.
 The integration is up to 21.0~MeV.
 Upper plot: 
 the vertical axis is the expected number of
 events in SK--I's 1496 day final data sample after all cuts.
 The solid and dashed lines correspond to the hep solar neutrino only the and
 $^8$B+hep 90\% C.L. upper limit.
 Lower plot: the ratio  of the two histograms 
 in the upper plot (dashed/solid).}
 \label{fig:hep1}
\end{figure}

Figure~\ref{fig:hep1} shows the expected integral energy distributions 
for solar neutrinos.
In the high energy region, the relative hep contribution is high, 
but the expected number of events is small 
because of the limited observation time. 
For this analysis, the best energy aperture for the
hep solar neutrinos was determined to be 18.0$\sim$21.0~MeV.
In this energy region 0.90 hep solar neutrino events are expected 
from the predicted BP2004 SSM rate.

\begin{figure}
 \includegraphics[width=8.5cm,clip]{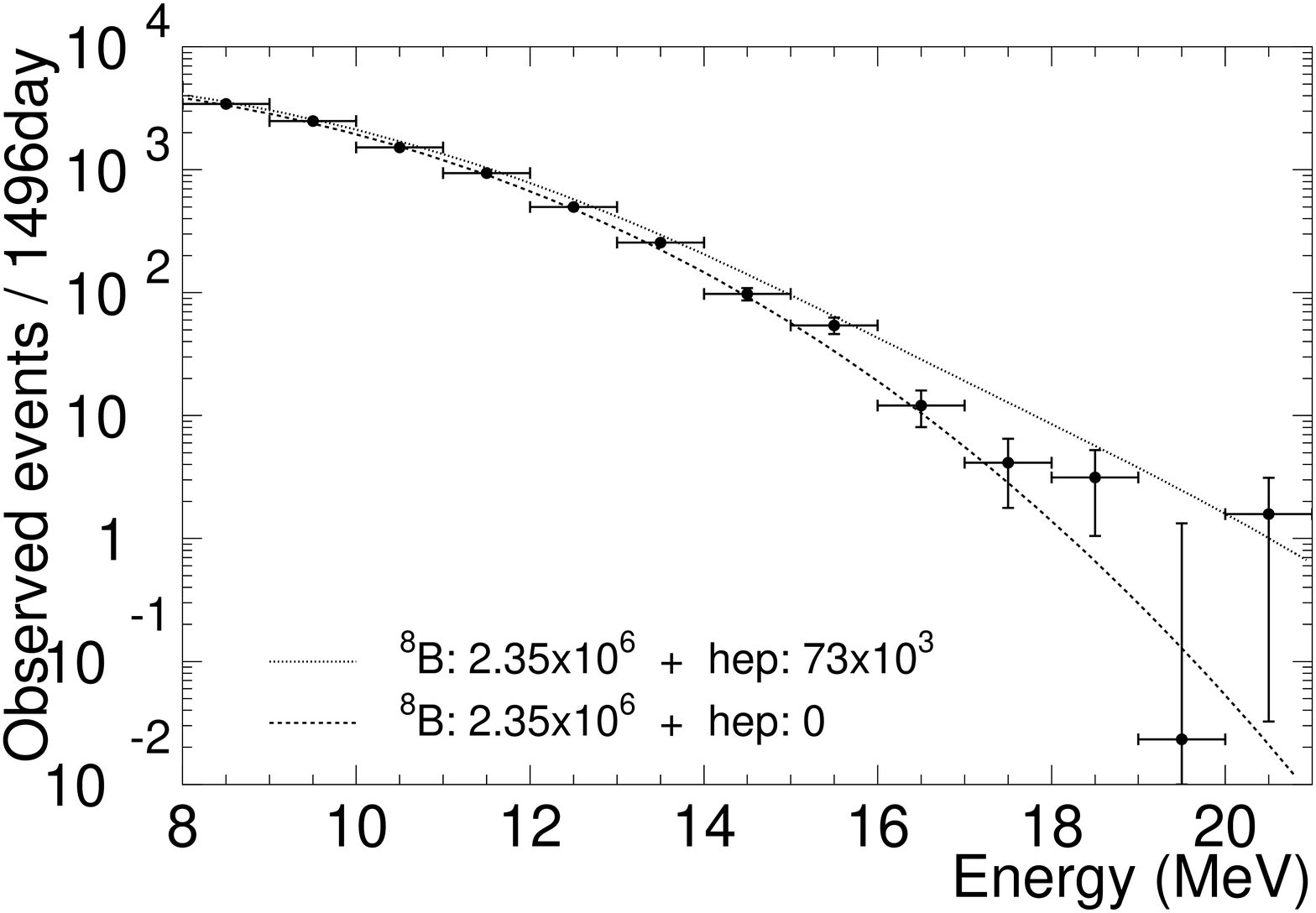}
 \caption{Energy spectrum of recoil electrons in the high energy region.
 The points show data with statistical error bars.
 The curves show expected spectra with various hep
 contributions to the best-fit $^8$B spectrum.
 The unit of the fluxes for the curves are $/cm^2/s$.
 The dotted and dashed curves show the spectrum with  73  
 and 0 ($\times 10^3/cm^2/s$) hep fluxes, respectively.}
\label{fig:hep2}
\end{figure}

Applying the same signal extraction method to the data events in
the 18.0$\sim$21.0~MeV region, we found $4.9 \pm 2.7$ 
solar neutrino signal events.
Assuming that all of these events are due to hep neutrinos, 
the 90\% confidence level upper limit of the hep neutrino flux 
was $73 \times 10^3$~cm$^{-2}$~s$^{-1}$. 
Figure~\ref{fig:hep2} shows the differential energy spectrum of solar
neutrino signals in the high energy region.

%
%
\section{Solar neutrino oscillation analysis}\label{sec:osc}
\newcommand{\plumin}[2]{^{+#1}_{-#2}}
\def\dmsq{\Delta m^2}
\def\tasq{\tan^2\theta}

\begin{figure}[tb]
\includegraphics[width=8.6cm]{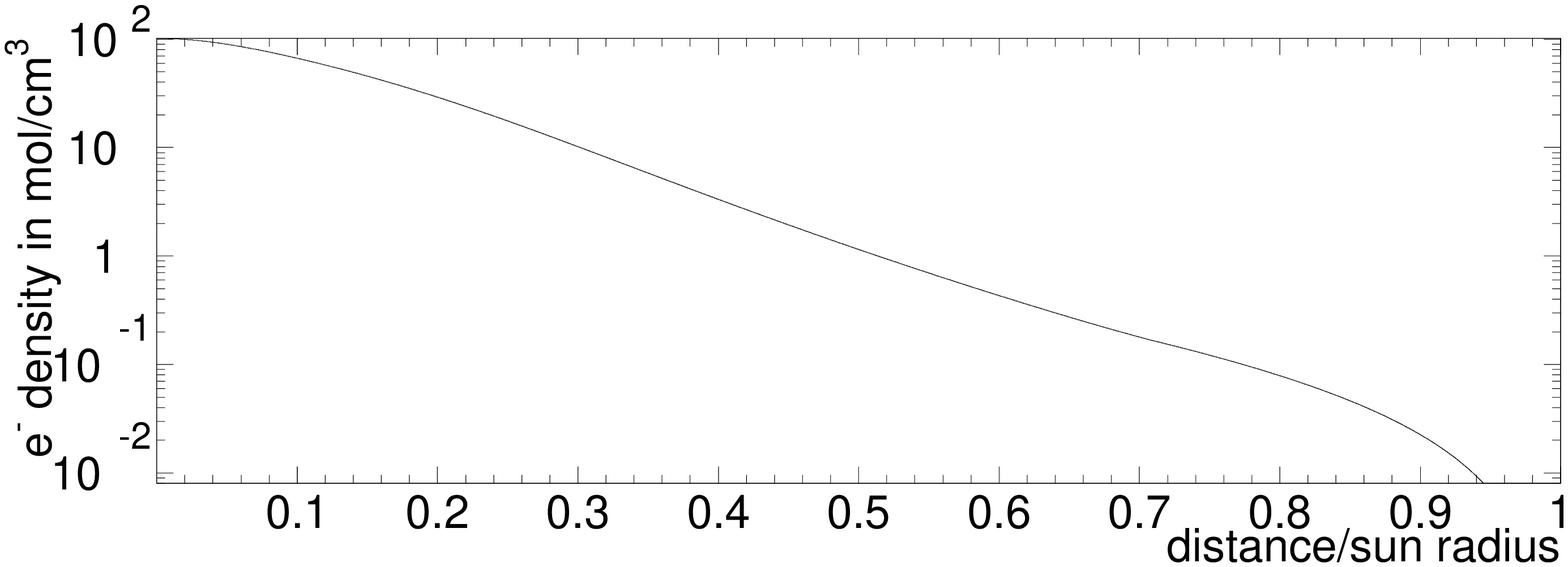}
\includegraphics[width=8.6cm]{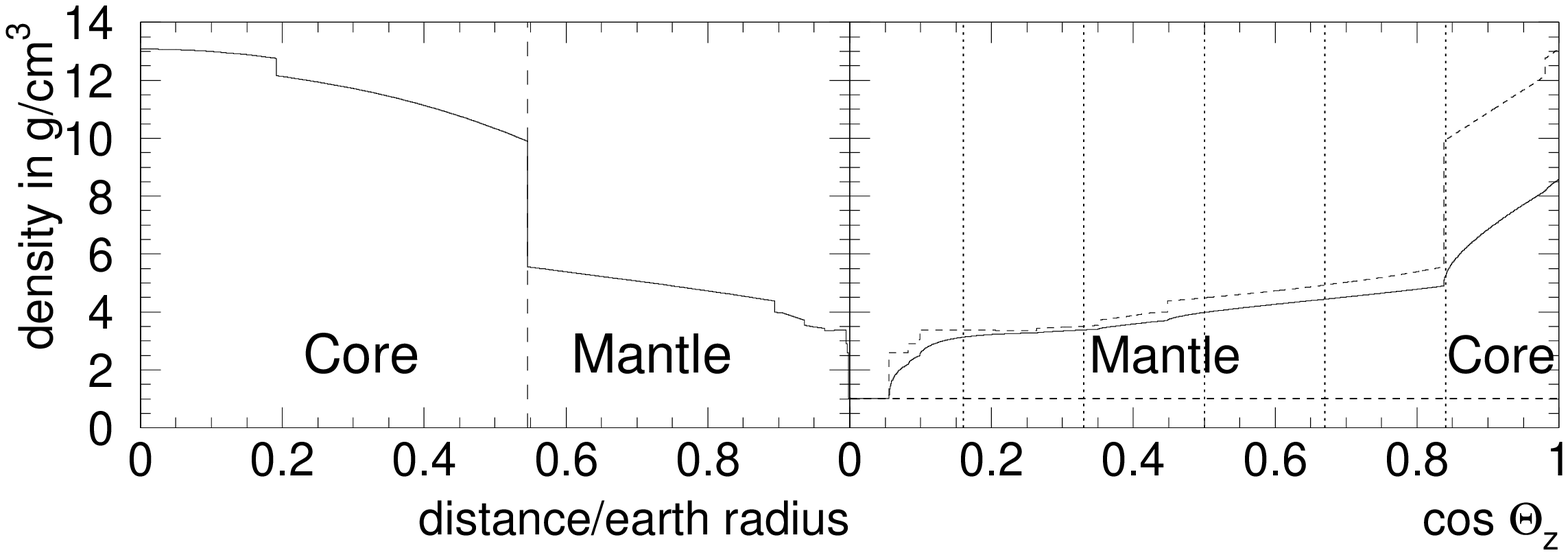}
\caption{The Sun's electron density profile
from the standard solar model~\cite{ssm} [SSM] as a
function of the distance to the center (top panel), 
and Earth's mass density profile (bottom panels).
The left plot shows the density as a function of the distance to the
Earth's center. The right plot shows the minimum (dashed horizontal line),
average (solid line) and maximum density (dashed line)
a solar neutrino of zenith angle $\theta_z$ encounters on its path
through the Earth. The electron density is obtained from this
by multiplying by 0.497~mol/cm$^3$ for the mantle and 0.468~mol/cm$^3$
for the core.}
\label{fig:density}
\end{figure}

\subsection{Introduction}

Two-neutrino oscillations are so far sufficient to explain and
describe all measured solar neutrino phenomena. The flavor
eigenstates $\nu_e$ and $\nu_x$ (where $\nu_x$ is either $\nu_\mu$ or
$\nu_\tau$ or an admixture of both) describe weak interactions of neutrinos
and electrons or nucleons. Only solar $\nu_e$'s can
participate in charged-current reactions, since the solar neutrino
energy is below the $\mu$ mass threshold. These flavor eigenstates
are related to the mass eigenstates $\nu_1$ and $\nu_2$ via the
unitary mixing matrix $U$, which for two neutrinos can be expressed
in terms of a single parameter, the weak mixing angle $\theta$:
\[
U=\begin{pmatrix}
\cos\theta &\sin\theta \cr -\sin\theta & \cos\theta
\end{pmatrix}
\]
In vacuum, the neutrino wavefunction oscillates in space with
a frequency of $\sqrt{p^2+m^2}\approx p+m^2/2p$ leading to an
oscillatory transition probability of the flavor eigenstates
\begin{equation}
p_{e\longrightarrow x}=p_{x\longrightarrow e}=\sin^22\theta\sin^2
\left(\pi\frac{L}{L_{\mbox{\small osc}}}\right)
\label{eq:osc}
\end{equation}
where the oscillation length 
$L_{\mbox{\small osc}}=\frac{\pi}{1.27}\frac{E}{\Delta m^2}$.
In a medium of matter density $\rho$, the
small angle scattering of the $\nu_e$ flavor differs
(due to the additional charged-current amplitude)
from the $\nu_x$ flavor; this can be described with a matter
potential $\Delta V(\rho)$.
The transition probability is still given by Eq.~(\ref{eq:osc})
but with the ``effective'' oscillation length and mixing angle
\[
L_{\mbox{\small eff}}=\frac{\pi}
{1.27\sqrt{\left(2\Delta V+\frac{\Delta m^2}{E}\cos2\theta\right)^2+
\left(\frac{\Delta m^2}{E}\sin2\theta\right)^2}}
\]
\[
\tan2\theta_{\mbox{\small eff}}=
\frac{\frac{\Delta m^2}{E}\sin2\theta}
{2\Delta V+\frac{\Delta m^2}{E}\cos2\theta}
\]
These relations are valid only for a constant matter
potential (and therefore constant matter density). We
calculate the oscillation probability by means of a
numerical simulation using the position-dependent
matter density profile in the Sun and Earth
(shown in Figure~\ref{fig:density}).

As noted by Mikheyev, Smirnov, and Wolfenstein [MSW]
\cite{msw}, the matter density in the Sun
(see Figure~\ref{fig:density}) is large enough for
a resonance ($\Delta V=-\frac{\Delta m^2}{2E}\cos2\theta$)
to occur, producing effective maximal mixing
($\theta_{\mbox{\small eff}}=\frac{\pi}{4}$) even
if the fundamental mixing $\theta$ is small.
The solar and terrestrial matter densities influence the transition probability
for solar neutrinos between $\Delta m^2/E\approx10^{-9}$~eV$^2$/MeV and
$\approx10^{-5}$~eV$^2$/MeV [MSW range]. For the so-called Large Mixing Angle
[LMA; $\theta\approx\pi/6$] MSW ``solution'' to the solar
neutrino problem, $\Delta m^2$ is
chosen so that the solar neutrino spectrum lies at the
``end'' of this range (near $10^{-5}$~eV$^2$/MeV). It explains all observed
solar neutrino interaction rates. In the case of the Small Mixing Angle
[SMA; $\theta\approx0.05$] MSW solution, which also explains all
observed rates, the solar neutrino spectrum must be placed near
$10^{-7}$~eV$^2$/MeV (closer to the center of the MSW range). Other
large angle solutions exist as well: the Low $\Delta m^2$ solution
[LOW] lies near the ``beginning''
of the MSW range ($10^{-8}$~eV$^2$/MeV), while the (quasi-) Vacuum [VAC]
solutions are ``below'' the MSW range ($10^{-12}$~eV$^2$/MeV
to $10^{-11}$~eV$^2$/MeV).

\begin{figure}[bt]
\includegraphics[width=8.8cm]{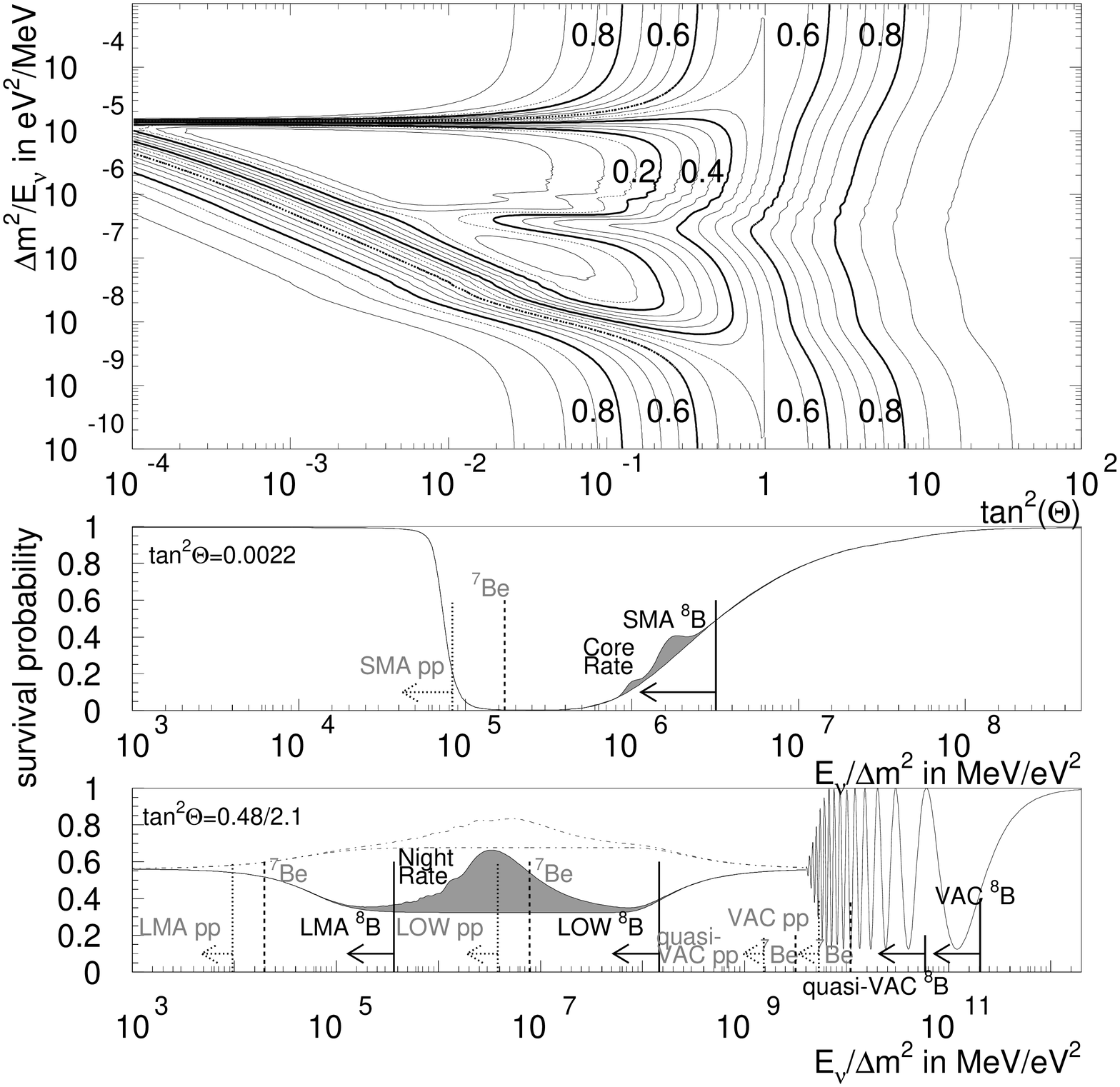}
\caption{Survival probability for $^8$B solar neutrinos. The top plot
displays contours of equal probability as a function of
$\Delta m^2/E$ and $\tan^2\theta$. The middle panel shows the
same probability as a function of $\Delta E/m^2$ only where a particular
(small) mixing angle has been selected. The shaded area contrasts the
difference between neutrinos passing through the core of the Earth
(which has the highest matter density) 
with neutrinos arriving
directly from the Sun.
The bottom panel shows a large mixing case. The line above the
shaded area depicts the average probability for neutrinos
passing through the Earth. Superimposed on these two lower panels 
are the locations of three neutrino branches 
($pp$ -- dotted, $^7$Be -- dashed, and $^8$B -- solid) for the SMA, 
LMA, LOW, VAC, and quasi-VAC solutions.}
\label{fig:prob}
\end{figure}

To calculate the solar neutrino interaction rate on Earth, three
steps are required:
(i) the probability $p_1$ ($p_2$) of a solar neutrino, which is born
as $\nu_e$ in the core of the Sun, to emerge at the surface as $\nu_1$
($\nu_2$), (ii) coherent or incoherent propagation of the $\nu_1,\nu_2$
admixture to the surface of the Earth,
(iii) the probability $p_{1e}$ ($p_{2e}$) for a $\nu_1$ ($\nu_2$) neutrino to
appear as $\nu_e$ in the detector (after propagation through part of the Earth,
if the Sun is below the horizon, using the PREM~\cite{prem}
density profile of the Earth as shown in Figure~\ref{fig:density}).
In and above the MSW range of $\Delta m^2/E$, the distance between the Sun and
the Earth is much larger than the vacuum oscillation length, so the
propagation (ii) can be assumed to be incoherent.
In that case, the total survival probability of the $\nu_e$ flavor is
\[
p_e=p_1\times p_{1e}+p_2\times p_{2e}=2p_1p_{1e}+1-p_1-p_{1e}
\]
where $p_1$ and $p_{1e}$ are computed numerically.
Below the MSW range, both the solar and terrestrial matter densities can
be neglected. However, the distance L between the Sun and the Earth
approaches the oscillation length, so (ii) must be done coherently,
and the survival probability of the $\nu_e$ flavor is 
\[
p_e=1-p_{e\longrightarrow x}=1-\sin^22\theta\sin^2
\left(\pi\frac{L}{L_{\mbox{\small osc}}}\right)
\]
using Eq.~(\ref{eq:osc}). Figure~\ref{fig:prob} shows (as an
example) the survival probability of $^8$B neutrinos. Other solar neutrino
branches may have slightly different probabilities, because the radial
distributions of the neutrino production location differ.

\begin{figure}[bt]
\includegraphics[width=8.8cm]{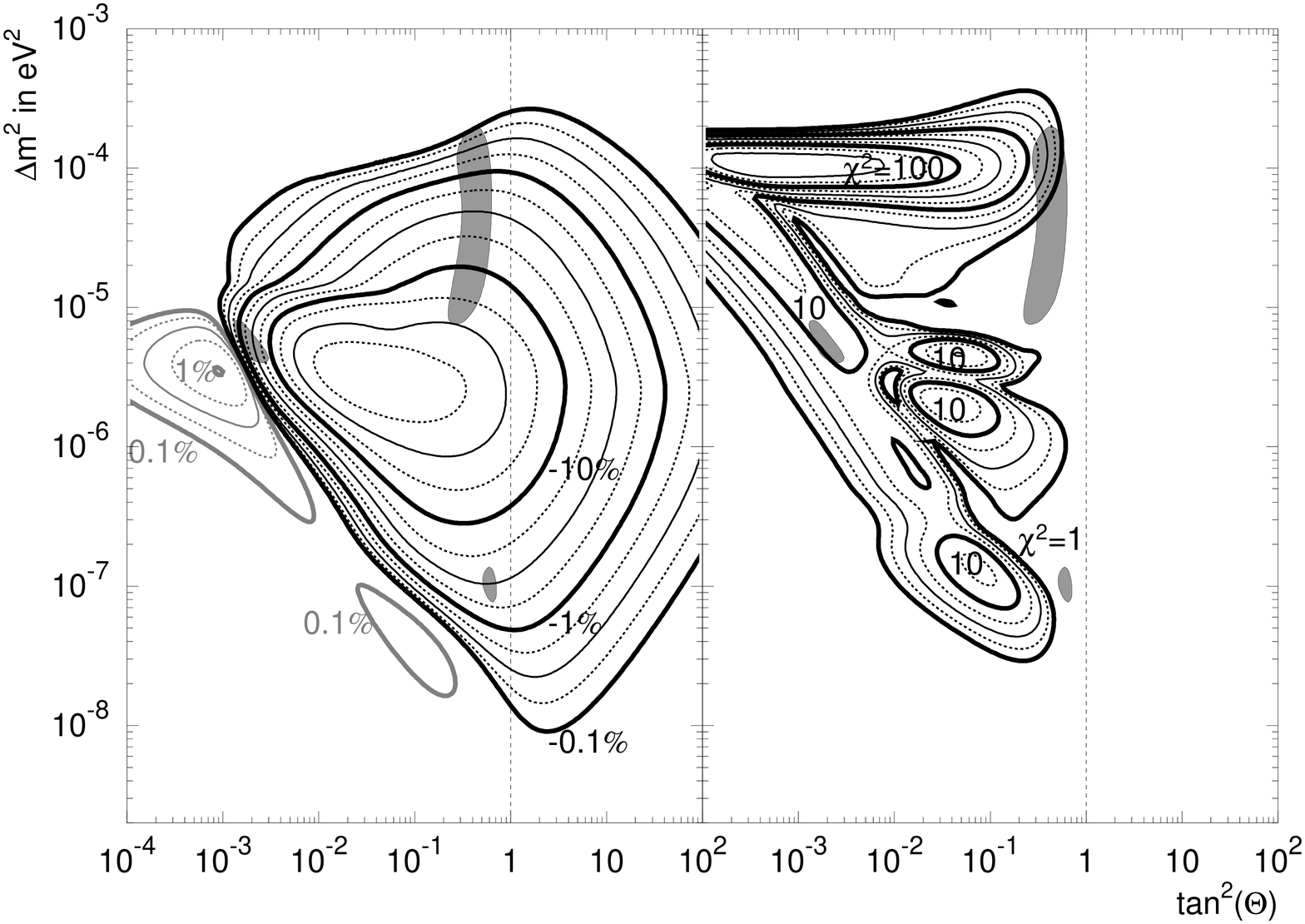}
\caption{Expected SK day/night asymmetry 
(left) and spectral distortion (right) in the MSW region. The scale of the
contours of equal asymmetry (distortion) is logarithmic.
The spectral distortion is measured in terms of a $\chi^2$
using the SK spectrum uncertainties (but not the values of
the SK spectrum).
The gray-shaded areas correspond to the LMA, SMA, and LOW
solutions.}
\label{fig:oscexpectmsw}
\end{figure}

\begin{figure}[bt]
\includegraphics[width=8.8cm]{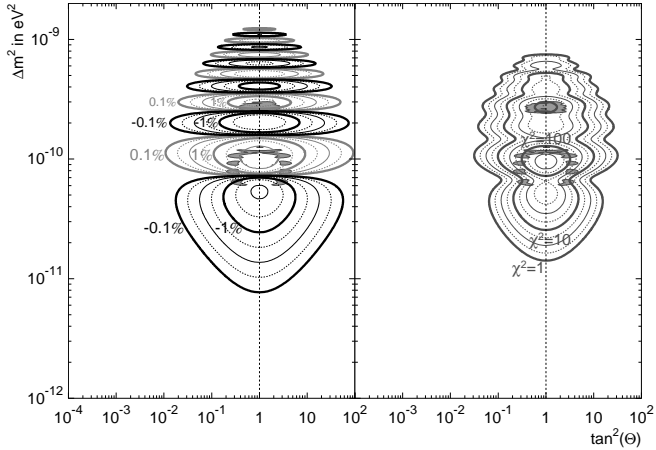}
\caption{Expected SK summer/winter asymmetry
(left) and spectral distortion (right) in the vacuum oscillation
region. The scale of the
contours of equal asymmetry (distortion) is logarithmic.
The spectral distortion is measured in terms of a $\chi^2$
using the SK spectrum uncertainties (but not the values of
the SK spectrum).
The gray-shaded areas correspond to the VAC
solutions.}
\label{fig:oscexpectvo}
\end{figure}

Neutrino oscillations impact SK data in three independent ways:
(i) Since electron neutrinos have a much larger elastic scattering
cross section than other flavors, neutrino oscillations
reduce the rate of solar neutrino interactions. (ii) The spectrum
of recoil electrons is distorted due to the energy dependence of
the survival probability. (iii) The influence of the Earth's matter
on the survival probability induces an apparent time dependence
of the solar neutrino interaction rate with a 24 hour period. The amplitude
of that time dependence is expressed in the day/night asymmetry
$\frac{D-N}{0.5(D+N)}$ where $D$ ($N$) is the averaged interaction
rate during the day (night). Due to the eccentricity of the Earth's orbit
the distance between Sun and Earth changes periodically.
Since the survival probability depends on this distance (if the
neutrinos propagate coherently), this leads to another time-variation
(with a 365 day period) expressed in the summer/winter asymmetry 
$\frac{S-W}{0.5(S+W)}$ where $S$ ($W$) is the averaged interaction
rate during the summer (winter), corrected for the $1/r^2$ dependence
of the solar neutrino intensity. The day/night and summer/winter
variations cannot both be present in the SK data, since above
a $\dmsq$ of about $10^{-8}$~eV$^2$ neutrinos propagate incoherently and below
$10^{-8}$ eV$^2$ the Earth's matter effects on the survival probability
are negligible. Figure~\ref{fig:oscexpectmsw} depicts
the expected SK day/night asymmetry and Figure~\ref{fig:oscexpectvo} the
summer/winter asymmetry, depending on
the oscillation parameters. They also show
the SK sensitivity to distortions of the recoil electron spectrum.

Using the survival probability $p_e$,
the neutrino interaction rate due to the neutrino
spectrum $I(E_\nu)$ at SK is
\[
r_{\mbox{\tiny osc}}=
\int_{E_0}^{E_1}\hspace*{-0.3cm}dE
\int_{E_\nu}\hspace*{-0.2cm}dE_\nu I(E_\nu)
\int_{E_e}\hspace*{-0.2cm}dE_e
R(E_e,E) (S_e(E_\nu,E_e)p_e+\hspace*{1cm}
\]
\vspace*{-0.5cm}
\[
\hspace*{6cm}S_x(E_\nu,E_e)(1-p_e))
\]
where $S_{e,x}(E_\nu,E_e)$ describe the probability that the
elastic scattering of a $\nu_{e,x}$ of energy $E_\nu$ with electrons 
produces a recoil electron of energy $E_e$. The Super--Kamiokande
detector response is given by $R(E_e,E)$, which describes the probability 
that a recoil electron of energy $E_e$ is reconstructed with energy
$E$. The rate $r=r_{\mbox{\tiny noosc}}$ expected without
oscillation is obtained by setting $p_e$ to 1.
Since $p_e$ is different for each neutrino species
(pp, $^8$B, {\it hep}, etc.),
the calculation must be repeated for each relevant species. Due to its
recoil electron energy threshold of 5.0~MeV, Super--Kamiokande--I is only 
sensitive to $^8$B and hep neutrinos. 
The oscillation analysis was performed by two methods:
the first method subdivides the data sample in both
recoil electron energy and solar zenith angle [Zenith Spectrum], while the
second method uses only energy bins and searches for
time variations by means of an unbinned likelihood
[Unbinned Time Variation].

\begin{figure}[tbp]
\includegraphics[width=8.8cm]{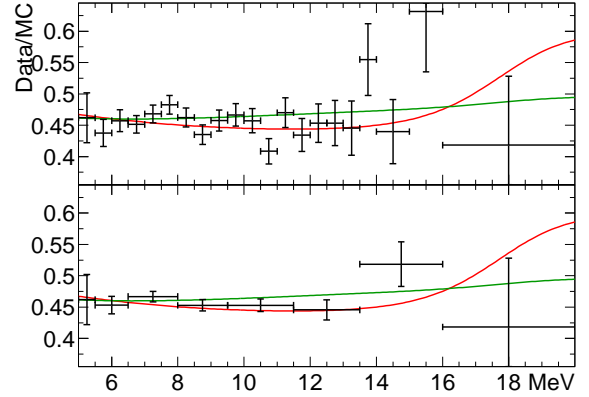}
\caption{Distortion of the recoil electron spectrum
using 21 (upper panel) and 8 (lower panel) bins.
The measured event rates in each bin are divided
by the event rates expected from MC assuming
a $^8$B $\nu$ flux of $5.076\times10^6/$cm$^2$s and
a hep $\nu$ flux of 32$\times10^6/$cm$^2$s.
Overlaid is a typical LMA solution
(red) and a LOW solution (green), where the assumed neutrino
fluxes were fit to best describe the data.}
\label{fig:oscspectrum}
\end{figure}

\subsection{Analysis of the zenith spectrum $\chi^2$}

\begin{figure}[tbp]
\includegraphics[width=8.8cm]{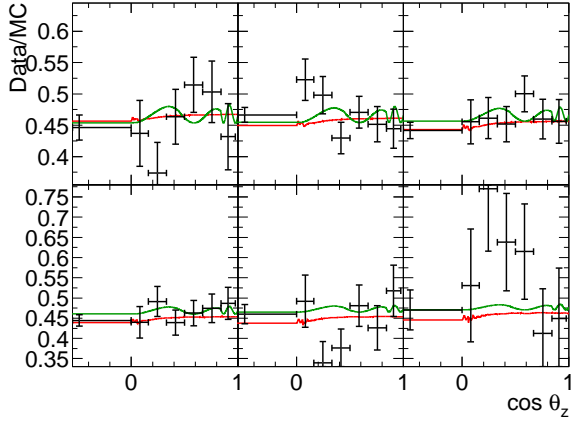}
\caption{Dependence of Data/MC
on solar zenith angle.
The MC assumes a $^8$B $\nu$ flux of $5.076\times10^6/$cm$^2$s and
a hep $\nu$ flux of 32$\times10^6/$cm$^2$s.
The six panels correspond to the energy ranges
5.5-6.5~MeV, 6.5-8.0~MeV, 8.0-9.5~MeV, 9.5-11.5~MeV,
11.5-13.5~MeV, and 13.5-16.0~MeV.
Overlaid is a typical LMA solution
(red) and a LOW solution (green), where the assumed neutrino
fluxes were fit to best describe the data.}
\label{fig:osczenith}
\end{figure}

If $B^{\mbox{\tiny osc}}_{i,z}$
($H^{\mbox{\tiny osc}}_{i,z}$) denote the expected
$^8$B (hep) neutrino induced rate in energy bin $i$ and
zenith angle bin $z$ and $D_{i,z}$ the measured rate, the quantities
$b$, $h$ and $d$ are defined as follows:
\[
b_{i,z}=\frac{B^{\mbox{\tiny osc}}_{i,z}}{B_{i,z}+H_{i,z}}
\mbox{ , }
h_{i,z}=\frac{H^{\mbox{\tiny osc}}_{i,z}}{B_{i,z}+H_{i,z}}
\]\[
\hspace*{5mm}\mbox{and}\hspace*{5mm}
d_{i,z}=\frac{D_{i,z}}{B_{i,z}+H_{i,z}}.
\]
For $d$, the rate calculations in the denominator use
a full Monte Carlo simulation
of the Super--Kamiokande detector.
Figure~\ref{fig:oscspectrum} shows the average
$d$ in 21 and 8 energy bins. In Figure~\ref{fig:osczenith},
the individual $d_{i,z}$ are shown for 6 solar
zenith angle bins.
Now all zenith angle bins are combined into vectors.
The rate difference vector
\[
\overrightarrow{\Delta_i}(\beta,\eta)=
\left(\beta\cdot \overrightarrow{b_i} +
      \eta\cdot  \overrightarrow{h_i}\right)
\times f(E_i,\delta_B,\delta_S,\delta_R)-\overrightarrow{d_i}
\]
allows for arbitrary total neutrino fluxes through the free
parameters $\beta$ and $\eta$.
The combined rate predictions
$\beta \overrightarrow{b_i}+\eta \overrightarrow{h_i}$
are modified by the energy-shape factors
\[
f(E_i,\delta_B,\delta_S,\delta_R)=
f_B(E_i,\delta_B)\times f_S(E_i,\delta_S)\times f_R(E_i,\delta_R)
\]
with $\delta_B$ describing the $^8$B neutrino spectrum shape uncertainty,
$\delta_S$ describing the uncertainty of the SK energy scale (0.64\%) and
$\delta_R$ describing the uncertainty of the SK energy resolution (2.5\%).
The $n\times n$ matrices $V_i$ ($n$ is the number of zenith angle bins)
describe statistical and energy-uncorrelated systematic uncertainties.
Those systematic uncertainties are assumed to be fully correlated in
zenith angle. 

To construct the $V_i$, we transform the rate vectors from the
zenith angle basis to a new basis: the first component gives the
average rate, the second component the day/night difference, the
third component the difference between the rate of the mantle 1 bin
and the other night bins, etc. This is achieved by multiplying
$\overrightarrow{\Delta_i}$ by the non-singular matrix
$S$ which depends on the livetimes of each zenith angle bin:
$\overrightarrow{\Delta_i}'=S\cdot\overrightarrow{\Delta_i}$
\[
S=\begin{pmatrix}
L_1/L_{1\rightarrow n} & L_2/L_{1\rightarrow n} &\cdots&  L_n/L_{1\rightarrow n} \cr
     -1                 & L_2/L_{2\rightarrow n} &\cdots&  L_n/L_{2\rightarrow n} \cr
      0                 &      -1                &\cdots&  L_n/L_{3\rightarrow n} \cr
   \vdots               &    \vdots              &\cdots&          \vdots         \cr
      0                 &    \cdots              & -1   &             1         \end{pmatrix}
\]
where $L_z$ is the livetime of zenith angle bin $z$ and 
$L_{z\rightarrow n}=\sum_{z'=z}^{n}L_{z'}$. The $\chi^2$ quadratic form 
$\overrightarrow{\Delta_i}^T\cdot V_i^{-1}\cdot \overrightarrow{\Delta_i}$
transforms as $V_i'=S\cdot V_i\cdot S^T$
($(S\cdot V_i\cdot S^T)^{-1}=S^{T-1}\cdot V_i^{-1}\cdot S^{-1}$).
Since in this new basis, the first component
of $\overrightarrow{\Delta_i}'$ is the livetime-averaged rate difference
of energy bin $i$, the energy-uncorrelated systematic uncertainties
$\sigma^2_{i,u}$ (assumed to be fully correlated in zenith angle) can
simply be added to the statistical uncertainties
$S\cdot V_{i,\mbox{\small stat}}\cdot S^T$:
\[
S\cdot V_i\cdot S^T=S\cdot V_{i,\mbox{\small stat}}\cdot S^T
+\begin{pmatrix}
         \sigma_{i,u}^2 & 0    & \cdots & 0 \cr
          0          & 0    & \cdots & 0 \cr
         \vdots      &\vdots& \cdots &\vdots\cr
          0          & 0    & \cdots & 0
\end{pmatrix},
\]
Of course, $V_{i,\mbox{\small stat}}$ is diagonal.
If $\sigma_{i,u}$ is asymmetric, two error matrices $V_{i+}$ and
$V_{i-}$ are constructed. The sign of $\Delta_{i,1}^\prime$ 
(livetime-averaged rate difference) decides, if
$\overrightarrow{\Delta_i}^T\cdot V_{i+}^{-1}\cdot\overrightarrow{\Delta_i}$
or
$\overrightarrow{\Delta_i}^T\cdot V_{i-}^{-1}\cdot\overrightarrow{\Delta_i}$
contributes to the $\chi^2$
\begin{equation}
\chi^2_0=
\sum_{i=1}^{m}
\overrightarrow{\Delta_i}^T\cdot V_i^{-1}\cdot\overrightarrow{\Delta_i}
\label{eqn:zenspchi2}
\end{equation}

For any given parameters $\delta_k$, $\chi^2_0$
is just a quadratic form in the neutrino flux factors
$\beta$ and $\eta$ and can be written as
\[
\chi^2_{0}=\chi^2_m+\overrightarrow{\Phi}^TC_0\overrightarrow{\Phi}
\hspace*{5mm}\mbox{with}\hspace*{5mm}\
\phi=\begin{pmatrix}
\beta-\beta_{\mbox{\small min}} \cr \eta-\eta_{\mbox{\small min}}
\end{pmatrix}
\]
where the $2\times2$ curvature matrix $C_0$ is the inverse of the
covariance matrix for $\beta$ and $\eta$. This matrix
$C_0$ must therefore be inversely proportional to the
combined uncertainty of all data bins:
\[
\frac{1}{\sigma_0^2}=\sum_{i=1}^{m}
\frac{1}{\sigma_{\mbox{\small stat},i}^2+\sigma_{i,u}^2}
\hspace*{5mm}\mbox{with}\hspace*{5mm}\
\frac{1}{\sigma_{\mbox{\small stat},i}^2}=\sum_{z=1}^{n}\frac{1}{\sigma_{\mbox{\small stat},i,z}^2}
\]
However, the energy-uncorrelated systematic uncertainties
$\sigma_{i,u}$ do not reflect the total uncertainty of the
SK rate. For example, the uncertainty in the
fiducial volume due to a systematic vertex shift cancels
for the spectrum shape and is therefore not included in
$\sigma_{i,u}$. If the spectrum data
is used to constrain the total rate, the total uncertainty
of that rate ($\sigma_0$) therefore neglects that part (called $\sigma_r$)
of the systematic uncertainty which cancels for the spectrum.
While $\sigma_r$ is fully correlated in both energy and zenith angle,
it is not covered by the energy-correlated uncertainties either,
which reflect only the uncertainties in the $^8$B neutrino spectrum,
the SK energy scale, and the SK energy resolution.
To take $\sigma_r$ into account $\chi^2_0$ is modified to
\[
\chi^2_1=\chi^2_m+\overrightarrow{\Phi}^TC_1\overrightarrow{\Phi}
\hspace*{5mm}\mbox{with}\hspace*{5mm}\
C_1=\frac{\sigma_0^2}{\sigma_0^2+\sigma_r^2}\times C_0
\]
which has the same minimum as $\chi^2_0$ but allows an enlarged
range of parameters $\beta, \eta$.
The total $\chi^2$ for the SK zenith spectrum shape is then
\begin{equation}
\begin{matrix}
\chi^2_{\mbox{\tiny SK}}=
\mbox{Min}(& \chi^2_1(\beta,\eta,\delta_B,\delta_S,\delta_R)
+\cr
&
\left(\frac{\delta_B}{\sigma_B}\right)^2
+\left(\frac{\delta_S}{\sigma_S}\right)^2
+\left(\frac{\delta_R}{\sigma_R}\right)^2
)
\end{matrix}
\label{eqn:skchi2}
\end{equation}
where all $\delta_k$ as well as $\beta, \eta$ are minimized.
To constrain the $^8$B flux, the term
$\left(\frac{\beta-1}
{\sigma_f}\right)^2$
is added.

\begin{figure}[tbp]
\includegraphics[width=8.8cm]{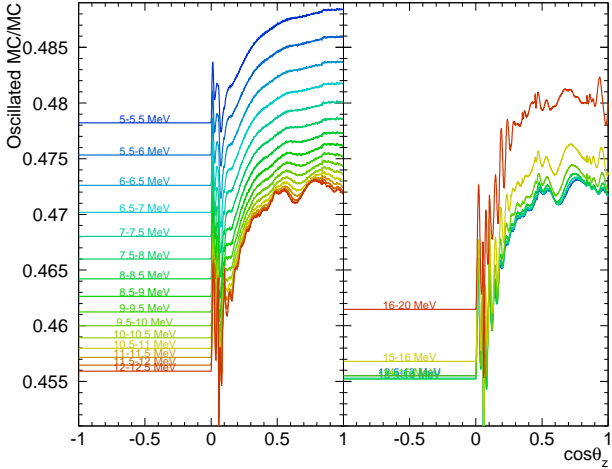}
\caption{Predicted solar zenith angle variation assuming a typical LMA
solution, $\dmsq=6.3\times10^{-5}$~eV$^2$ and $\tasq=0.52$ (which is the 
best fit to the SK energy spectrum, rate, and day/night variation).}
\label{fig:lmashape}
\end{figure}

\subsection{Unbinned time-variation analysis}

In a different approach to combine time-variation and spectrum
constraints we form the total likelihood $\cal L$
modifying the solar signal factors $s_{ij}$
of Eq.~(\ref{eqn:likelihood}) to
\begin{equation}
s_{ij}=p(\cos\theta_{ij},E_j)\times \frac{r_i(t_{j})}{r^{\mbox{\tiny av}}_i}
\label{eqn:likelihood2}
\end{equation}
where $t_j$ is the event time, 
$r_i(t)$ the predicted time dependence of the solar neutrino
interaction rate in energy bin $i$,
and $r^{\mbox{\tiny av}}_i$ the predicted time-averaged rate.
Figure~\ref{fig:lmashape}
shows the expected solar zenith angle dependence
$r_i(\cos\theta_z)$ in each energy bin for a typical LMA solution.
The recoil electron spectrum enters this likelihood through
the weight factors $Y_i$ of Eq.~(\ref{eqn:likelihood}), so the likelihood
needs to add terms like those in Eq.~(\ref{eqn:skchi2}) to
account for energy-correlated systematic uncertainties.
The definition of energy bins is the same as in Eq.~(\ref{eqn:likelihood}):
This analysis uses 21 energy bins compared to 8 energy bins
for the zenith spectrum.

\begin{figure}[tbp]
\centerline{\includegraphics[width=8.8cm]{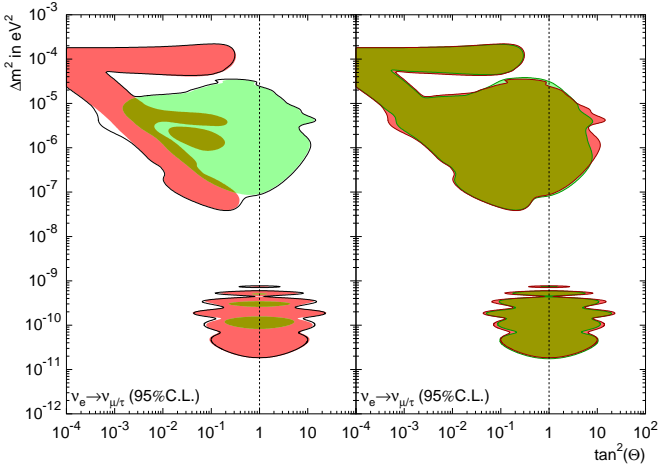}}
\caption{Left: excluded regions from the SK spectrum (red),
SK rate time variations (green) and both combined (solid line)
at 95\% C.L..
Right: excluded regions from the SK zenith spectrum (green) 
compared to the SK spectrum combined with the time variation likelihood 
(red) at 95\%C.L..}
\label{fig:skoscfree}
\end{figure}

Due to the large number of solar neutrino candidates, the maximization
of this likelihood is too slow to be practical. We therefore
split it into
\[
\log {\cal L}=(\log {\cal L}-\log {\cal L}_{\mbox{\tiny av}})+
\log {\cal L}_{\mbox{\tiny av}}
\label{eqn:likelihood5}
\]
where ${\cal L}_{\mbox{\tiny av}}$ is the likelihood without the time
variations, which means that it depends only on the average rates in each
recoil electron energy bin. We can cast this in terms of a
$\chi^2=-2\log {\cal L}$:
\begin{equation}
\chi^2=\Delta\chi^2_{\mbox{\tiny tv}}+\chi^2_{\mbox{\tiny av}}
\hspace*{0.2cm}
\mbox{with }
\Delta\chi^2_{\mbox{\tiny tv}}=
-2(\log {\cal L}-\log {\cal L}_{\mbox{\tiny av}})
\label{eqn:likelihood4}
\end{equation}
The first term of $\Delta\chi^2_{\mbox{\tiny tv}}$
uses the time-dependent solar signal factors
of Eq.~(\ref{eqn:likelihood2}) while the second term is the
same as Eq.~(\ref{eqn:likelihood}).
The spectrum weights $Y_i$ occurring in that equation are determined
from the oscillated predicted spectrum.
This predicted spectrum is formed using
the $^8$B and hep neutrino fluxes 
as well as the systematic uncertainty parameters 
$\delta_B$, $\sigma_B$ and $\delta_S$.
The values of these five parameters result
from a fit to the (time-averaged) SK spectrum data.
The dark-gray areas in Figure~\ref{fig:skoscfree} are excluded at 95\% C.L.
by this $\Delta\chi^2_{\mbox{\tiny tv}}$.

All systematic uncertainties are assumed to be fully correlated in
zenith angle or any other time variation variable.
Since the sensitivity is dominated by statistical uncertainties,
this assumption is not a serious limitation.
We found the statistical likelihood in each energy bin $i$ to be very close to
a simple Gaussian form, so $\chi^2_{\mbox{\tiny av}}$ takes the same form
as Eq.~(\ref{eqn:skchi2}) but replacing Eq.~(\ref{eqn:zenspchi2})
with
\begin{equation}
\chi^2_0=\sum_{i=1}^{N_{\mbox{\tiny bin}}}
\left(\frac{\Delta_i}{\sigma_i}\right)^2
\label{eqn:likelihood6}
\end{equation}
where $\Delta_i$ has only one zenith angle bin
(between $-1<\cos\theta_z<1$)
\[
\Delta_i(\beta,\eta)=
\left(\beta\cdot b_i+\eta\cdot  h_i\right)
\times f(E_i,\delta_B,\delta_S,\delta_R)-d_i
\]
and $\sigma_i^2=\sigma_{i,\mbox{\tiny stat}}^2+\sigma_{i,u}^2$
is the total energy-uncorrelated uncertainty in that bin.
The regions colored in light-gray of Figure~\ref{fig:skoscfree} are
excluded at 95\% C.L. using $\chi^2_{\mbox{\tiny av}}$ without the
term constraining the $^8$B neutrino flux.

\begin{figure}[tbp]
\includegraphics[width=8.8cm]{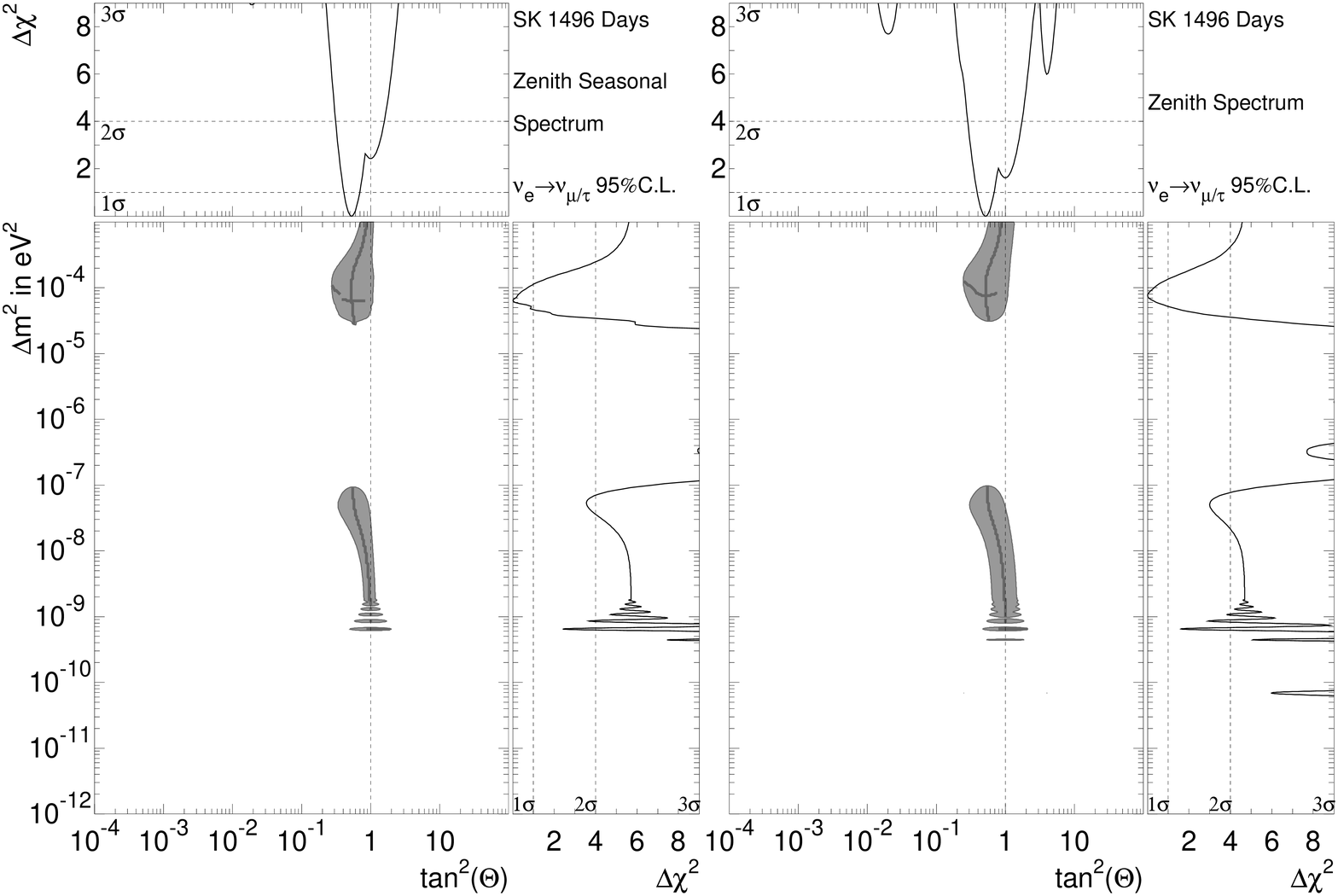}
\caption{Allowed areas from the rate-constrained
SK unbinned time-variation analysis
(left) and the SK zenith spectrum (right)
at 95\% C.L..
The graphs at the top (and right) show the $\chi^2$ difference
as a function of $\tasq$ ($\dmsq$) alone where
the $\dmsq$ ($\tasq$) is chosen to minimize $\chi^2$. The lines
inside the contours show which $\dmsq$ ($\tasq$) is chosen.}
\label{fig:skosc}
\end{figure}
\begin{figure*}[tbp]
\centerline{(a)\hspace*{8.6cm}(b)\hspace*{8.6cm}}
\centerline{\includegraphics[width=8.8cm]{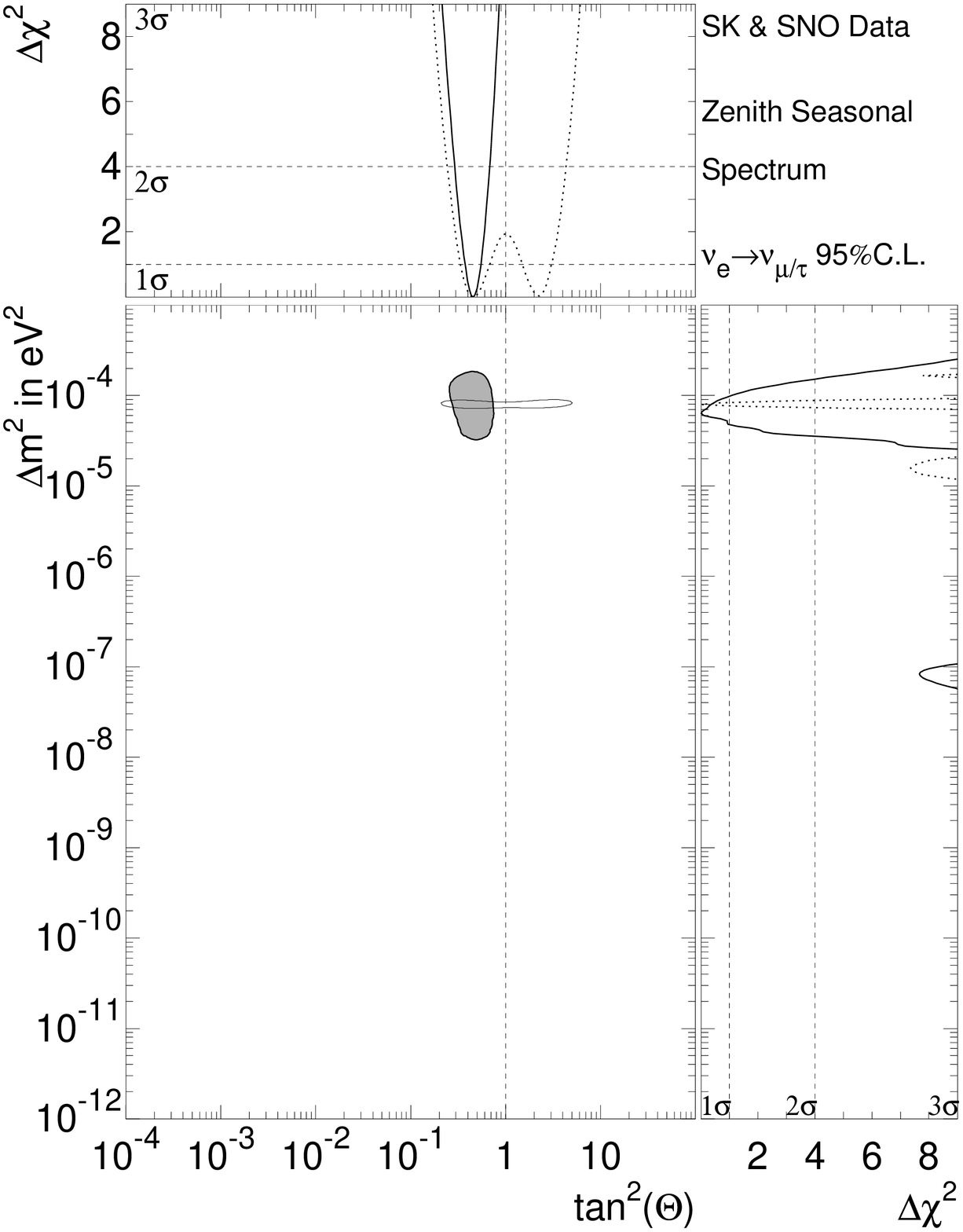}\hspace*{0.5cm}
            \includegraphics[width=8.8cm]{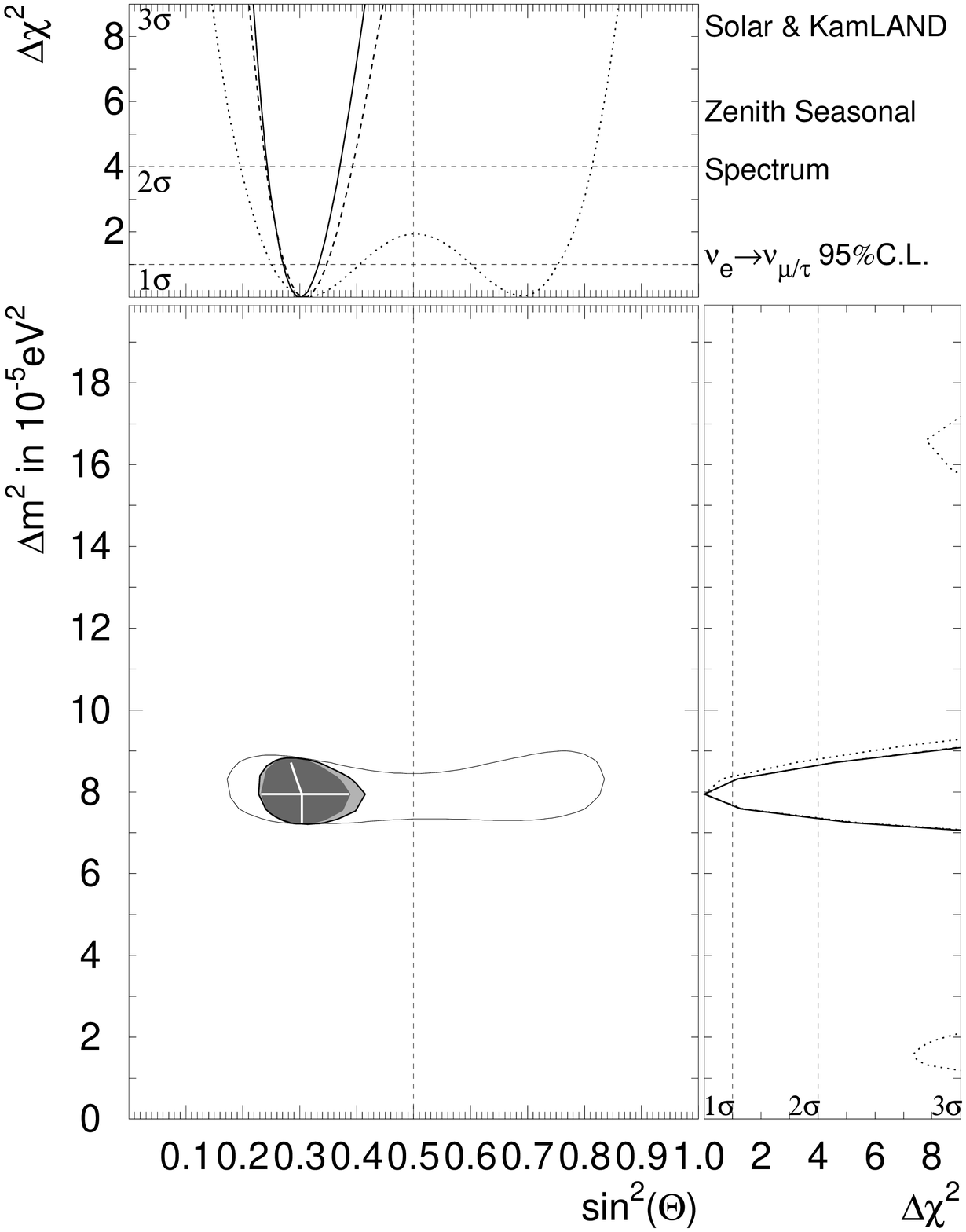}}
\caption{(a) Allowed area at 95\% C.L from the combination of
SK and SNO.
The graphs at the top (and right) show the $\chi^2$ difference
as a function of $\tasq$ ($\dmsq$) only: the solid line is
the SK/SNO fit, the dotted line results from KamLAND~\cite{kamland}
data.
The lines inside the contours show which $\dmsq$ ($\tasq$) is chosen.
(b) Allowed area at 95\% C.L. from the combination of SK/SNO (gray)
and all solar data (dark gray), and from the combination of solar and
KamLAND data (light gray).
Note the linear scale of both axes. Only LMA--I remains allowed.
Overlaid is the region allowed by KamLAND alone .}
\label{fig:globosc}
\end{figure*}

\subsection{Oscillation constraints from SK}

The combined excluded areas
(from $\chi^2=\chi^2_{\mbox{\tiny av}}+\Delta\chi^2_{\mbox{\tiny tv}}$)
at 95\% C.L. are shown inside the solid line in the left panel of 
Figure~\ref{fig:skoscfree}. The right panel of the same figure
compares these areas to the excluded areas from the zenith spectrum
analysis. Both contours are quite similar, however the unbinned time-variation
analysis has more stringent limits on the time-variation
(note the region $10^{-6}$~eV$^2<\dmsq<10^{-5}$~eV$^2$ at
$\tasq\approx10^{-3}$ and $\tasq>1$). Due to the larger number of energy
bins, the likelihood analysis has also larger excluded areas in the
vacuum region and at $\dmsq\approx10^{-4}$~eV$^2$.
If we include the last term in Eq.~(\ref{eqn:likelihood6}),
we get the allowed regions (95\%C.L.) shown in Figure~\ref{fig:skosc}
which depend on the total $^8$B neutrino flux measurement of
SNO~\cite{snosalt}.
The best fit is in the LMA region at $\tasq=0.52$
and $\dmsq=6.3\times10^{-5}$~eV$^2$ ($\dmsq=7.6\times10^{-5}$~eV$^2$)
for the unbinned time-variation analysis (zenith spectrum analysis) 
where a day/night asymmetry of -2.1\%(-1.5\%) is expected and
$-1.8\pm1.6\%$ ($-1.7\pm1.6\%$) is found. (Here, the uncertainties do not
include systematic effects).
The $\chi^2$ is 17.3 for 20 degrees of freedom (63\% C.L.).
The $^8$B flux is fit to $4.91\times10^6/$cm$^2$s.
The $\chi^2_{\mbox{\tiny av}}$ of the SK spectrum and rate is 18.5 for
20 degrees of freedom (55\% C.L.).
The $\chi^2$ analysis of the zenith spectrum gives a
minimum $\chi^2$ of 39.0 with 43 degrees of freedom (65\% C.L.).
The $^8$B flux is fit to $4.86\times10^6/$cm$^2$s.
Figure~\ref{fig:skosc} also shows the
$\chi^2$ as a function of $\tasq$ alone where a $\dmsq$ is chosen
for each $\tasq$ to minimize $\chi^2$:
SK data excludes small mixing at more than $3\sigma$.
SK data also disfavors $\dmsq>10^{-3}$eV$^2$ and
$2\times10^{-9}$eV$^2<\dmsq<3\times10^{-5}$eV$^2$
(see the plot in the right panel, where $\chi^2$
is minimized with respect to $\tasq$). Again, the unbinned time-variation
analysis has more stringent oscillation constraints
and favors the LMA region more strongly.

\subsection{Combined oscillation constraints from several experiments}
Stronger constraints on $\dmsq$ result from the combination of the
SK measurements with other solar neutrino data.
The 95\% C.L. allowed region of Figure~\ref{fig:globosc}(a) 
include in addition to the SK data
the SNO measurements of the charged-current
interaction rate (and day/night asymmetry)
of solar electron neutrinos with deuterons~\cite{snoncdn};
only LMA solutions survive. Overlaid are the contours allowed
by the KamLAND reactor neutrino spectrum~\cite{kamland}.
Figure~\ref{fig:globosc}(b) shows the allowed contours
of a combined fit to SK, SNO and KamLAND data. SK and SNO
remove the ambiguities in $\dmsq$ and $\tasq$ of KamLAND 
and tighten the constraint on the mixing angle.
When the charged-current rates measured by Homestake, GALLEX,
and SAGE~\cite{othersolar} are included as well, the allowed
LMA solutions are further
reduced. However, in this case the fit relies on the SSM predictions of the
pp, pep, CNO and $^7$Be neutrino fluxes.

\begin{figure}[tbp]
\includegraphics[width=8.cm]{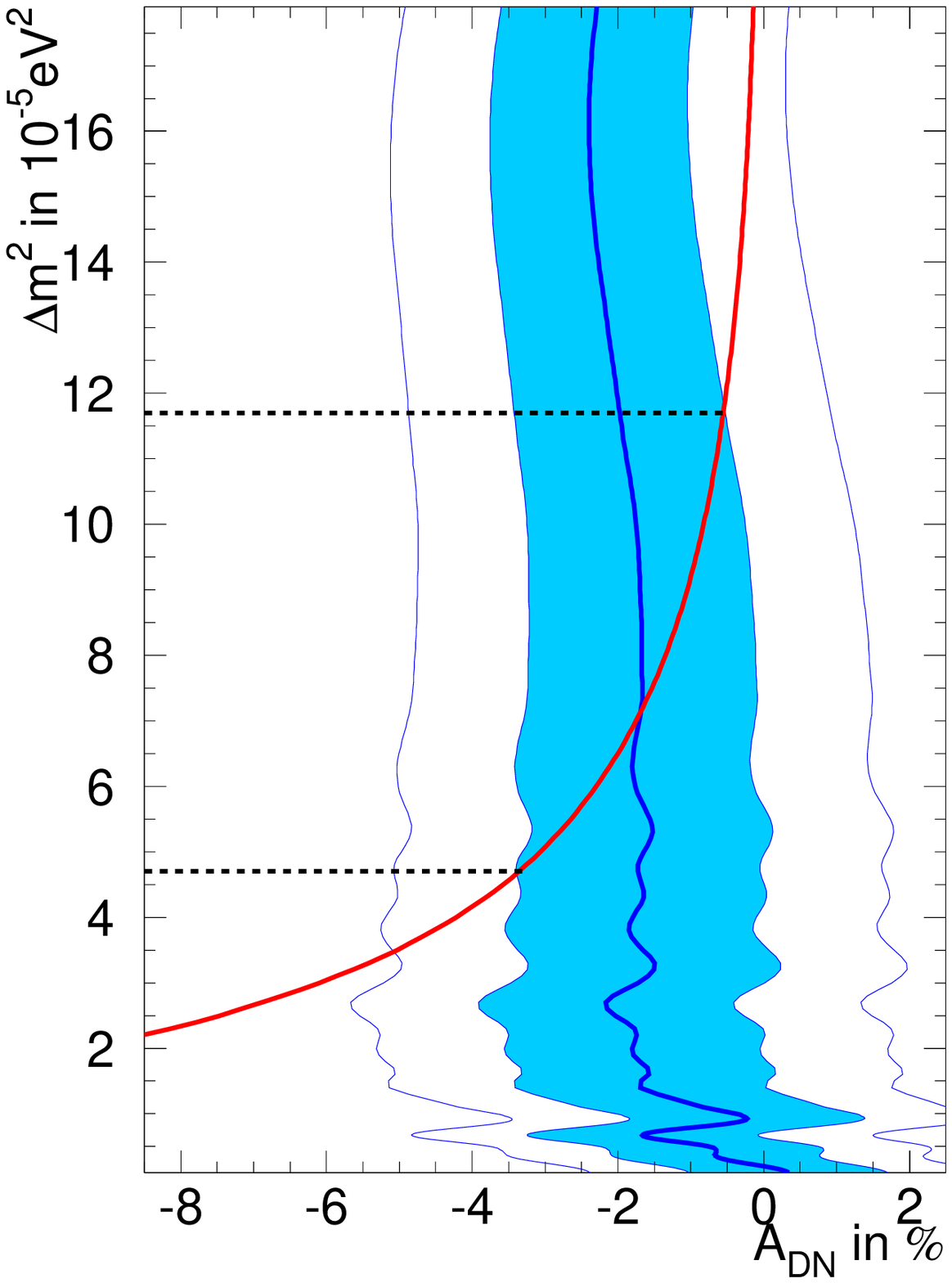}
\caption{Day/Night amplitude fit depending on
$\dmsq$. The fit depends only very weakly on
the mixing angle which is assumed to be $\tasq=0.44$.
Overlaid (red line) is the expected amplitude
and (blue lines) the $2\sigma$ boundaries of
the amplitude fit.}
\label{fig:oscdnplot}
\end{figure}

\subsection{Solar day/night effect}
Even though KamLAND provides by far the best constraint
on the solar $\dmsq$ it is interesting to study the solar
constraints as well. The upper bound arises from the
ratios of the electron--neutrino elastic scattering rate
in SK, the charged-current neutrino interactions
in SNO and Homestake, and the neutral-current neutrino
interactions in SNO. The lower bound is due to the solar
neutrino day/night effect. As described in~\cite{dnpaper},
we fit the amplitude of this variation to our data and compare
it to the expected amplitude.
For this fit, only the amplitude of the
day/night variation is varied, the {\it shape} is fixed
to the calculated shape for a particular $\dmsq$
and $\tasq$ (see Figure~\ref{fig:lmashape} for a
typical LMA solution). Demanding consistency between expected and
observed amplitude, we obtain a range of
$5\times10^{-5}$~eV$^2<\dmsq<12\times10^{-5}$~eV$^2$ at the
$1\sigma$ level and a lower bound of $3\times10^{-5}$~eV$^2<\dmsq$
at the $2\sigma$ level ($\tasq=0.44$). This agrees very
well with the KamLAND measurement of $7.9\times10^{-5}$~eV$^2$.
Conversely, if we confine $\dmsq$ to the $3\sigma$ range
$7\times10^{-5}$~eV$^2<\dmsq<9\times10^{-5}$~eV$^2$ allowed
by KamLAND, the amplitude fit varies only very slightly.
In that range the measured SK day/night variation amplitude
corresponds to a day/night asymmetry of
-1.7\%$\pm$1.6\%(stat)$^{+1.3}_{-1.2}$(syst)$\pm$0.04\%($\dmsq$)
while the expected asymmetry ranges from -1.7\% to -1.0\%.
At the SK best-fit $\dmsq=6.3\times10^{-5}$~eV$^2$, the day/night
amplitude fit corresponds to the asymmetry
-1.8\%$\pm$1.6\%(stat)$^{+1.3}_{-1.2}$(syst).


%
%
\section{Conclusion}
Solar neutrino measurements using neutrino-electron
scattering in the Super--Kamiokande detector are described.
We obtained 1496 effective days of data in the time
period of May 31st 1996 through July 15th 2001. The analysis threshold was 
6.5~MeV for the first 280 days, and 5.0~MeV for the remaining 1216 days.
The observed interaction rate corresponds to
a $^{8}$B solar electron-neutrino flux of
$2.35 \pm 0.02 \pm 0.08 \times 10^{6}$cm$^{-2}$sec$^{-1}$.
We searched for periodic time variations of this rate and found
only the expected seasonal variation caused by the eccentricity
of the earth's orbit. The energy spectrum of the recoiling electron is
consistent with an undistorted solar $^{8}$B neutrino spectrum.
Based on these results the solar neutrino oscillation analysis
imposes strong constraints on the oscillation parameters, selecting
large mixing and favoring the Large Mixing Angle solution. The combination
with data from other experiments confirms the Large Mixing Angle solution
and further reduces the uncertainty in the oscillation parameters.

\begin{acknowledgments}
The authors acknowledge the cooperation of the Kamioka Mining and 
Smelting Company.
Super--Kamiokande has been built and operated from funding 
by the Japanese Ministry of Education, Culture, Sports, Science and
Technology, and the United States Department of Energy, and the U.S. National
Science Foundation. Some of us have been supported by funds from the Korean
Research Foundation (BK21) and the Korea Science and Engineering Foundation,
the State Committee for Scientific Research in Poland (grants 1P03B08227 and
1P03B03826), and Japan Society for the Promotion of Science.


\end{acknowledgments}
%
%

\end{document}